\documentclass[twocolumn]{aastex631}
\usepackage{array} % required for text wrapping in tables
\usepackage{amsmath}

\newcolumntype{P}[1]{>{\centering\arraybackslash}p{#1}}

\expandafter\def\expandafter\normalsize\expandafter{%
    \normalsize%
    \setlength\abovedisplayskip{2pt}%
    \setlength\belowdisplayskip{4pt}%
    \setlength\abovedisplayshortskip{4pt}%
    \setlength\belowdisplayshortskip{4pt}%
}
\begin{document}

\title{SHELLQs-JWST: Revealing the Spectra of Extended Emission in 12 $\rm{z}>6$ Quasar Host Galaxies using the JWST NIRSpec Fixed Slit}

 \author[0000-0002-2099-0254]{Camryn L. Phillips}
 \email{cp5619@princeton.edu}
 \affiliation{Department of Astrophysical Sciences, Princeton University, 4 Ivy Lane, Princeton, NJ 08544, USA }
 
 \author[0000-0002-0106-7755]{Michael A. Strauss}
 \affiliation{Department of Astrophysical Sciences, Princeton University, 4 Ivy Lane, Princeton, NJ 08544, USA }

\author[0000-0003-2984-6803]{Masafusa Onoue}
\affiliation{Waseda Institute for Advanced Study (WIAS), Waseda University, 1-21-1, Nishi-Waseda, Shinjuku, Tokyo 169-0051, Japan}
%\affiliation{Center for Data Science, Waseda University, 1-6-1, Nishi-Waseda, Shinjuku, Tokyo 169-0051, Japan}
\affiliation{Kavli Institute for the Physics and Mathematics of the Universe (Kavli IPMU, WPI), UTIAS, Tokyo Institutes for Advanced Study, University of Tokyo, Chiba, 277-8583, Japan}

\author[0000-0001-8917-2148]{Xuheng Ding}
\affiliation{Kavli Institute for the Physics and Mathematics of the Universe (Kavli IPMU, WPI), UTIAS, Tokyo Institutes for Advanced Study, University of Tokyo, Chiba, 277-8583, Japan}
\affiliation{Center for Data-Driven Discovery, Kavli IPMU (WPI), UTIAS, The University of Tokyo, Kashiwa, Chiba 277-8583, Japan}

\author[0000-0002-0000-6977]{John D. Silverman}
\affiliation{Kavli Institute for the Physics and Mathematics of the Universe (Kavli IPMU, WPI), UTIAS, Tokyo Institutes for Advanced Study, University of Tokyo, Chiba, 277-8583, Japan}
\affiliation{Center for Data-Driven Discovery, Kavli IPMU (WPI), UTIAS, The University of Tokyo, Kashiwa, Chiba 277-8583, Japan}
\affiliation{Department of Astronomy, School of Science, The University of Tokyo, 7-3-1 Hongo, Bunkyo-ku, Tokyo, 113-0033, Japan}
\affiliation{Center for Astrophysical Sciences, Department of Physics \& Astronomy, Johns Hopkins University, Baltimore, MD 21218, USA}

 \author[0000-0002-7402-5441]{Yoshiki Matsuoka}
 \affiliation{Research Center for Space and Cosmic Evolution, Ehime University, Matsuyama, Ehime 790-8577, Japan}

 \author[0000-0001-9452-0813]{Takuma Izumi}
 \affiliation{National Astronomical Observatory of Japan, Osawa, Mitaka, Tokyo 181-8588, Japan}
  \affiliation{Department of Astronomy, School of Science, Graduate University for Advanced Studies (SOKENDAI), Mitaka, Tokyo 181-8588, Japan}

\author[0009-0007-0864-7094]{Junya Arita}
\affiliation{Department of Astronomy, School of Science, The University of Tokyo, 7-3-1 Hongo, Bunkyo-ku, Tokyo, 113-0033, Japan}

\author[0000-0003-4569-1098]{Kentaro Aoki}
\affiliation{Subaru Telescope, National Astronomical Observatory of Japan, Hilo, HI 96720}

\author[0000-0002-9850-6290]{Shunsuke Baba}
\affiliation{Institute of Space and Astronautical Science (ISAS), Japan Aerospace Exploration Agency (JAXA), 3-1-1 Yoshinodai, Chuo-ku, Sagamihara, Kanagawa 252-5210, Japan}

\author[0000-0001-6186-8792]{Masatoshi Imanishi}
\affiliation{National Astronomical Observatory of Japan, Osawa, Mitaka, Tokyo 181-8588, Japan}
\affiliation{Department of Astronomy, School of Science, Graduate University for Advanced Studies (SOKENDAI), Mitaka, Tokyo 181-8588, Japan}

\author[0000-0003-3954-4219]{Nobunari Kashikawa}
\affiliation{Department of Astronomy, School of Science, The University of Tokyo, 7-3-1 Hongo, Bunkyo-ku, Tokyo, 113-0033, Japan}
\affiliation{Research Center for the Early Universe, The University of Tokyo, 7-3-1 Hongo, Bunkyo-ku, Tokyo 113-0033, Japan}

\author[0000-0002-3866-9645]{Toshihiro Kawaguchi}
\affiliation{Graduate School of Science and Engineering, University of Toyama, Gofuku 3190, Toyama 930-8555, Japan}

\author[0000-0003-1700-5740]{Chien-Hsiu Lee}
\affiliation{Hobby-Eberly Telescope, McDonald Observatory, UT Austin}

\author[0009-0003-5438-8303]{Mahoshi Sawamura}
\affiliation{Department of Astronomy, School of Science, The University of Tokyo, 7-3-1 Hongo, Bunkyo-ku, Tokyo, 113-0033, Japan}
\affiliation{National Astronomical Observatory of Japan, Osawa, Mitaka, Tokyo 181-8588, Japan}

\author[0000-0002-3531-7863]{Yoshiki Toba}
\affiliation{National Astronomical Observatory of Japan, Osawa, Mitaka, Tokyo 181-8588, Japan}
\affiliation{Department of Physics, Nara Women’s University, Kitauoyanishi-machi, Nara, Nara630-8506, Japan}
\affiliation{Academia Sinica Institute of Astronomy and Astrophysics, 11F of Astronomy-Mathematics Building, AS/NTU, No.1, Section 4, Roosevelt Road, Taipei 10617, Taiwan}
\affiliation{Research Center for Space and Cosmic Evolution, Ehime University, Matsuyama, Ehime 790-8577, Japan}

\author[0000-0002-7633-431X]{Feige Wang}
\affiliation{Steward Observatory, University of Arizona, 933 N Cherry Avenue, Tucson, AZ 85721, USA}

\author[0000-0001-5287-4242]{Jinyi Yang}
\affiliation{Steward Observatory, University of Arizona, 933 N Cherry Avenue, Tucson, AZ 85721, USA}

\begin{abstract}
We present an analysis of the rest frame optical JWST NIRSpec Fixed Slit spectra of extended host galaxy emission in 12 quasars from the Subaru High-z Exploration of Low-Luminosity Quasars (SHELLQs) sample at redshifts $\rm{6.0<z<6.4}$. The spatial point spread function is modeled primarily by a sum of two Gaussians as a function of wavelength and is used to fit and subtract the quasar from the 2D spectra, leaving only extended galaxy emission which we analyze. Ten of 12 systems show spatially extended line emission and five of 12 systems show an extended stellar continuum. From the extended $\rm{[OIII]5008}$ emission line, we measure a $132\pm19\:\rm{km/s}$ ionized outflow in one system and $52\pm12\:\rm{km/s}$ rotation, suggesting a coherent disk, in another. From the extended narrow $\rm{H}\alpha$ emission, which we hypothesize is ionized by star-forming regions rather than the quasar, we measure star formation rates ranging from $\sim7$ to $111\:\rm{M_\odot/yr}$, the majority of which are consistent with the star-forming main sequence at $\rm{z}\approx6$. The positions of our host galaxies on the $\rm{log_{10}[OIII]5008/H\beta}$ vs. $\rm{log_{10}[NII]6584/H\alpha}$ (R3N2) Baldwin-Phillips-Terlevich (BPT) diagram indicate ionization rates typical of AGN activity in the low-redshift universe, but are consistent with the placement of similar $\rm{z}\approx6$ quasar host galaxies, suggesting that the R3N2 line ratios cannot distinguish AGN and star-formation powered line emission at high redshifts. We conclude from the consistency between our quasar host sample with $\rm{z}\sim6$ galaxies that the presence of a low-luminosity AGN causes little significant change in the properties of galaxies at $z \approx 6$ on $10\:\rm{Myr}$ timescales.%We present an analysis of the rest frame optical JWST NIRSpec Fixed Slit spectra of extended host galaxy emission in 12 quasars from the Subaru High-z Exploration of Low-Luminosity Quasars (SHELLQs) sample at redshifts $\rm{6.0<z<6.4}$. A model consisting primarily of a sum of two Gaussians is used to fit and subtract the spatial point spread function of the quasar as a function of wavelength, leaving only extended galaxy emission. Ten of 12 systems show spatially extended line emission and five of 12 systems show an extended stellar continuum. From the $\rm{[OIII]5008}$ emission line, we measure a $132\pm19$ km/s ionized gas outflow in one system and $52\pm12$ km/s rotation, suggesting a coherent disk, in another. From the extended $\rm{H}\alpha$ emission, which we hypothesize is ionized by star-forming regions rather than the narrow-line region of the quasar, we measure star formation rates ranging from $\sim7$ to $111\: \rm{M_\odot/yr}$, the majority of which are consistent with the star-forming main sequence at $\rm{z} \approx 6$. The positions of our host galaxies on the $\rm{log_{10}[OIII]5008/H\beta}$ vs. $\rm{log_{10}[NII]6584/H\alpha}$ (R3N2) Baldwin-Phillips-Terlevich (BPT) diagram indicate ionization rates typical of AGN activity in the low-redshift universe, but are consistent with the placement of similar $\rm{z}\approx6$ quasar host galaxies from the literature, suggesting that the traditional R3N2 line ratios cannot distinguish AGN and star-formation powered line emission at high redshifts. We conclude from these consistencies between our quasar host sample with $\rm{z}\sim6$ galaxies in the literature that the presence of a low-luminosity AGN in a galaxy causes little to no significant variation from other $\rm{z}>6$ galaxies on the $10 \:\rm{Myr}$ timescales probed by our measurements.%, and that AGN feedback or quenching of host star formation occurs on longer timescales.
\end{abstract}

\section{Introduction} \label{sec:intro}
The study of quasar host galaxies provides unique insights into the connection between supermassive black holes (SMBHs) and the galaxies in which they reside. In the local universe, the masses of SMBH are known to correlate with the bulge mass and velocity dispersion of their hosts, despite the many orders of magnitude separating their masses and sizes \citep{Magorrian1998,Gebhardt2000,Ferrarese2000,Haring2004, Kormendy2013ARA&A..51..511K}. This implies an interaction between SMBH and host galaxy, leading many to theorize that outflows and heating from the SMBH provide thermal and kinetic feedback to the galaxy, halting the collapse of cold molecular gas into stars, while others think these same outflows might launch shocks which instigate stellar growth \citep{Croton2006, Fabian2012ARA&A..50..455F, Kormendy2013ARA&A..51..511K, Schaye2015, Weinberger2017, Sturm2018, Laha2021NatAs...5...13L}. Still others have suggested that processes which invigorate stellar growth, such as galaxy mergers, also funnel material to the center of the galaxy, concurrently feeding the central black hole \citep{DiMatteo2005Natur.433..604D, Hopkins2005, Hopkins2006, Storchi-Bergmann2007, Storchi-Bergmann2019}. Whether these processes acted similarly at both high and low redshifts is a question of open discussion. In all cases, the interaction mechanism between the SMBH and the galaxy is expected to occur while the SMBH is active, making quasar hosts a prime target for high redshift studies of SMBH-galaxy coevolution. 

However, quasar hosts are difficult to observe because their light is drowned out by the overwhelming luminosity of the quasars at their centers. Before the James Webb Space Telescope (JWST), starlight from quasar hosts was completely inaccessible at high redshifts due to $(1+\rm{z})^4$ surface brightness dimming and the wavelength coverage necessary to observe starlight at high redshifts, largely limiting z $>$ 6 studies of quasar hosts to other measures of host galaxy properties such as dust and gas emission at $\rm{mm}$ wavelengths \citep[e.g.,][]{Decarli2018, Venemans2020, Yang2020c, Neeleman2021, Wang2021ApJ...907L...1W}.

With the launch of JWST in 2021 \citep{Rigby2023PASP..135d8001R}, the high redshift ($\rm{z} > 6$) universe is more accessible than ever before. Because the Earth's atmosphere is bright in the near-infrared (NIR), we cannot easily observe faint objects in the infrared from the ground. As a result, we needed a powerful space-based infrared telescope like JWST, whose 6.5-meter diameter mirror allows for high resolution and broad wavelength coverage of the near- and mid-infrared, to explore the rest-frame optical and ultraviolet (UV) properties of galaxies in the first billion years of the universe. 

The rest frame UV/optical spectra of active galactic nuclei (AGN) and their host galaxies have been well characterized in the local universe by ground-based instruments such as the Sloan Digital Sky Survey (SDSS) \citep{York2000} and space-based instruments such as the Hubble Space Telescope (HST) \citep{Bahcall1997, VandenBerk2001, Dunlop2003, Kauffmann2003, Kormendy2013ARA&A..51..511K}. AGN+host spectra are characterized by nebular emission lines, absorption lines, and continuum. Emission lines result from ionized nebulae, with the ionizing radiation being attributed to the AGN accretion disk or young O and B stars, diagnosed by the ionization ratios of common metal lines such as $\rm{[OII]3727}, \rm{[OIII]5008}, \rm{[NII]6585}$, and $\rm{[SII]6730}$ to the hydrogen lines. The total continuum is the sum of the stellar and thermal accretion disk continua, the former of which peaks in the rest frame near UV/optical except for heavily dust-obscured systems, whereas the latter continues to increase at shorter wavelengths. \cite{Kauffmann2003} have shown that unobscured AGN at low redshift tend to reside in massive $(\rm{M_\star>10^{10} M_{\odot}})$, early-type galaxies and that the hosts of luminous quasars tend to have younger stellar populations than less luminous quasars. Photometric studies such as \cite{Silverman2008, Silverman2009, Silverman2025arXiv250723066S} have similarly shown that X-ray selected AGN at $\rm{z}<1$ are more likely to reside in massive bulge-dominated galaxies with enhanced star formation rates and bluer colors. 

The application of rest-frame UV/optical spectroscopy to high-redshift galaxies and AGN has revolutionized our understanding of early galaxy evolution. Already, JWST has found faint, unobscured AGN to be more prevalent in the early universe than expected from extrapolation of the $\rm{z}\sim6$ quasar luminosity function \citep{Matsuoka2018ApJ...869..150M} to lower luminosities \citep{Shen2020MNRAS.495.3252S, Pacucci2023, Matthee2024ApJ...963..129M, Greene2024ApJ...964...39G}, with many residing in relatively undermassive stellar hosts \citep{Stone2024} that suggest rapid SMBH growth due to heavy seeds or super-Eddington accretion \citep{Inayoshi2016MNRAS.459.3738I,Larson2023,Furtak2024}. A new class of objects, little red dots (LRDs), are spectral mysteries that do not conform to the usual AGN+host+dust models and appear to be relatively common at z $>$ 3 but largely vanish at lower redshifts \citep{Matthee2024, Ma2024ApJ...975...87M, Ma2025arXiv250408032M, Lin2025arXiv250710659L, naidu2025arXiv250316596N, rusakov2025arXiv250316595R}. GNz-11, one of the highest redshift galaxies known to date, may host a SMBH accreting at 5 times the Eddington limit, with stellar mass and SMBH mass implying very rapid formation \citep{Tacchella2023ApJ...952...74T}, and UHZ-1 ($\rm{z}_{\text{spec}}=10.073$), which was detected in X-rays and observed with JWST, has a SMBH-to-galaxy mass ratio 2-3 orders of magnitude higher than local values \citep{Goulding2023b}. Indeed, all of these AGN host systems are already pushing the bounds of our theories of early galaxy formation, particularly in the moderate- and low-luminosity quasar regime. Because high-redshift samples are biased towards the most luminous quasars, the study of such low- and moderate-luminosity quasars and their host galaxies is vital to creating a holistic picture of AGN in the high-redshift universe.

In this context, we present a study of the 2D spectra of twelve moderate-luminosity unobscured z $>$ 6 quasars and their hosts, using similar methods of spatio-spectral decomposition to \cite{Jahnke2007MNRAS.378...23J} and \cite{Roche2014MNRAS.443.3795R}. These twelve targets were selected from the Subaru High-z Exploration of Low-Luminosity Quasars (SHELLQs) sample (see $\S$\ref{subsec:targets} for a detailed description) for the JWST Cycle 1 General Observers program 1967 (PI: M. Onoue). The targets were observed with the NIRCam F356W and F150W filters and the NIRSpec F290LP G395M filter-grism combination as part of that program. Photometric analysis of all 12 objects has been published in \cite{Ding2023Natur.621...51D, ding2025shellqsjwstunveilshostgalaxies} and spectroscopic analysis of two objects (J2236+0032, J1512+4422) has been published in \cite{Onoue2024arXiv240907113O}. This paper's investigation of the host galaxies' spectroscopic properties will be a companion piece to those and future SHELLQs analyses. Using NIRSpec Fixed Slit (FS) observations, we extract quasar-subtracted host galaxy spectra from extended emission and infer several properties of the host galaxies such as their star formation rates, ionized line ratios, and dust extinction. By comparing to star-forming galaxies and AGN at redshifts 0 $<$ z $\lesssim$ 8, we hope to better understand the evolution of AGN host galaxies through cosmic time. This paper investigates the spectral properties of the extended host galaxy emission, where present, of the 12 systems from GO program 1967. \cite{ding2025shellqsjwstunveilshostgalaxies} investigate the photometric properties of the host galaxies using Sérsic profile fitting of NIRCam photometry, and \cite{OnoueInPrep} investigate the full quasar+host spectra from NIRSpec spectroscopy, inferring SMBH properties.

The paper is laid out as follows: In \S \ref{sec:data}, we describe the data selection and collection methods. In \S \ref{sec:methods} we describe the parameterization and application of the NIRSpec Fixed Slit Point Spread Function (PSF) to remove the quasar and reveal extended stellar emission. In \S \ref{sec:data analysis} we describe the fitting and modeling of the data and the extraction of galactic parameters of interest such as star formation rate. In \S \ref{sec:results} we discuss the implications of the values calculated in the analysis section and compare our results to the literature, as well as make direct comparisons to the values calculated by \cite{ding2025shellqsjwstunveilshostgalaxies} and \cite{OnoueInPrep}. Finally, in \S \ref{sec:conclusion} we summarize our findings and contemplate future avenues of inquiry. We use cosmological parameters $\rm{H_0=70~km \: s^{-1}\: Mpc^{-1}}$, $\rm{\Omega_M=0.3}$, and $\Omega_\Lambda=0.7$ \citep{Planck2016}, and all magnitudes are in the AB system \citep{Oke1983}.

\section{Data} \label{sec:data}
\subsection{Targets}\label{subsec:targets}
%Our targets are a subsample of the Subaru High-z Exploration of Low Luminosity Quasars (SHELLQs) sample, which consists of relatively low-luminosity (\(M_{1450}>-26\)), high-redshift (\(z > 5.7\)) quasars discovered by the Hyper Suprime-Cam Subaru Strategic Program (HSC-SSP) \citep{Matsuoka2015, Matsuoka2016, Matsuoka2018PASJ...70S..35M, Matsuoka2018ApJS..237....5M, Matsuoka2019}. HSC-SSP is an optical wide-field multiband imaging survey \citep{Aihara2018TheDesign} conducted using the Hyper Suprime-Cam (HSC) \citep{Miyazaki2018} mounted on the Subaru 8.2 meter telescope on Maunakea in Hawai'i. The SHELLQs sample objects were identified as quasars using color cuts to select dropout systems, with maximum apparent magnitudes of \(z_{AB} < 24.5\) and \(y_{AB} < 25\) \citep{Matsuoka2016}. To date, 162 of these quasars have been spectroscopically confirmed through ground-based follow-up, placing them at redshifts of \(5.7 < z < 7.1\) and reaching a rest-frame UV absolute magnitude limit of \(M_{1450}>-26\)  \citep{Matsuoka2022, Matsuoka2023ApJ...949L..42M}. 
The SHELLQs sample consists of relatively low-luminosity (\(\rm{M}_{1450}>-26\)), high-redshift (\(\rm{z} > 5.7\)) quasars \citep{Matsuoka2016ApJ...828...26M, Matsuoka2018ApJS..237....5M, Matsuoka2018PASJ...70S..35M, Matsuoka2019ApJ...883..183M} discovered by the Hyper Suprime-Cam Subaru Strategic Program (HSC-SSP) \citep{Aihara2018TheDesign}. The SHELLQs sample was selected as dropout sources with maximum apparent magnitudes of \(\rm{z_{AB}} < 24.5\) and \(\rm{y_{AB}} < 25\) \citep{Matsuoka2016ApJ...828...26M} and was a prime parent sample for targets to follow up with JWST imaging and spectroscopy \citep{Matsuoka2016ApJ...828...26M, Izumi2019, Onoue2021}. To date, $\sim180$ of these quasars have been spectroscopically confirmed, placing them at redshifts of \(5.7 < \rm{z} < 7.1\) and reaching a rest-frame UV absolute magnitude limit of \(\rm{M}_{1450}>-21\)  \citep{Matsuoka2022ApJS..259...18M, Matsuoka2023ApJ...949L..42M, Matsuoka2025arXiv250821229M}. 

 The subsample of SHELLQs quasars observed with JWST in Cycle 1 were ten quasars presented in \cite{Matsuoka2018ApJ...869..150M} with redshifts $6.18<\rm{z}<6.4$ and \(\rm{M}_{1450}>-24\), plus the two faintest \(\rm{z}>6\) SHELLQs quasars known at the time of the proposal, J1146$-$0005 and J0911+0152  \citep{Matsuoka2018ApJS..237....5M, Matsuoka2018PASJ...70S..35M}. The target information for the twelve systems is listed in Table \ref{table:target_info}, below. Before JWST, no z $>$ 5 quasars had host starlight detections, as their host galaxies were too faint for current ground-based instruments \citep{Goto2012, Decarli2018}. As a result, the faintest quasars were specifically chosen in the hope that they would not drown out the light from their host galaxies, allowing the host parameters, especially stellar luminosity and mass, to be measured \citep{Ding2023Natur.621...51D, ding2025shellqsjwstunveilshostgalaxies}. Together, these twelve quasars cover a redshift range of \(6.07 <\rm{z}< 6.40\) and a rest-frame UV absolute magnitude range of $-24.0 < \rm{M}_{1450} < -21.5$ \citep{Matsuoka2022ApJS..259...18M, ding2025shellqsjwstunveilshostgalaxies}. Eight of the twelve systems also have Atacama Large Millimeter Array (ALMA) follow-up data, allowing investigation of their star formation rates and the dynamics of their cold dust and gas using the $\rm{[CII]}158~\mu m$ line \citep{IzumiInPrep}.
    
    The NIRCam photometry and NIRSpec Fixed Slit spectroscopy of all twelve sources were collected between November 2022 and January 2024. Target acquisition was performed using the usual Wide aperture target acquisition (WATA) setup and FULL array. The F140X filter was used for target acquisition and alignment to the slit in all cases except for the three faintest targets (J0911+0152, J1146$-$0005, J1512+4422), for which the CLEAR filter was used. Data were read using NRSIR2RAPID mode, except in the three faintest targets, for which NRSIRS2 mode was used. All the JWST data used in this paper can be found in MAST: \dataset[10.17909/vk7s-me07]{http://dx.doi.org/10.17909/vk7s-me07}. 

\begin{table*}
    \centering
    \caption{Details of Target information}\label{table:target_info}
    \begin{tabular}{cccccccc}
    \hline\hline
     (1)&(2)& (3)& (4)& (5)& (6)  & (7)& (8)\\
     Full ID&Target ID & RA & Dec & $\rm{z}$& $\rm{M}_{1450}$& Exposure time& Date of \\
     && (deg) & (deg) & &  & (s)& observation\\
    \hline
     HSC J091114.27+015219.4&J0911+0152 & 137.80947 & 1.87204 & 6.07 & -22.1   & 8840.868& 2023-11-28\\
     HSCS J142517$-$001540& J1425$-$0015 & 216.32379 & $-$0.26137 & 6.18 &-23.4   & 2013.267& 2023-01-23\\
     HSC J091833+013923&J0918+0139 & 139.63821 & 1.65648 & 6.19 & -23.7   & 2013.267& 2022-12-25\\
     HSC J151248.71+442217.5&J1512+4422 & 228.20298 & 44.37154 & 6.18 & -23.1   & 7965.534& 2023-02-14\\
     HSC J084408.61$-$013216.5&J0844$-$0132 & 131.03589 & $-$1.53795 & 6.18 & -23.7   & 3107.433& 2022-11-28\\
     HSC J021721.59$-$020852.6 &J0217$-$0208 & 34.33999 & $-$2.14798 & 6.2 & -23.2  & 3107.433& 2023-07-22\\
     HSCS J084431$-$005254&J0844$-$0052 & 131.13166 & $-$0.88185 & 6.25 & -23.7   & 2013.267& 2022-11-16\\
 HSC J114648.42+012420.1& J1146+0124 & 176.70173 & 1.40557 & 6.27 & -23.7   & 3107.433&2023-06-22\\
     HSCS J152555+430324&J1525+4303 & 231.48244 & 43.05667 & 6.27 & -23.9   & 3107.433& 2023-03-18\\
     HSC J114658.89$-$000537.7&J1146$-$0005 & 176.74536 & $-$0.09382 & 6.3 & -21.5   & 8840.868& 2023-06-22\\
     HSC J225538.04+025126.6&J2255+0251 & 343.90849 & 2.85740 & 6.34 & -23.9  & 2013.267& 2022-10-28\\
     HSC J223644.58+003256.9&J2236+0032 & 339.18575 & 0.54911 & 6.4 & -23.8  & 2013.267& 2022-10-30\\
    \hline
    \end{tabular}
    \tablecomments{Table adapted from \cite{ding2025shellqsjwstunveilshostgalaxies} with permission. Column 1: Full Object ID as given in \cite{Matsuoka2018ApJS..237....5M, Matsuoka2018PASJ...70S..35M}.  Column 2: Abbreviated Object ID, used hereafter to refer to individual objects in our sample. Columns 3 and 4: J2000 R.A. and decl. coordinates. Column 5: Spectroscopic redshift based on $\rm{Ly}\alpha$ and Lyman break. Column 6: The $\rm{M_{1450}}$ values measured in \cite{Matsuoka2018PASJ...70S..35M} and \cite{Matsuoka2018ApJS..237....5M}}
    %Column 7: The Eddington Ratios are estimated and will be reported in Onoue et al. in prep. 
    %Column 7: The number of PSF stars collected from the corresponding JWST NIRCam field of view in F150W and F356W.}
\end{table*}

\subsection{NIRSpec G395M Spectroscopy}\label{subsec:G395Mspec}
The NIRSpec Spectrograph in Fixed Slit observing mode \citep{Jakobsen2022} is a long slit spectrograph that uses a combination of grating-prism (grism) and filter system. We use the G395M grism and F290LP filter, providing observed wavelength coverage of 2.87 to 5.27$~\mu \rm{m}$ and a resolution of 1000, or $\Delta \rm{v} = 300$ km/s in physical units. For the median redshift of the sample ($\rm{z} \approx 6.2$), this corresponds to rest wavelengths of \(0.41\:\rm{\mu m}\) to \(0.75\:\rm{\mu m}\). Slit S200A2, which has a 0$''$.2 by 3$''$.2 rectangular footprint on the sky, was used with a three-nod dither pattern centered on the quasar and no subpixel dithering. The exposure time for each source was chosen so that the quasar+host continuum around $\rm{H}\alpha$ would be detected with $\rm{S/N} > 7$ per pixel. The three-nod dithering pattern places the source in three positions on the detector in order to reduce individual pixel effects. The dithers are used as backgrounds for each other, causing the final 2D spectrum to have ``photo negatives'' $+0''.8737$ above and $-1''.0832$ below the primary dither position. These negatives limit the spatial extent of the spectrum to the $\sim$8 nearest pixels given the instrument pixel scale of $0''.1/\rm{pix}$. We discuss the effects and handling of these spatial limits in \S \ref{subsec:spatial extent} below. These negatives, which are commonly seen mirrored on each side of 2D spectra in the literature, are cropped in all figures herein.

As mentioned above, in this study we rely on the spatial and spectral information provided by long-slit spectroscopy. Because the slit extends $1''.6$ in either direction from the centered quasar, the extracted 2D spectra contains both the quasar and any extended host galaxy emission that lies along the slit. The quasar is expected to dominate in only the central few pixels; combined with the spatial limits imposed by the dithering pattern, this leaves two $0''.2$ by approximately $0''.5$ regions on either side of the quasar where the host extended emission is expected to dominate (see panel $(d)$ of Figures \ref{fig:J0911 spectrum} - \ref{fig:J2236 spectrum} for visualization of this geometry). Because the quasar hosts had previously been undetected in the discovery spectra taken by ground-based telescopes, the orientation of the slit could not be aligned a priori to the longest axes of the galaxies, and thus the orientations are effectively random. %In almost all cases, we see no signs of extended emission beyond the central except \textit{As a result, any light from the outer 8 pixels (halfway between the centers) may be dampened by any extended emission being subtracted and encroaching on the coadded spectrum. We see no signs of extended emission beyond 5 pixels from center, so we continue without accounting for this possible overlap, except in particular J11465 and J844132. }

    We use the \texttt{jwst} v.1.17.1 pipeline code for all calibration and data reduction except for removal of 1/f noise, cosmic rays, and snowballs, which are removed using custom additions to the pipeline, described in \cite{Onoue2024arXiv240907113O}. The \texttt{jwst} pipeline has three Stages which produce three levels of output, of which we use the last two. From Stage 2, we get three background-subtracted level 2 output files, one for each dither. These three inputs are resampled onto a uniform grid and coadded together in Stage 3 to produce the fully calibrated level 3 output. Resampling uses the 2D spectrum's WCS (World Coordinate System) and distortion information to straighten and re-align the trace, with uncertainties propagated using the Drizzle algorithm \citep{Fruchter2002PASP..114..144F}. We mask by hand bad pixels and cosmic rays which were not caught by the outlier removal step of the pipeline, though these only account for approximately 1$\%$ of pixels and have little impact on our analysis even when included. %The narrower spatial extent of the PSF in the individual dithers compared to the rectified and resampled final spectra is seen in the bottom part Figure \ref{fig:dither wiggles}. Further discussion of the effect of broadening on the spatial PSF can be found in section \ref{sec:methods}.
   
\subsection{NIRCam Photometry}\label{subsec:photometry}
     The targets were imaged in the F150W and F356W bands, covering observed wavelengths $1.33~\mu \rm{m}$ to $1.67~\mu \rm{m}$ and $3.14 ~\mu \rm{m}$ to $3.98~\mu \rm{m}$, respectively. The F356W band overlaps our spectra, while the F150W band was chosen to lie blueward of the 4000$~\mathring{\rm{A}}$ break (redshifted to $2.8~\mu \rm{m}$ at $\rm{z}\approx6$) in order to better constrain star formation histories of the host galaxies using the combination of the two filters. Photometric decomposition and fitting were performed by \cite{ding2025shellqsjwstunveilshostgalaxies} using the \texttt{Galight} code \citep{Ding2020}, revealing the host emission (see panel $(d)$ of Figures \ref{fig:J0911 spectrum} - \ref{fig:J2236 spectrum} below or Figure 1 of \cite{Ding2023Natur.621...51D}). The details of these fits are presented in \cite{ding2025shellqsjwstunveilshostgalaxies}, along with the galaxy parameters derived from them. In our analysis, we use the quasar-subtracted host photometry and stellar masses produced by these fits. Of the twelve targets, eleven have host galaxies detected in the F356W band, and seven are detected in the F150W band. % Ten to 20 field stars were used for each target to characterize the PSF of the NIRCam detector. The quasar was then fit using this PSF while the host was simultaneously fit using a Sérsic model, and the quasar was subtracted off to 
     
\section{Extracting the Host Galaxy Spectra} \label{sec:methods}
Our goal is to remove all flux originating from the quasar in the long-slit NIRSpec data, while leaving as much of the galaxy light as possible. Our method relies on the different spatial profiles of point and extended sources in a long-slit detector. We model the pure point source continuum and emission lines as a function of wavelength, and then subtract off that model to reveal the residual extended emission, which we assume is entirely light from the galaxy. We herein describe the model used to characterize a point source in the detector and the application of that model to our data. 

\subsection{Characterizing the NIRSpec Fixed Slit Spatial Point Spread Function} \label{subsec:characterizing PSF}
    The PSF defines the response of an optical system to an unresolved point source. Since we want to reveal spatially extended emission in the 2D NIRSpec spectra beyond the point source, we will need to characterize the point source as a function of wavelength in order to remove it. At the time of writing, there is no publicly available numerical or analytical model of the NIRSpec Fixed Slit spatial (i.e., cross-dispersion) PSF as a function of wavelength. We therefore create our own numerical mode for the spatial PSF using public data from calibration star ID 1808347 (2MASS J18083474+6927286, also known as TYC 4433-1800-1; JWST project ID 1128), which is an A3V type star with a K-band Vega magnitude of 11.53 that was observed using the same instrument configuration as our targets (i.e., NIRSpec FS, G395M grism, F290LP filter,  S200A2 slit, 3-point nod dithering pattern). It was observed several months before our observations began; however, current thermal modeling shows no significant variation in the NIRSpec detectors or optics in this time period \citep{optics2023jwst.rept.8497L, thermal2024}. The data for star 1808347 were reduced using the same process as for the target quasars, using the same version of the \texttt{jwst} pipeline and the same custom $\rm{1/f}$ noise removal script \citep{Onoue2024arXiv240907113O}. No other observations of stars using the above instrument configuration have been made public since the launch of JWST. As a result, our PSF model is limited to a single star, and we acknowledge the resulting limitation in model accuracy. 
    
    Our primary concern is to carefully and accurately fit the wings of the PSF, where we expect the host to dominate, as a function of wavelength. As we will describe below, we fit the PSF using a functional form (rather than simply smoothing and scaling the 2D spectrum of star 1808347) to understand the wavelength dependence of the individual model parameters and to better separate the PSF from the extended host and background noise, which have different wavelength dependencies than the PSF. We remind the reader that, except when otherwise noted, these fits are performed on the level 2 data for each dithered exposure which have been background subtracted but not rectified or coadded. 
    
    We first attempt to characterize the cross-dispersion profile at a given wavelength $\lambda$ using a single Gaussian model \(\rm{(PSF_{1G}(x, \lambda)=\frac{A(\lambda)}{\sqrt{2\pi}\sigma(\lambda)}\exp{\frac{-(x-\mu(\lambda))^2}{2\sigma(\lambda)^2}} )}\), where $\rm{x}$ is the cross-dispersion position, in pixels, and $\lambda$ is the observed wavelength. The NIRSpec pixel scale is $0''.1/\rm{pix}$, and the slit tilt for our instrument configuration is $\sim2^\circ.0$, which is close enough to $0^\circ$ that we treat any tilt effects as negligible. The single Gaussian model is fitted separately to each wavelength (column) of the 2D spectrum, requiring only that the amplitude, $\rm{A}$, and width, $\sigma$, be positive and the centroid, \(\mu\), vary by fewer than 3 pixels between consecutive wavelengths to enforce continuity. All fitting is performed using the \texttt{LevMarLSQFitter} function of the \texttt{astropy v.5.3.4} package, which uses the Levenberg-Marquardt algorithm and the least-squares statistic \citep{Astropy_2022}.
\begin{figure*}
    \centering
    \includegraphics[width=1\linewidth]{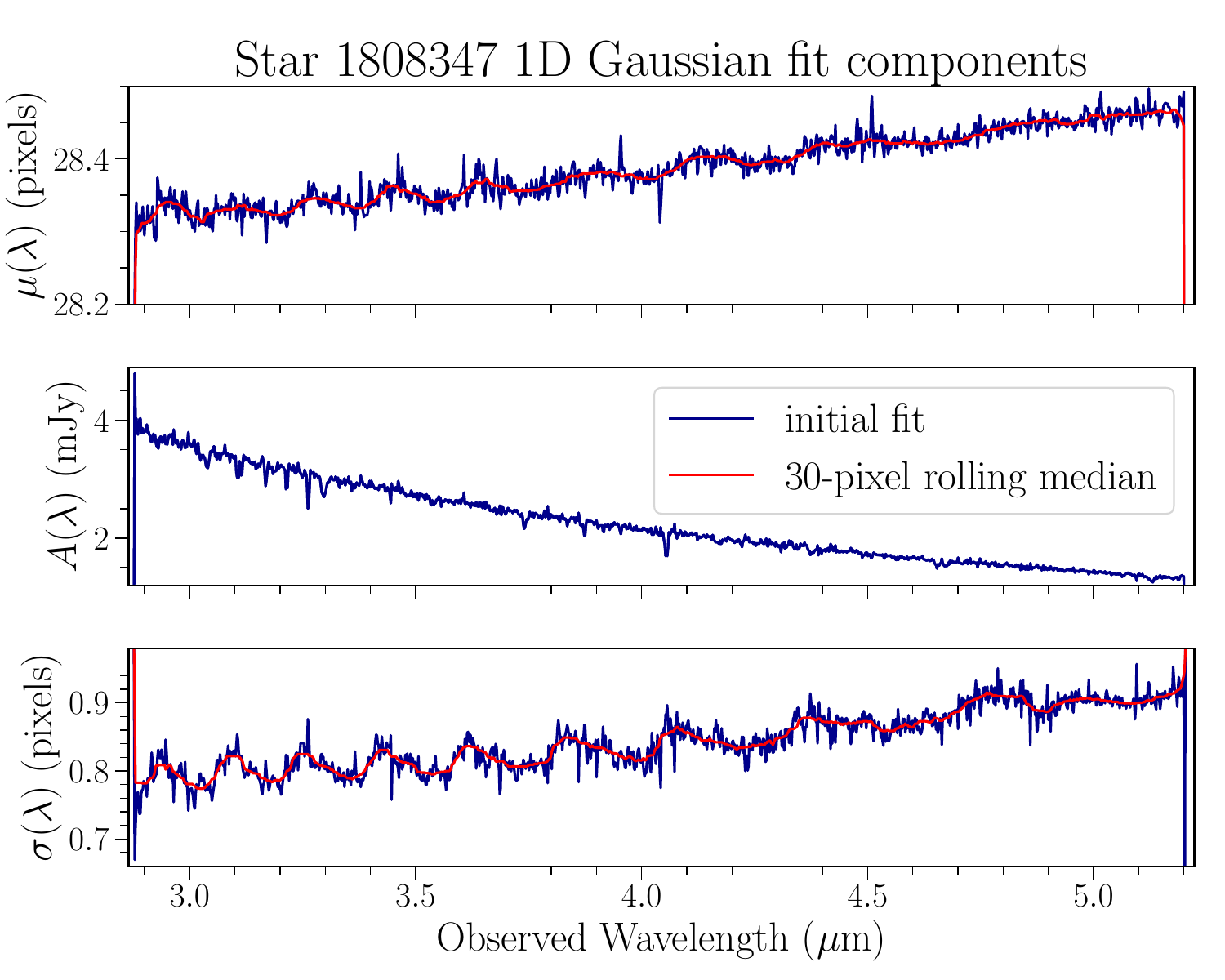}
    \caption{Example of the wavelength dependence of the centroid ($\mu$), amplitude ($\rm{A}$), and width ($\sigma$) of a single Gaussian fit $\rm{(PSF_{1G})}$ to the level 3 spatial profile of star 1808347 at each wavelength, where the blue line represents the parameters of the initial fit to the data and the red line is the same data smoothed by a 30 pixel rolling median. The centroid and width increase with wavelength, and both show a distinct quasi-periodic pattern of wiggles, which are discussed in $\S$ \ref{subsec:width variation}. We fit the level 3 (rectified) rather than the level 2 (unrectified) 2D spectrum here to better illustrate the variation of the centroid with wavelength, since the dynamic range of $\mu$ is too large in level 2 data.}
    \label{fig:variation with wavelength}
\end{figure*}

    The dependence of the three parameters with respect to wavelength of the single Gaussian model to the star is shown in Figure \ref{fig:variation with wavelength}. The width ($\sigma$) and centroid ($\mu$) vary considerably with wavelength due to noise, so we performed a 30-pixel rolling median on the centroid and width to smooth these variations, where the 30-pixel smoothing width is chosen because it reduces noise-level variations while leaving large scale variations due to real detector effects unchanged. The amplitude is exempt from smoothing since it is well constrained by the extracted 1D spectrum, and we expect variations on scales much less than 30 pixels. The width and centroid show quasi-periodic variation with wavelength, which is due to undersampling of the PSF, as we discuss in \S \ref{subsec:width variation} below. The single Gaussian models for each wavelength column with smoothing are then combined to create a 2D spectrum model.

\begin{figure*}
    \centering
    \includegraphics[width=1\linewidth]{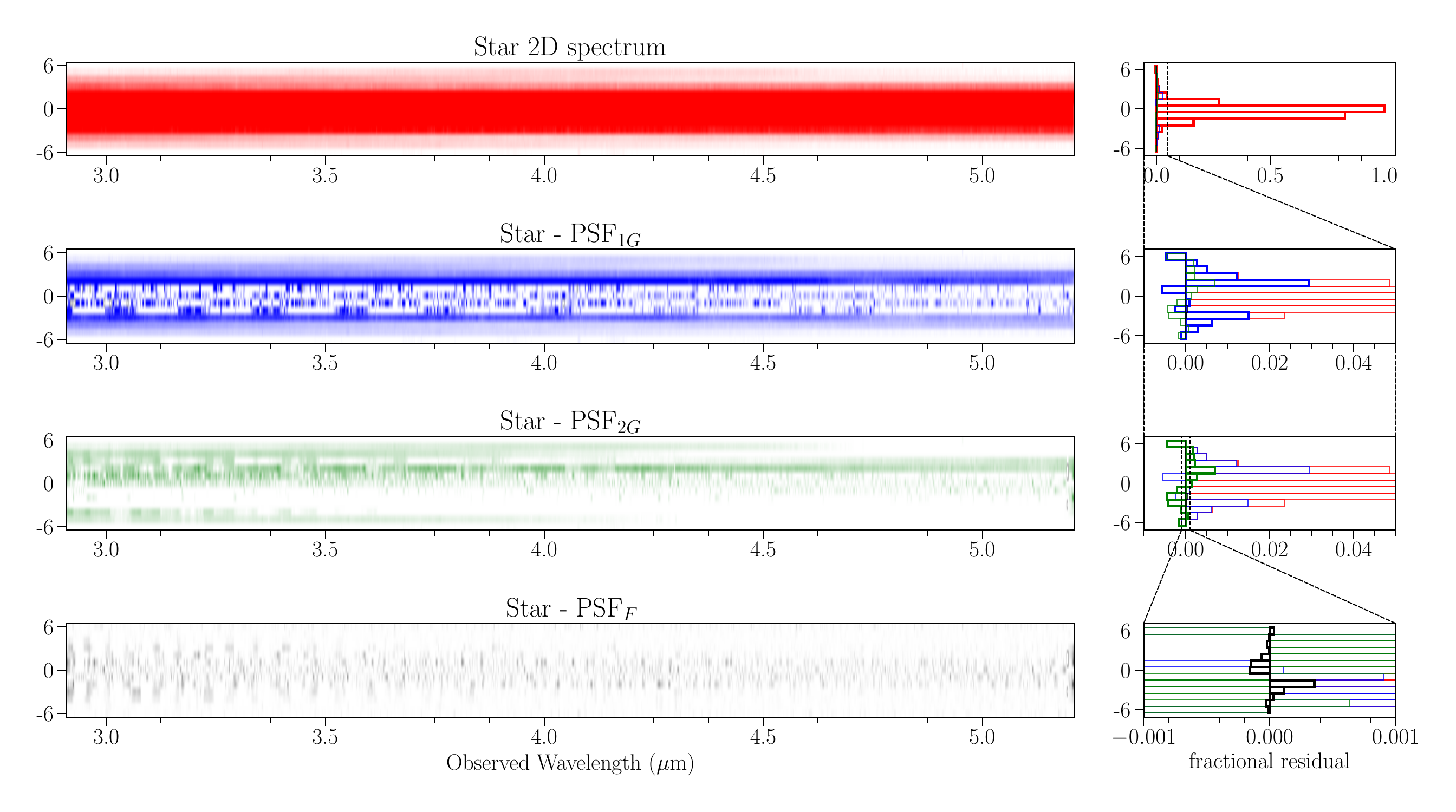}
    \caption{2D (\textit{left}) and 1D (\textit{right}) level 3 residuals of three models for the PSF along the spatial direction across the full wavelength range of the detector. All 2D spectra are normalized to the maximum value of the calibration star's 2D spectrum and are presented using the same stretch, although different colors, with all values less than zero shown in white. The y-axes for both 2D spectra and 1D residuals are pixels in the cross-dispersion direction of the detector, with the PSF center placed at zero. The 1D residuals are plotted horizontally in order to match the orientation of the 2D spectra. The residual profile averaged over all wavelengths are shown for all four plots, with the profile of interest for each row bolded. \textit{First row:} The left panel shows the 2D spectrum of the calibration star 1808347 used as a PSF, and the right panel shows the average spatial profile of the calibration star summed over all wavelengths and normalized to a peak of unity. In this and the following panels, red represents the star profile. \textit{Second row:} Single Gaussian model $\rm{(PSF_{1G})}$ residuals. While the central four spatial pixels are fit well, the right panel (which has been enlarged compared to the right panel of the first row) shows that the wings of the profile have residuals on the order of 4$\%$, highlighted in blue. \textit{Third row:} Double Gaussian model $\rm{(PSF_{2G})}$ residuals. Residuals are reduced to the order of $\approx1\%$, as shown in green in the right panel. The profiles of the Single Gaussian model (blue) and star (red) are shown at lowered opacity in the background of the right panel for comparison. \textit{Fourth row:} Final model $\rm{(PSF_{F})}$ residuals, showing that the full PSF model leaves uncorrelated and small 2D residuals, which, when summed over all wavelengths (black), have an amplitude of less than $0.05\%$.} %The 2D residuals shown in the left panel of the third row are treated as constant and appended to the double Gaussian model in order to create the full, final PSF model. This final model by definition will have residuals of zero, so is not shown.}
    \label{fig:2DPSF}
    \end{figure*}

    To test the efficacy of the single Gaussian model, we subtract the 2D model from the 2D spectrum of star 1808347 (Figure \ref{fig:2DPSF}) then recombine the three dithers using the level 3 \texttt{jwst} pipeline. Since the star has no extended emission, a perfect model should show residuals consistent with zero. Figure \ref{fig:2DPSF} shows the 2D and cumulative 1D residuals of the PSF models we tested compared to the original spectrum of star 1808347. The 2D residuals of the single Gaussian model can be seen in the second row of Figure \ref{fig:2DPSF}, where the first row is the original spectrum of the star. Although the single Gaussian model fits the central region of the PSF, the wings of the profile show significant residuals that, if ignored, would obscure or bias any extended galactic emission. We therefore proceed to a double Gaussian model:

    \begin{equation}
    \begin{aligned}
        \rm{PSF_{2G}(x, \lambda)} = \frac{\rm{A}_{1}(\lambda)}{\sqrt{2\pi}\sigma_{1}(\lambda)}\exp\left(\frac{-(x-\mu(\lambda))^2}{2\sigma_{1}(\lambda)^2}\right) + \\
        \rm{\frac{A_{2}(\lambda)}{\sqrt{2\pi}\sigma_{2}(\lambda)}\exp\left(\frac{-(x-\mu(\lambda))^2}{2\sigma_{2}(\lambda)^2}\right)}.
    \end{aligned}
    \label{eq:PSF2G}
    \end{equation}

   The double Gaussian model is defined by five parameters at a given wavelength, \(\rm{A_1}\), \(\rm{A_2}\), \(\sigma_1\),  \(\sigma_2\), and \(\mu\), which are the respective amplitudes ($\rm{A_1, A_2}$)  and widths ($\sigma_1, \sigma_2$) of the primary and secondary Gaussian and the shared centroid ($\mu$) for both Gaussians. However, if all five parameters are allowed to vary freely at the same time, the fit fails and the model becomes unphysical. We therefore proceed through three rounds of fitting, R1, R2, and R3, described below, to gradually constrain the parameters. As in the single Gaussian model, the amplitudes and widths are required to be positive, and the centroid is required to vary by fewer than 3 pixels between consecutive wavelengths. Additionally, upper and lower limits are placed on the widths to prevent the fitting algorithm from approaching the unphysical limits \(\sigma_{1,2} \rightarrow \infty\) and \(\sigma_{1,2} \rightarrow 0\).

\begin{figure*}
    \centering
    \includegraphics[width=1\linewidth]{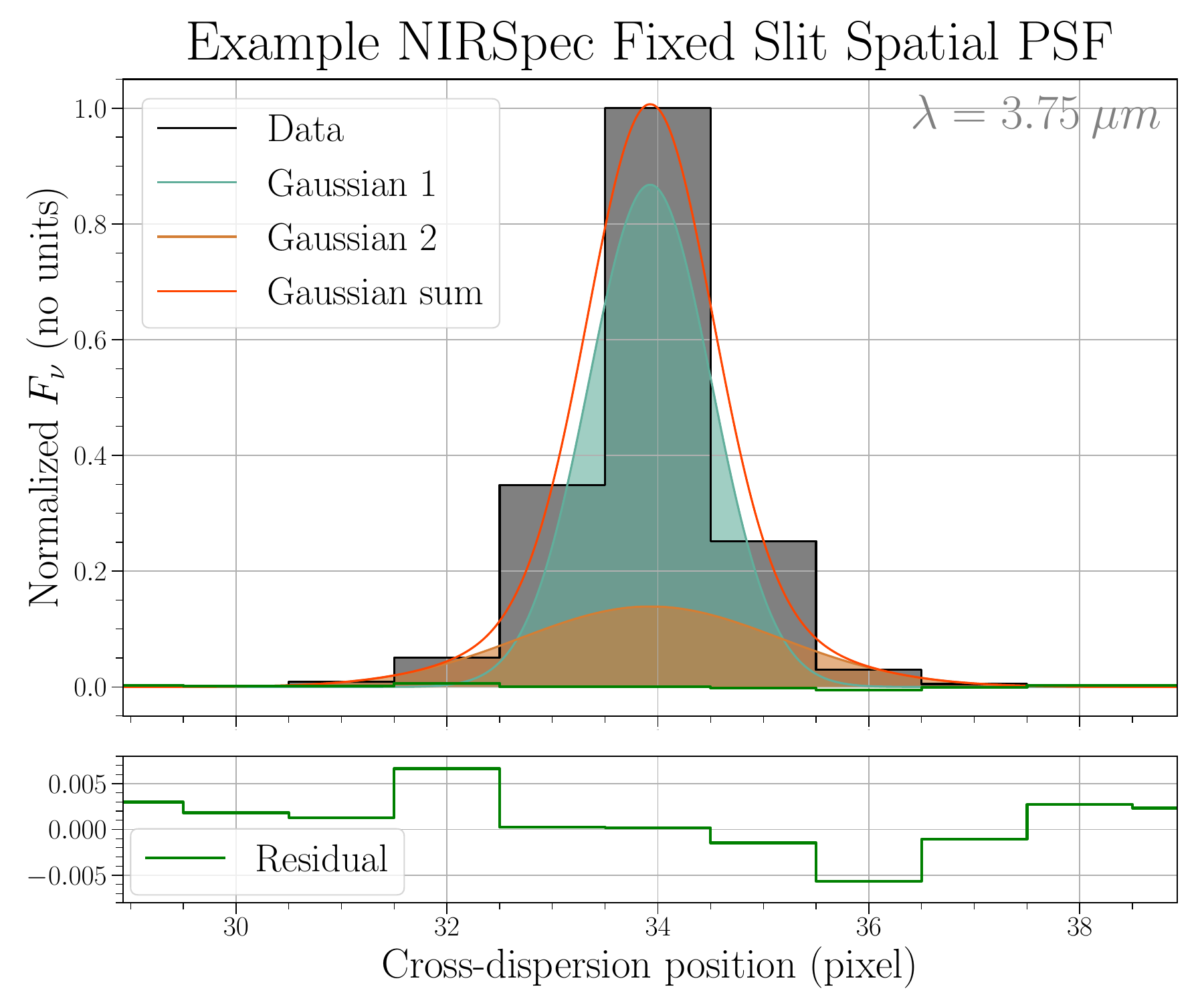}
    \caption{Example of the double Gaussian model fit to a single wavelength of the 2D spectrum of calibration star 1808347. Only the pixels being fit by the model are shown. The data are shown as a black histogram with gray fill. The green line with green fill shows the shape of the primary Gaussian $(\rm{A_1}, \sigma_1, \mu)$, while the pink line with pink fill shows the shape of the secondary Gaussian $(\rm{A_2}, \sigma_2, \mu)$. The sum of the two Gaussians (i.e. the full double Gaussian model) is shown by the blue line. The residual of the double Gaussian model fit subtracted from the data is shown as a black histogram, which lies very close to zero and is also shown in the zoomed-in panel below the main figure. The residuals across all wavelengths are smoothed in the wavelength direction using a 30-pixel rolling median and are added to the double Gaussian model to create the final PSF model. We emphasize that the primary and secondary Gaussians have no physical meaning and their form was chosen purely to best fit the PSF. It is therefore incorrect to interpret the secondary Gaussian as representing a more spatially extended physical source.}
    \label{fig:example gaussian}
\end{figure*}

In the first round of fitting (R1), the five free parameters \(\rm{A_1}\), \(\rm{A_2}\), \(\sigma_1\), \(\sigma_2\), and \(\mu\) are constrained only by the input bounds listed in Equations \ref{eq:u_input}, \ref{eq:sigma1_input},  \ref{eq:sigma2_input}, and \ref{eq:A1_A2_input}, below. An example of a fit of the double Gaussian model to a single wavelength column is shown in Figure \ref{fig:example gaussian}. The inequality \(\rm{A_1 > A_2}\) is enforced by requiring that $\rm{A_1}$ be greater than half the maximum value at that wavelength ($\rm{F_{max}}$), and $\rm{A_2}$ be less than that. The inequality \(\sigma_2 > \sigma_1\) is not directly enforced but is encouraged by their respective upper and lower bounds (see Equations \ref{eq:sigma1_input}, \ref{eq:sigma2_input}) to allow greater flexibility in fitting. 
 \begin{equation}
\rm{\mu \: \mathrm{input} =
  \left\{\begin{array}{@{}l@{}}
    \Delta \mu < 3, \quad \mathrm{(R1)}\\
    \rm{fixed,}\quad \mathrm{(R2, R3)}
  \end{array}\right.\,}
  \label{eq:u_input}
\end{equation}
\begin{equation}
\rm{\sigma_{1} \: \mathrm{input} = 
  \left\{\begin{array}{@{}l@{}}
    0.1 < \sigma_{1}< 1.5,\quad \mathrm{(R1)}\\
    \rm{fixed},\quad \mathrm{(R2, R3)}
  \end{array}\right.\,}
  \label{eq:sigma1_input}
\end{equation}
\begin{equation}
\rm{\sigma_{2} \: \mathrm{input} =
  \left\{\begin{array}{@{}l@{}}
    0.4 < \sigma_{2}< 30, \quad \mathrm{(R1)}\\
    \rm{fixed},\quad \mathrm{(R2, R3)}
  \end{array}\right.\,}
  \label{eq:sigma2_input}
\end{equation}
\begin{equation}
 \rm{A_{2} \: \mathrm{input} =
  \left\{\begin{array}{@{}l@{}}
    \rm{A_{2} < \frac{F_{max}}{2} < A_{1}}, \quad \mathrm{(R1, R2)}\\
    \rm{A_2 =B A_1,\quad} \mathrm{(R3)}
  \end{array}\right.\,}
 \label{eq:A1_A2_input}
\end{equation}
    At the end of the first round, both widths $(\sigma_1, \sigma_2)$ and the centroid \(\mu\) are smoothed using a 30-pixel rolling median like that in Figure \ref{fig:variation with wavelength} and fixed to their smoothed values. 
%\textit{Need to update with slight changes to "rounds", ie we do no bounds, then width and cent fixed, then B fixed (for the star)}    
In the second round of fitting (R2), the widths $(\sigma_1, \sigma_2)$ and centroid  $(\mu)$ are fixed to their smoothed values, leaving the amplitudes $(\rm{A_1, A_2})$ as free parameters. Using the resulting best fit values for the amplitudes $(\rm{A_1, A_2})$, we define the scaling factor $\rm{B}$ at each wavelength bin as $\rm{B = \frac{A_{2}}{A_{1}}}$. $\rm{B}$ is the ratio of the integrated areas of the primary and secondary Gaussian and defines the shape of the PSF. In the third and final round of fitting (R3), all variables, including the newly defined $\rm{B}$ which is smoothed using a rolling 30-pixel median, are fixed except for $\rm{A_1}$. This reduces the free variables at a given wavelength to a single parameter, $\rm{A_1}$, which is fit to produce the complete double Gaussian model.

    From this point forward, we refer to $\rm{A_1}$, $\rm{A_1}$, etc. as the primary, secondary, etc. amplitude of the PSF fit to a \textit{generic} point source, while $\rm{A_{1,\star}}$, $\rm{A_{2,\star}}$, etc. will refer to the amplitudes of the PSF fit \textit{specifically} to the 2D spectrum of star 1808347. The 2D and 1D residuals of the double Gaussian PSF model compared to the spectrum of star 1808347 are shown in the third row of Figure \ref{fig:2DPSF}. As Figure \ref{fig:2DPSF} shows, the double Gaussian model characterizes the PSF very well, leaving residuals on the order of $\sim1\%$ relative to the amplitude of the original star spectrum.  We smooth the remaining $\sim1\%$ residual in the wavelength direction using a 30-pixel rolling median in order to avoid overfitting and scale the smoothed residual by $\rm{\frac{A_1}{A_{1,\star}}}$. This scaled and smoothed residual is added to the double Gaussian model to produce the final generic PSF model: %In order to fully characterize the PSF, we treat this $\sim1\%$ residual as a constant array $C$ and append it to the double Gaussian model, where $C$ is defined as the 2D normalized residual at each wavelength of the double Gaussian model shown in row three of Figure \ref{fig:2DPSF}, scaled to the 2D spectrum being fit (i.e. \(C =  \mathrm{2G \: residual}*F_{\nu, \mathrm{data}}/F_{\nu,\mathrm{star}}\)). The final model for the PSF is:

\begin{equation}
\begin{split}
&\rm{PSF_{F}(x, \lambda)} =\\
&\rm{A_1}(\lambda)\left(\frac{1}{\sqrt{2\pi}\sigma_{1}(\lambda)}\right.\exp\left(\frac{-(\rm{x}-\mu(\lambda))^2}{2\sigma_{1}(\lambda)^2} \right) +\\
&\frac{B(\lambda)}{\sqrt{2\pi}\sigma_2(\lambda)}\:\exp\left(\frac{-(\rm{x}-\mu(\lambda))^2}{2\sigma_{2}(\lambda)^2}\right)+\left.\rm{\frac{C(x, \lambda)}{A_{1,\star}(\lambda)}}\right).
\end{split}
\label{eq:PSF final}
\end{equation}

\noindent
where $\mu$, $\sigma_1$, $\sigma_2$, and $\rm{B}$ are all smoothed using a 30-pixel rolling median, $\rm{A_1}$ and $\rm{A_{1,\star}}$ are the primary amplitudes of the generic and specific (star 1808347) PSF fits as described above, and $\rm{C(x, \lambda)= [F_\star-PSF_{2G, \star}]_{\text{smoothed}}}$ is the residual defined as the difference between the 2D stellar spectrum and its best fit double Gaussian model, smoothed by a rolling 30-pixel median in the wavelength direction. 

\subsection{PSF Width Variation}\label{subsec:width variation}
    \begin{figure*}
    \centering
    \includegraphics[width=1\linewidth]{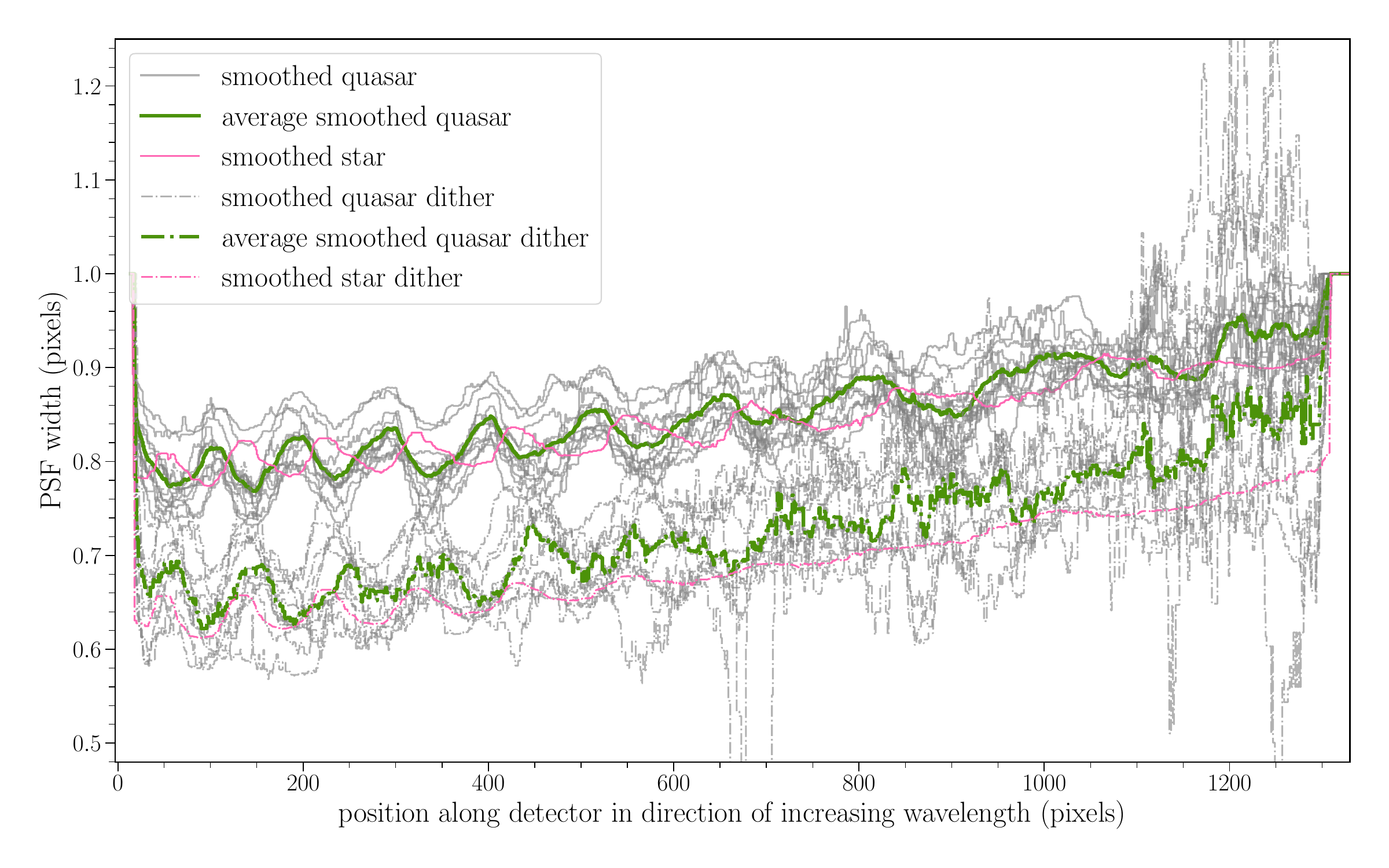}
    \caption{PSF width variation with wavelength. The smoothed data for each of the 12 individual quasars is shown in gray, while the average of the individual dithers for all twelve quasars is shown in green. Solid lines are combined and rectified level 3 data, and dash-dotted lines are for individual level 2 dithers (one per quasar). The pink lines represent the PSF width of star 1808347 for the level 2 (dash-dot) and level 3 (solid) data. All widths are from the Single Gaussian model $\rm{(PSF_{1G})}$ to illustrate aspects of the PSF which would be more complex to plot in the Double Gaussian model. We highlight in this figure two distinct and important phenomena: the width variation with wavelength (i.e. wiggles), and the broadening of the PSF in the level 3 data. The wiggles are present across all sources, implying this is a systematic rather than a physical effect. The rectified and co-added level 3 PSF width is on average broader by $\sim0.15$ pixels than the individual level 2 dithers. To avoid undue broadening of the PSF, we model and subtract off the PSF in the individual level 2 dithers before re-entering them into the regular Stage 3 \texttt{jwst} pipeline.}
    \label{fig:dither wiggles}
\end{figure*}
We find that the Stage 3 resampling and rectification steps of the $\texttt{jwst}$ pipeline artificially broaden the PSF width by a fraction of a pixel at all wavelengths. This is seen in Figure \ref{fig:dither wiggles}, where the width of the single-Gaussian fit to the spatial PSF for a single level 2 (gray, dash-dot) and level 3 (gray, solid) dither is shown for each quasar as well as the calibration star (pink, dash-dot and solid, respectively) as a function of wavelength. The rectified and resampled level 3 data are $\sim0.15$ pixels broader than the level 2 data for the quasars and the calibration star. Since we plan to model and then subtract the PSF in the 2D spectra in order to reveal extended emission, we perform the PSF analysis and subtraction on the three individual, un-resampled level 2 dithers of each target. We then reintroduce the PSF-subtracted 2D spectra to the Stage 3 pipeline to produce level 3 data products. 

The average of the level 2 and level 3 quasar PSF widths are shown in bold green. As can be seen by the average smoothed widths, the quasar profiles follow the same quasi-periodic pattern (i.e. wiggles) in all observations, though it is shifted approximately half a phase between the individual level 2 dithers and the resampled level 3 data. These wiggles are a result of undersampling of the PSF, as the PSF width is at a maximum when the trace on the detector is split between two pixels in the spatial direction, and thinnest when the trace is centered directly on a pixel. The wiggles also appear in the smoothed single-Gaussian PSF fit width of the calibration star, shown in pink. The wiggles have a shorter period at shorter wavelengths and have a maximum amplitude of about 0.1 pixels for both level 2 and level 3 data. This variation in the PSF width across the Fixed Slit detector is also seen by \cite{Espinoza_2023}, and the NIRSpec IFU community is facing a similar undersampling effect \citep{LiuW2024, Dumont2025WIggleWICKED}. The effects of the Fixed Slit spatial PSF have been partially accounted for in several Trans-Neptunian Object and asteroid-focused papers \citep{Rivkin2023PSJ.....4..214R, Thomas2025PSJ.....6..115T}, but, to our knowledge, no numerical or analytical models of the PSF width variation with respect to wavelength have been published. %Interestingly, the variation in the level 3 data of the star is offset $\sim$ half a phase from the average level 3 data of the quasars, but is only offset by $\sim$ a quarter of a phase in the level 2 data. We speculate that the phase may vary with time of source type, but have no definitive explanation.

%We believe our modeling method may be the first published explicit characterization of this PSF variation in the NIRSpec FS 2D spectra, though we hope that a JWST-team approved numerical or analytical model may be made public in the future. 

\subsection{Applying the Point Spread Function to Quasars}\label{subsec:applying PSF}
    In order to apply the PSF characterized above to the observed quasars, we assert that the quasars are point sources and that all point sources have identical PSF shapes in the NIRSpec detector. We additionally assert that the host galaxies (if detected) are \textit{not} point-like sources and \textit{cannot} be modeled by the PSF, so that any excess emission after PSF subtraction may only be attributed to the host galaxy. Since the brightest parts of our galaxies are spatially unresolved, the galaxy centers will also have point-source components which are often co-located with the quasar PSFs. Since we attribute all point-source flux to the quasar, we expect the galaxy to be over-subtracted in all cases, with the most severe over-subtraction occurring when the galaxy is centrally concentrated and the host-to-quasar light ratio is high.

    For the initial characterization of the PSF using the calibration star, three iterations of fitting (R1, R2, R3) were performed on each level 2 dither with constraints placed by the previous fits, resulting in three PSF models, one for each dither position. For subsequent modeling of the quasars, only two iterations of fitting are performed per level 2 dither, the first with the centroid, $\mu$, and primary amplitude, $\rm{A_1}$, as free parameters at each wavelength $\lambda$, then with $\mu$ fixed to its best fit value smoothed by a 30 pixel rolling median and $\rm{A_1}$ as the only free parameter. All components of the PSF shape (i.e. $\sigma_1$, $\sigma_2$, $\rm{B}$, and $\rm{\frac{C}{A_{1,\star}}}$) are predetermined from the calibration star PSF modeling described in $\S$ \ref{subsec:characterizing PSF}. The final model is subtracted from the 2D spectrum to reveal any extended emission from the host galaxy. 
%     \begin{equation}
%\mu \: \mathrm{input} =
%  \left\{\begin{array}{@{}l@{}}
%    \Delta \mu < 3 \quad \mathrm{(R1:quasar)}\\
%    fixed\quad \mathrm{(R2:quasar)}
%  \end{array}\right.\,
%  \label{eq:qu_input}
%\end{equation}
%\begin{equation}
%    A_2 = \frac{\sigma_1}{\sigma_2}B A_1\quad \mathrm{(R1, R2:quasar)}
%    \label{eq:qA1_A2_input}
%\end{equation}

\begin{figure*}
    \centering
    \includegraphics[width=1\linewidth]{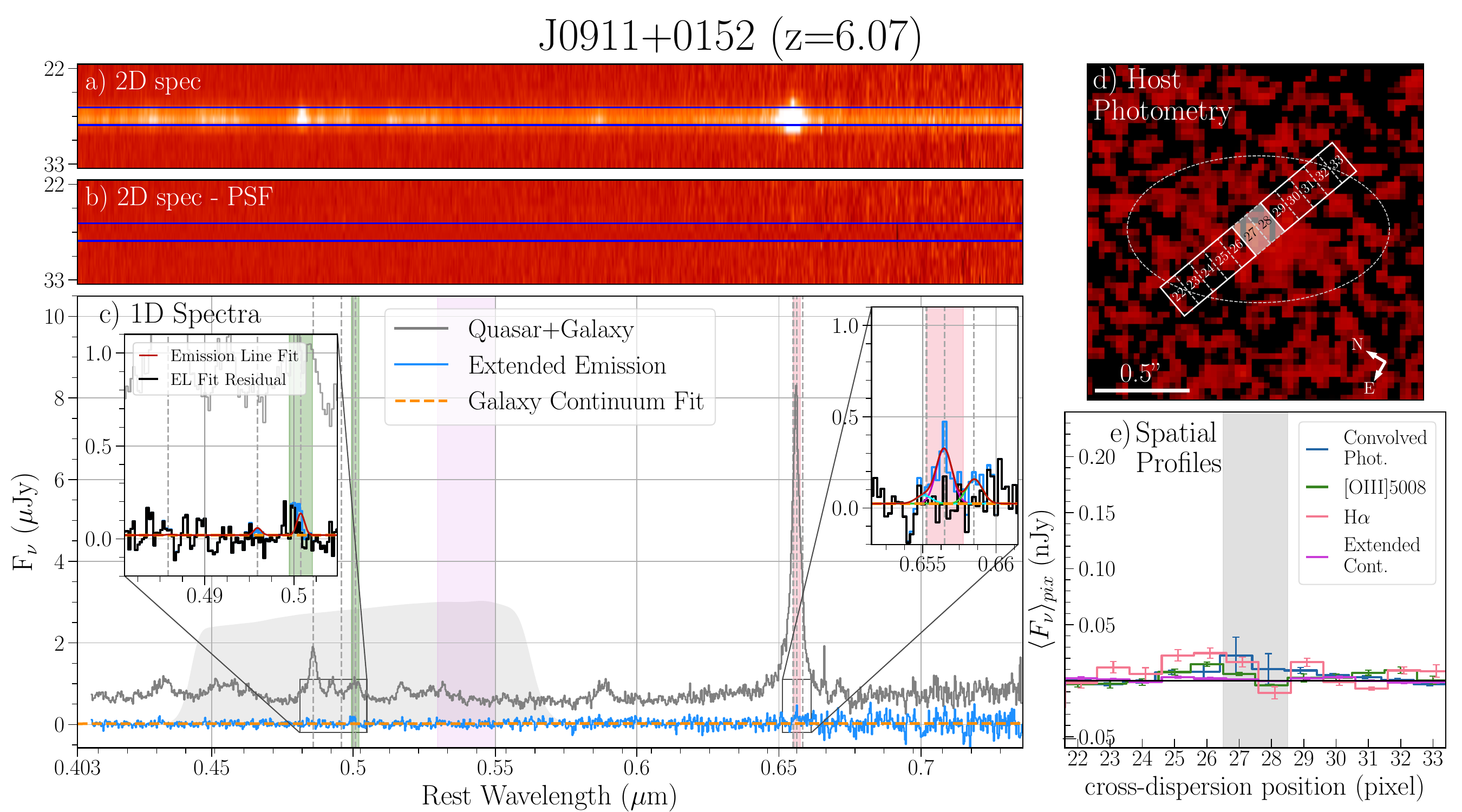}
    \caption{See $\S$ \ref{sec:individual objects} for discussion of the plot contents.
   Panels $(a)$ and $(b)$: Show the 2D spectrum before $(a)$ and after $(b)$ PSF subtraction for J0911$+$0152. Note that the aspect ratio of the figure means that the pixels are not square. The y-axes span $1''.2$ (12 pixels). The x-axes match that of panel $(c)$. The edges of the central two pixels in which the quasar dominates are marked by horizontal blue lines. Panel $(c)$: 1D quasar + galaxy spectrum (gray) and PSF subtracted spectrum (light blue). The wavelengths of prominent emission lines are marked with vertical dashed lines, and the linear continuum fit to the extended galactic emission is shown as a bold, orange dashed line. The shaded regions show the wavelength ranges over which the continuum and the emission line spatial profiles are averaged to produce the extended continuum (magenta), $\rm{[OIII]}5008$ (green), and $\rm{H}\alpha$ (pink) profiles shown in panel $(e)$. The response function of the F356W filter as a function of wavelength, normalized to half the plot height, is shown in light gray for illustrative purposes. In both inset plots, the best fit emission line model is plotted as a smooth red line, and the emission line fit residual is represented by a black histogram. The individual $\rm{[NII]}6551$, $\rm{[NII]}6585$, and $\rm{H}\alpha$ line fits are plotted in cyan, magenta, and lime green, respectively. Panel $(d)$: $1''.78 \times 1''.78$ postage stamp cutout of the NIRCam F356W image of the host galaxy with the quasar point source and any nearby galaxies subtracted off, taken from \cite{ding2025shellqsjwstunveilshostgalaxies}. The position and angular size at the time of observation of the S200A2 slit is overlaid on the image with the central 12 NIRSpec pixels used for data analysis numbered for comparison with panels (\textit{a}), (\textit{b}), and (\textit{e}). The central two pixels where the quasar dominates are shaded gray. The two $0''.2$ by $0''.5$ regions outlined in white define the modified slit in which we perform our flux calculations. Also overlaid is the ellipse from which we extract the total host galaxy flux used to calculate the star formation rate for the whole galaxy ($\S$\ref{subsec:star formation rates}). Panel $(e)$: Average spatial profiles per NIRSpec pixel of the extended continuum (magenta), $\rm{[OIII]}5008$ (green) and $\rm{H}\alpha$ (pink) emission lines, and the photometry convolved with the S200A2 Fixed Slit aperture (blue), each offset by 0.1 pixels to prevent overlap of error bars. The central two pixels in which the quasar dominates and the galaxy is over-subtracted is shaded light gray.}
    \label{fig:J0911 spectrum}
\end{figure*}

\begin{figure*}
    \centering
    \includegraphics[width=1\linewidth]{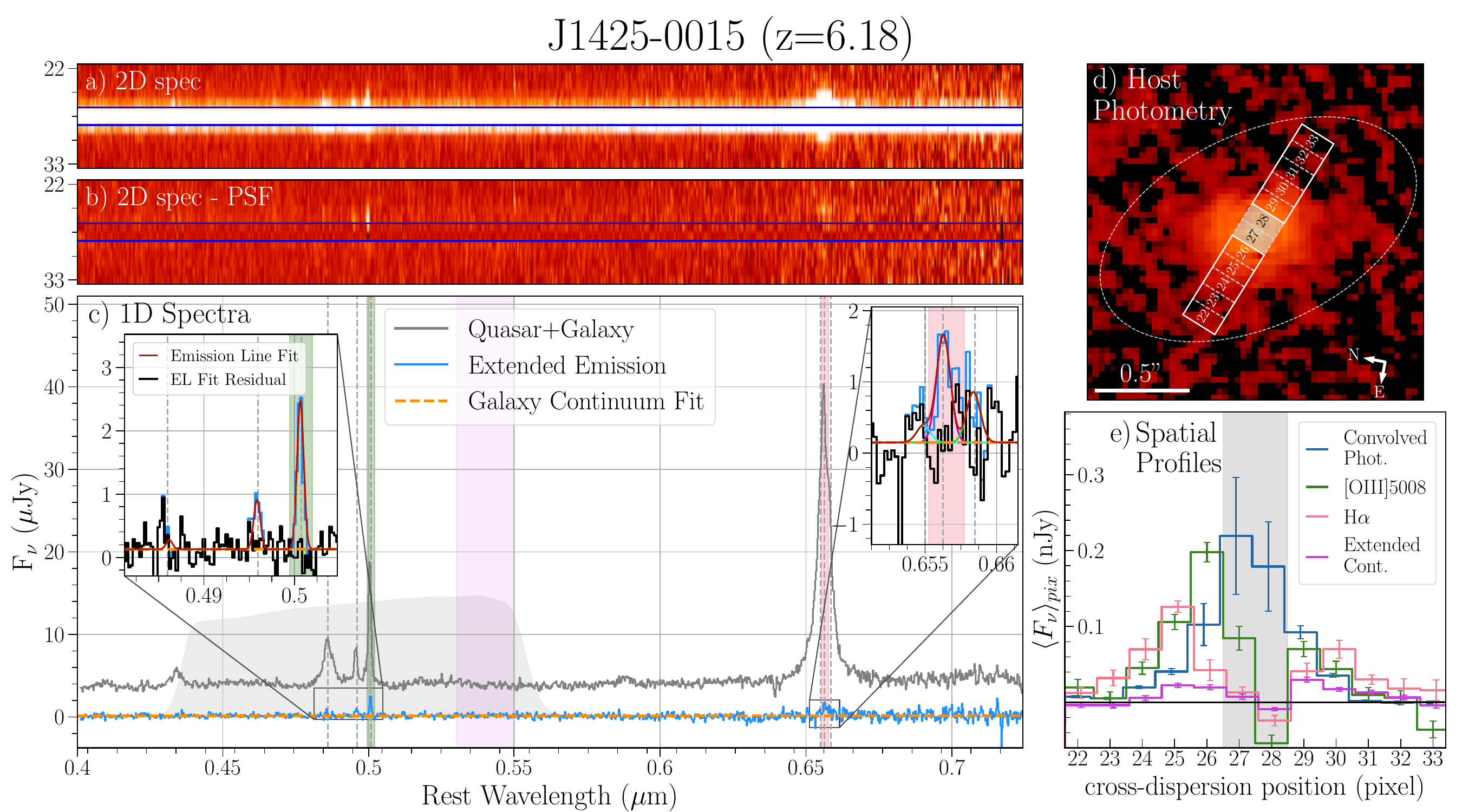}
    \caption{As in Figure \ref{fig:J0911 spectrum} for J1425$-$0015.}
    \label{fig:J1425 spectrum}
\end{figure*}

\begin{figure*}
    \centering
    \includegraphics[width=1\linewidth]{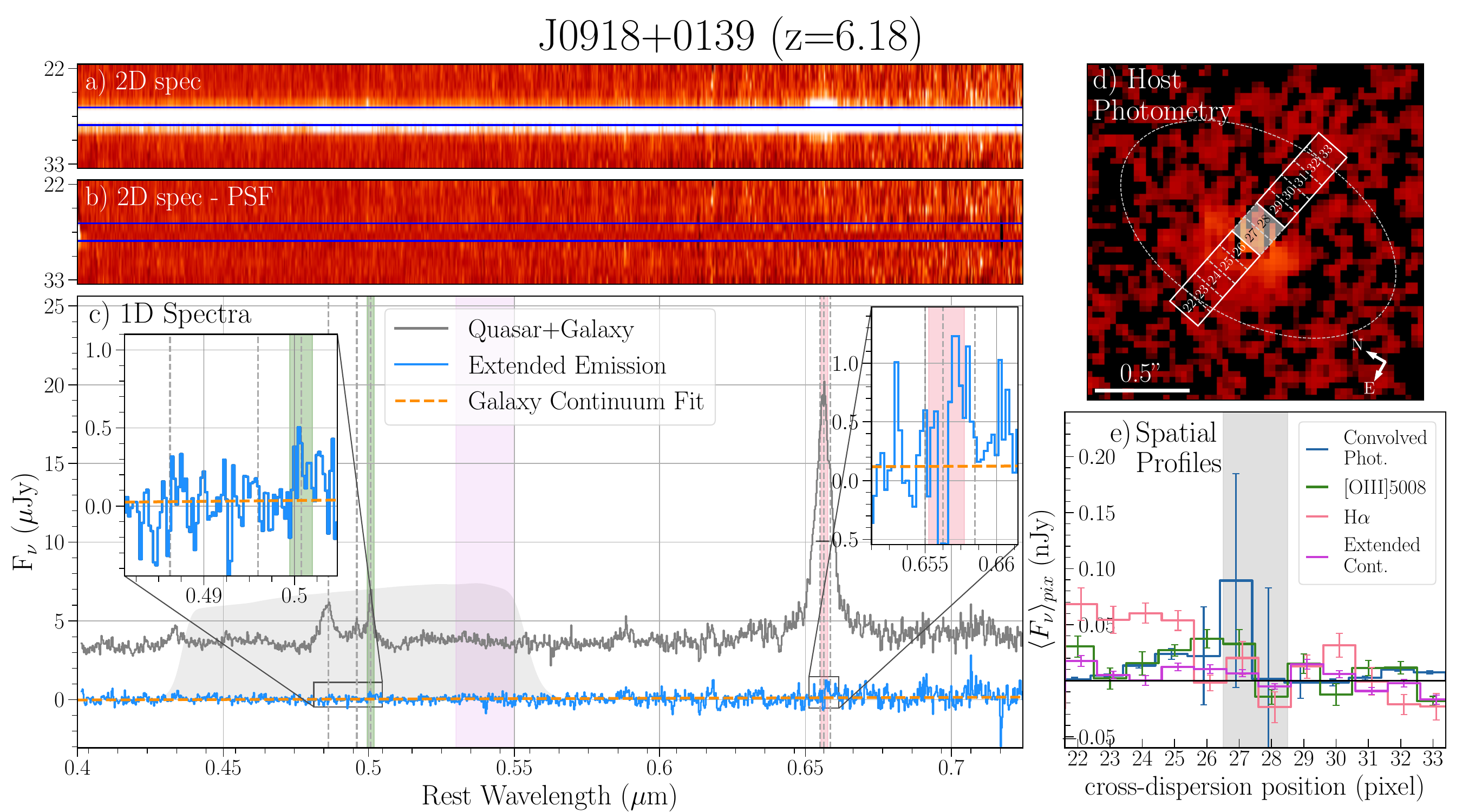}
    \caption{As in Figure \ref{fig:J0911 spectrum} for J0918+0139.} 
    \label{fig:J0918 spectrum}
\end{figure*}

\begin{figure*}
    \centering
    \includegraphics[width=1\linewidth]{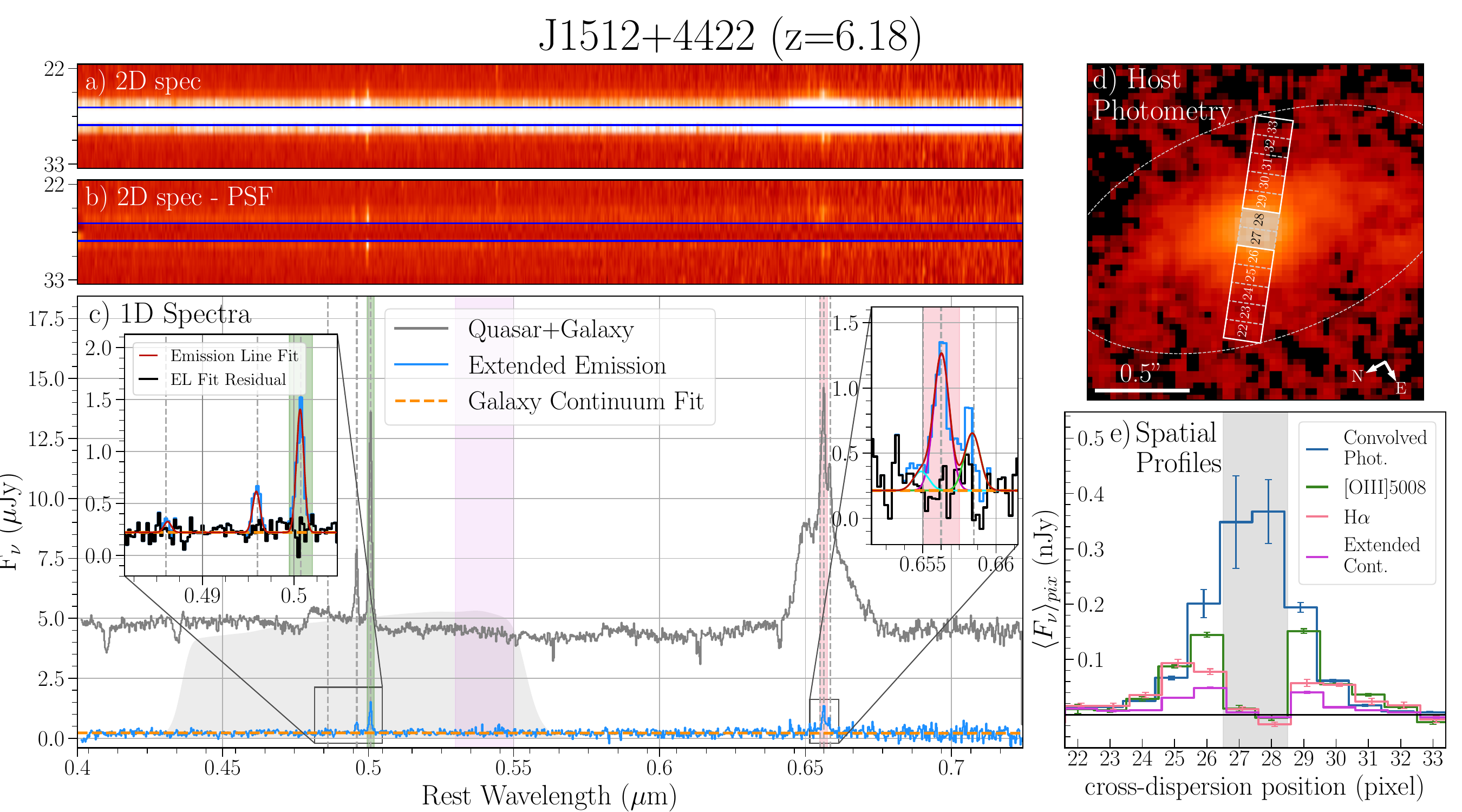}
    \caption{As in Figure \ref{fig:J0911 spectrum} for J1512+4422.}
    \label{fig:J1512 spectrum}
\end{figure*}

\begin{figure*}
    \centering
    \includegraphics[width=1\linewidth]{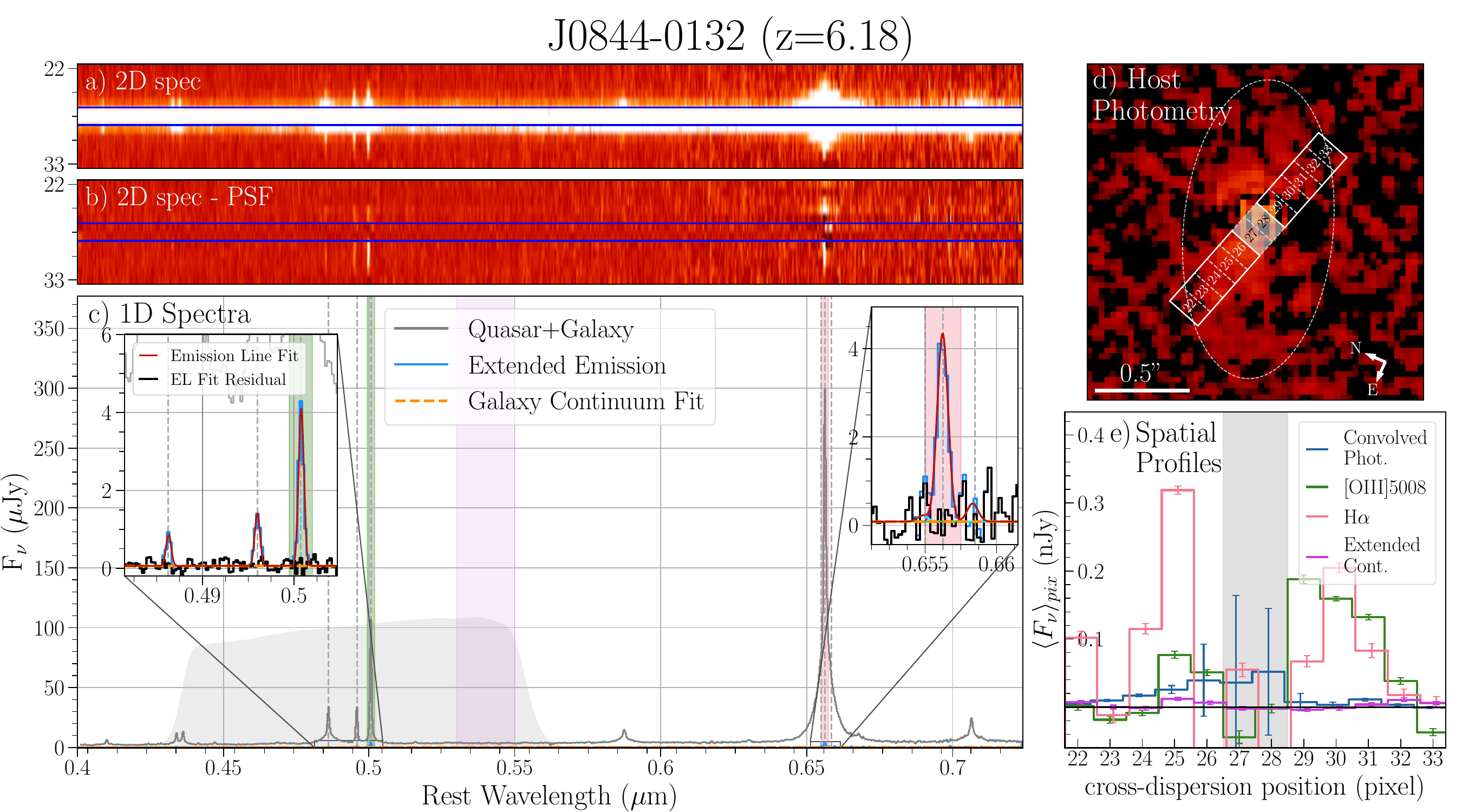}
    \caption{As in Figure \ref{fig:J0911 spectrum} for J0844$-$0132.}
    \label{fig:J844132 spectrum}
\end{figure*}

\begin{figure*}
    \centering
    \includegraphics[width=1\linewidth]{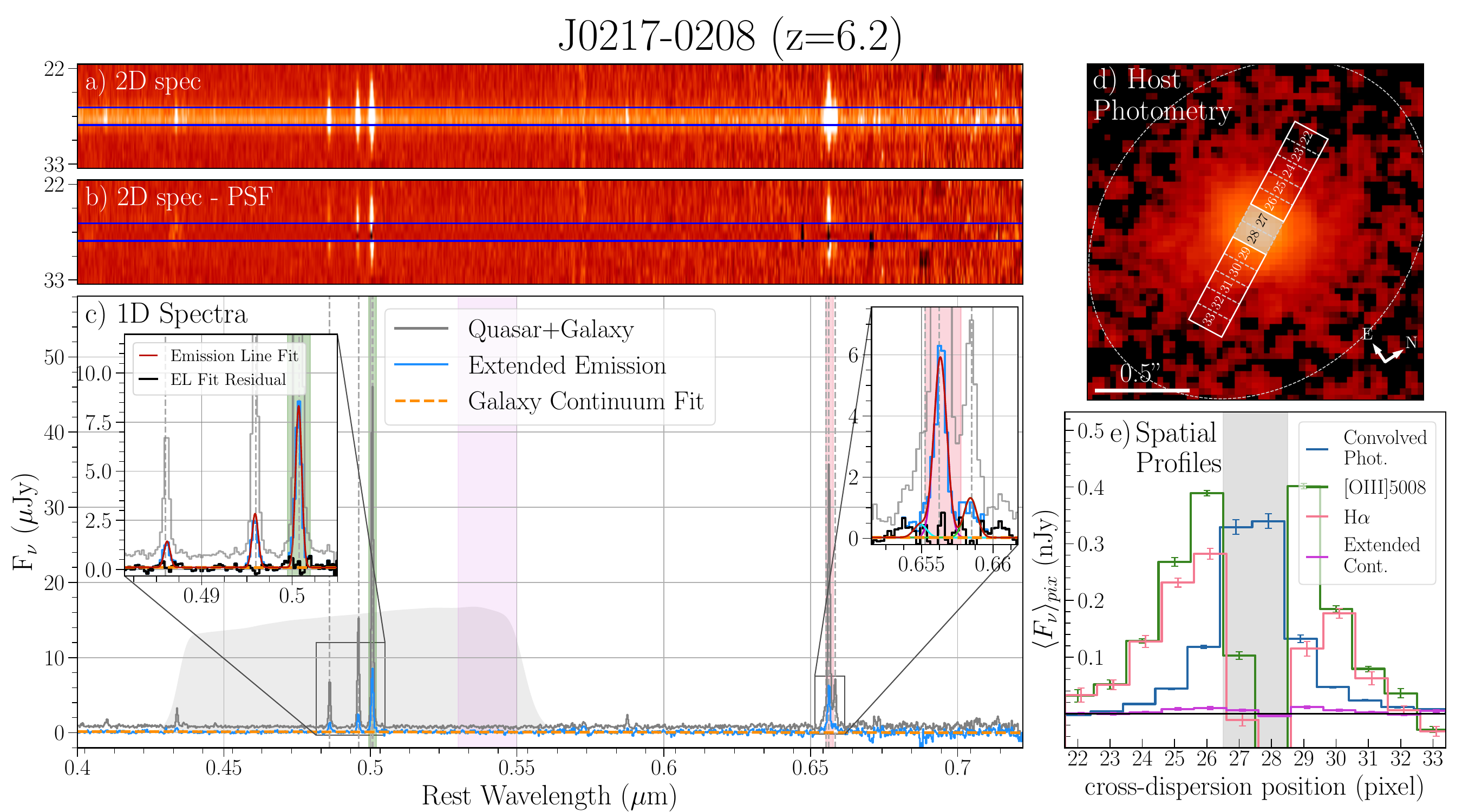}
    \caption{As in Figure \ref{fig:J0911 spectrum} for J0217$-$0208.}
    \label{fig:J0217 spectrum}
\end{figure*}

\begin{figure*}
    \centering
    \includegraphics[width=1\linewidth]{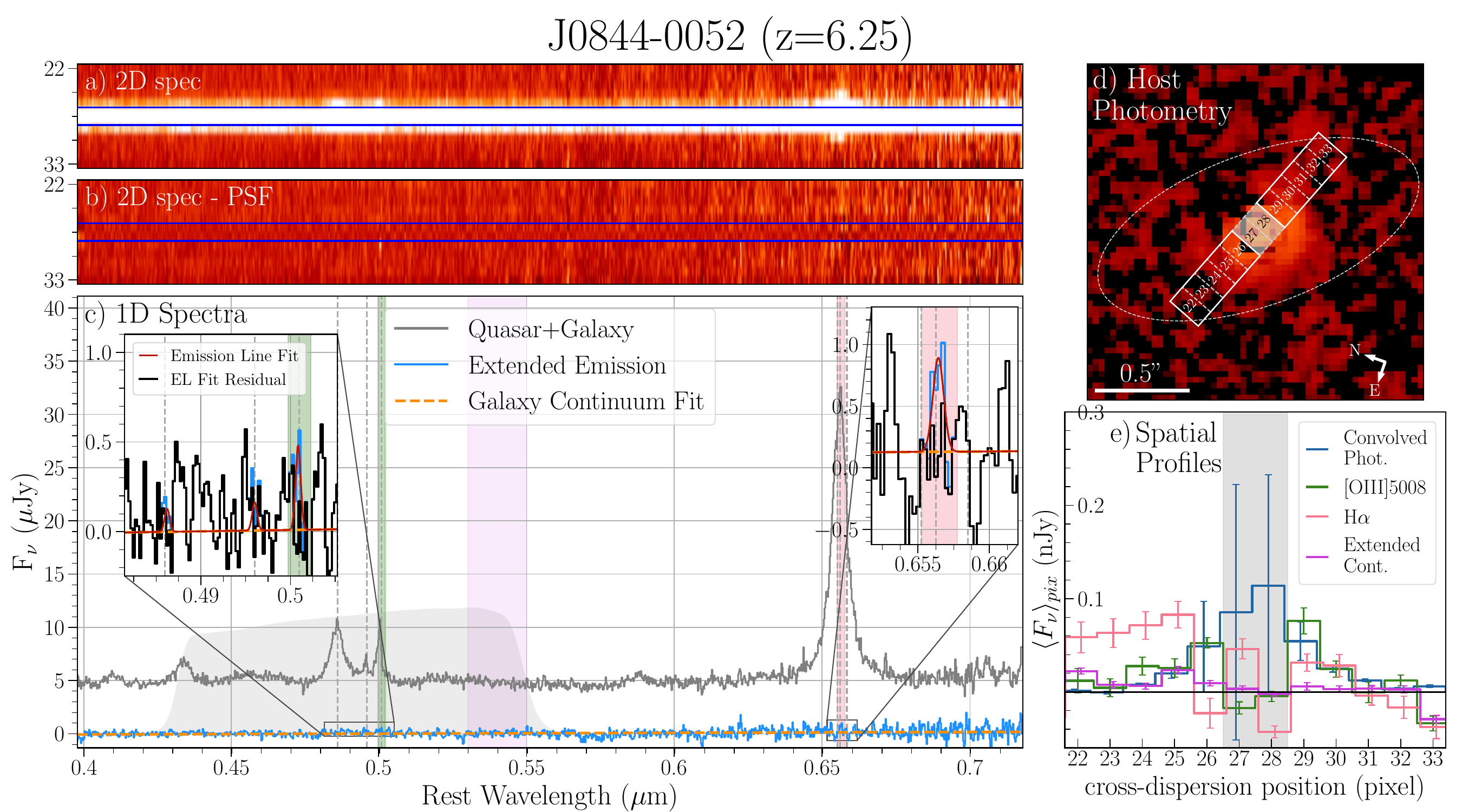}
    \caption{As in Figure \ref{fig:J0911 spectrum} for J0844$-$0052.}
    \label{fig:J84452 spectrum}
\end{figure*}

\begin{figure*}
    \centering
    \includegraphics[width=1\linewidth]{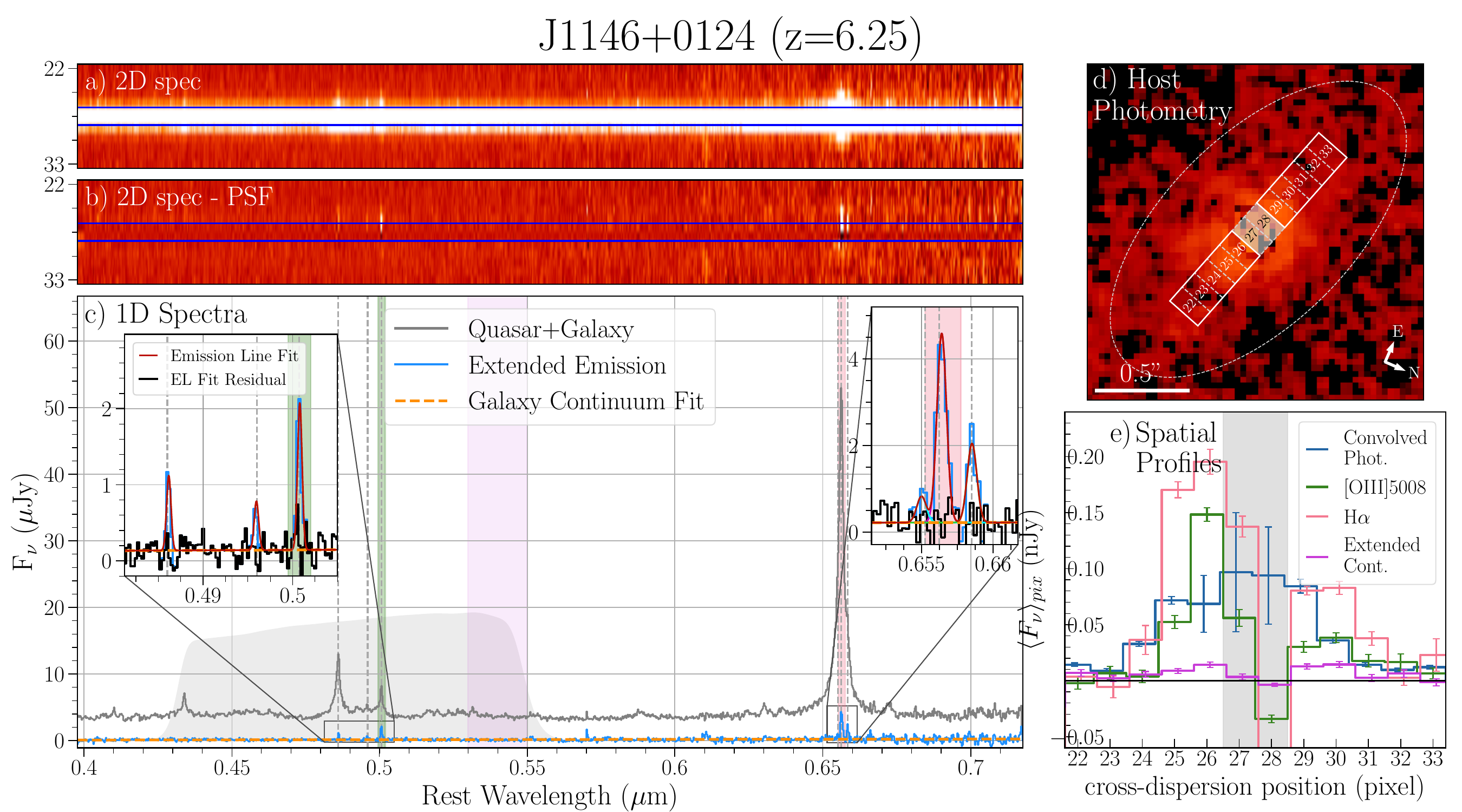}
    \caption{As in Figure \ref{fig:J0911 spectrum} for J1146+0124.}
    \label{fig:J1146124 spectrum}
\end{figure*}

\begin{figure*}
    \centering
    \includegraphics[width=1\linewidth]{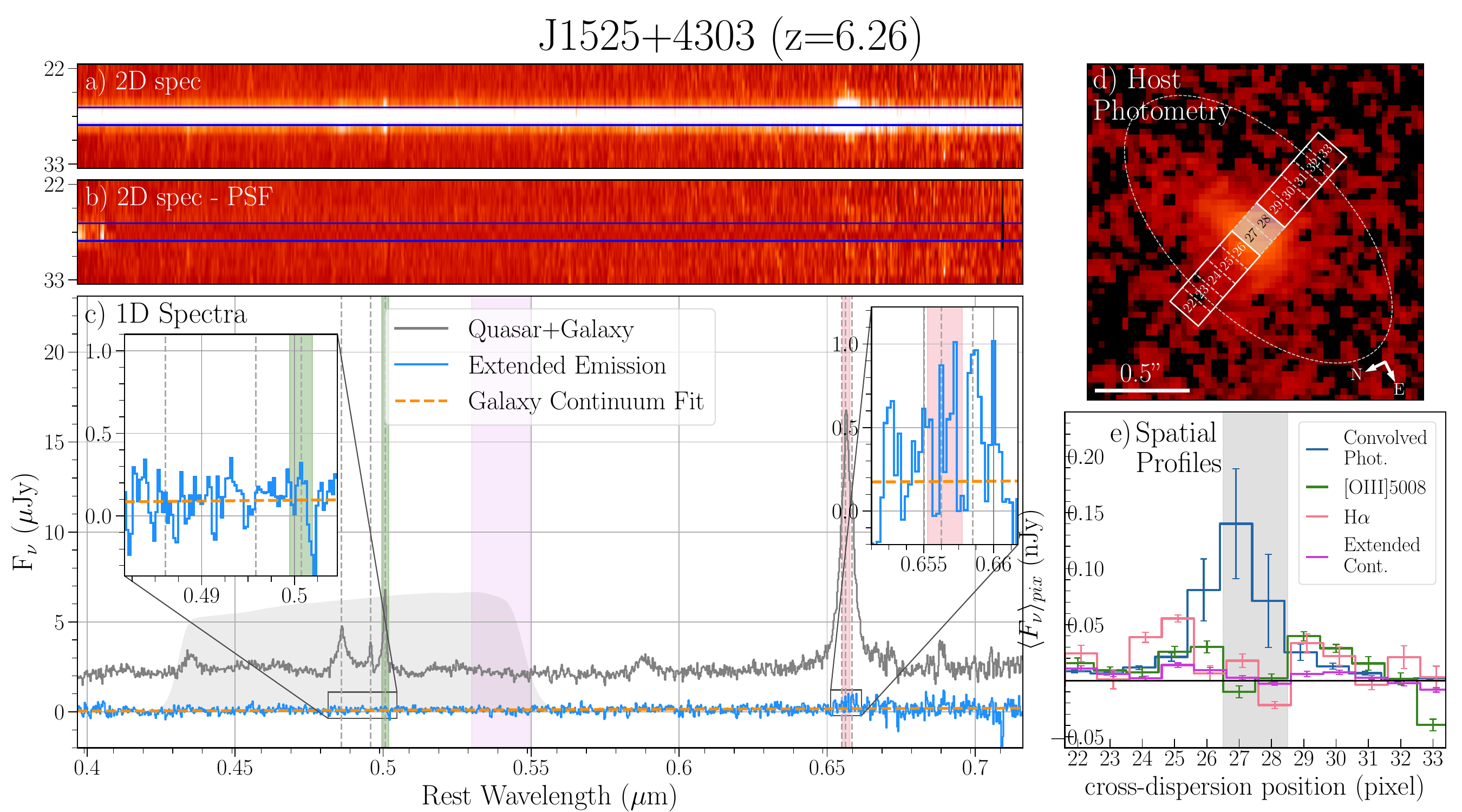}
    \caption{As in Figure \ref{fig:J0911 spectrum} for J1525+4303.}
    \label{fig:J1525 spectrum}
\end{figure*}

\begin{figure*}
    \centering
    \includegraphics[width=1\linewidth]{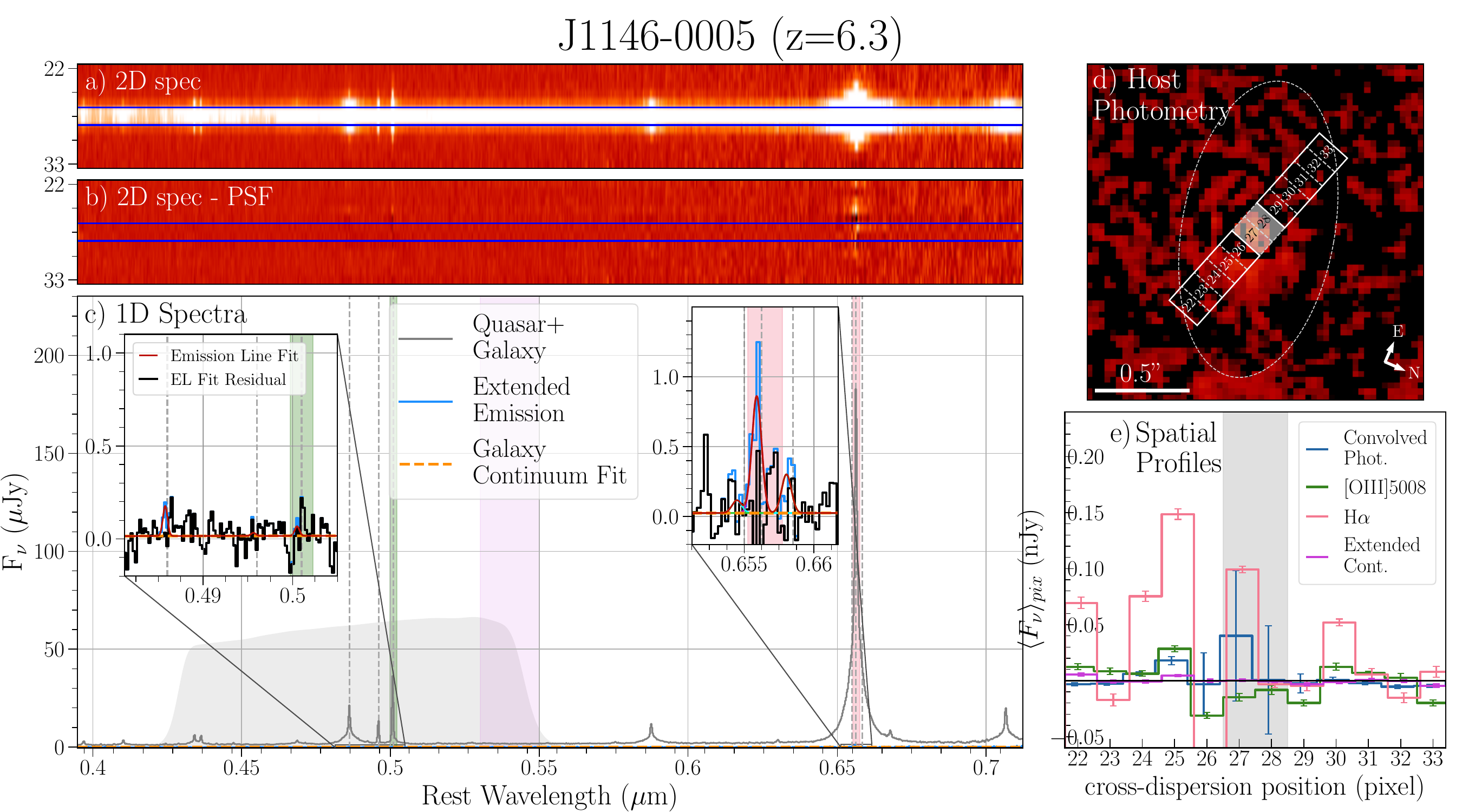}
    \caption{As in Figure \ref{fig:J0911 spectrum} for J1146$-$0005.}
    \label{fig:J11465 spectrum}
\end{figure*}

\begin{figure*}[!ht]
    \centering
    \includegraphics[width=1\textwidth]{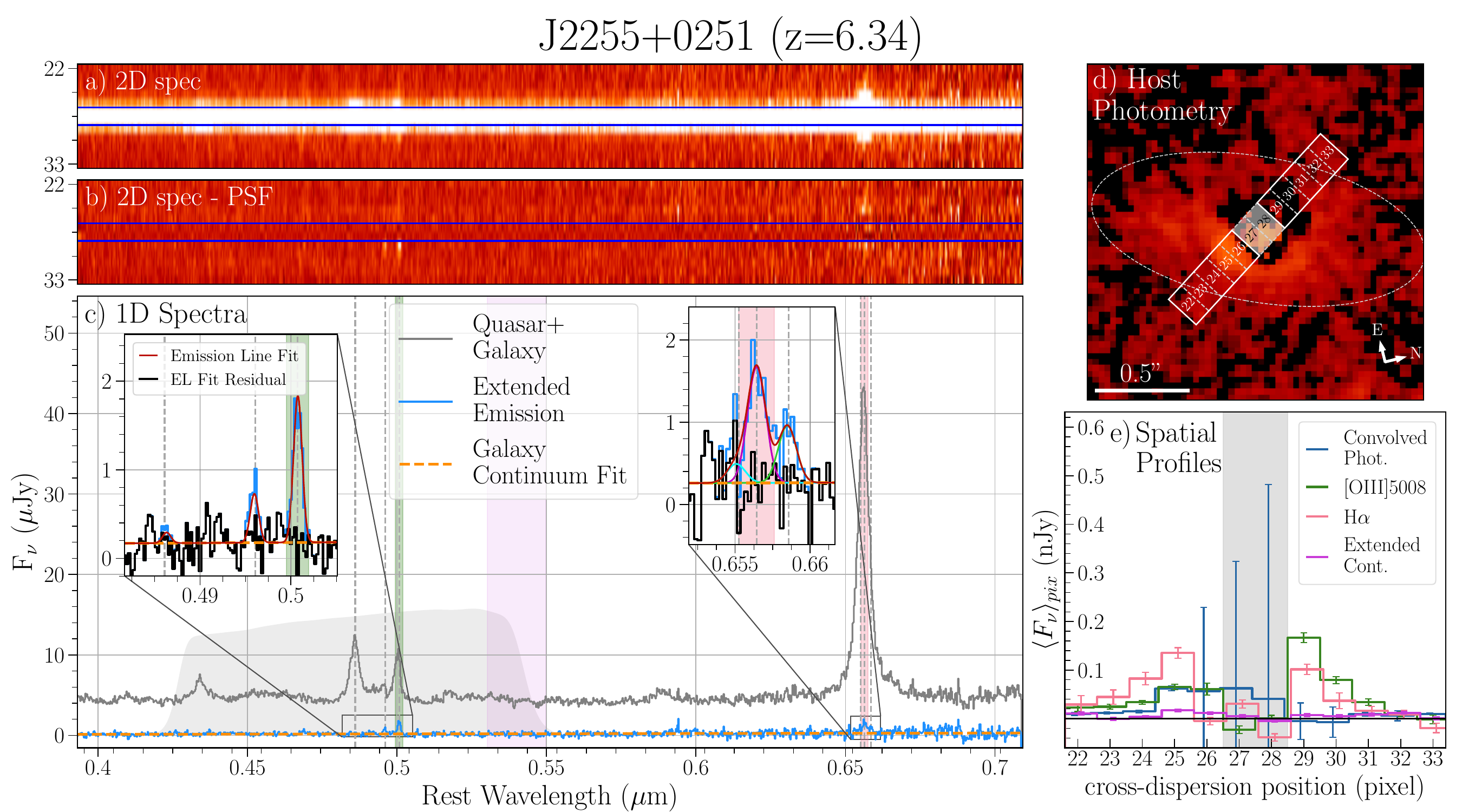}
    \caption{As in Figure \ref{fig:J0911 spectrum} for J2255+0251.}
    \label{fig:J2255 spectrum}
\end{figure*}

\begin{figure*}
    \centering
    \includegraphics[width=1\linewidth]{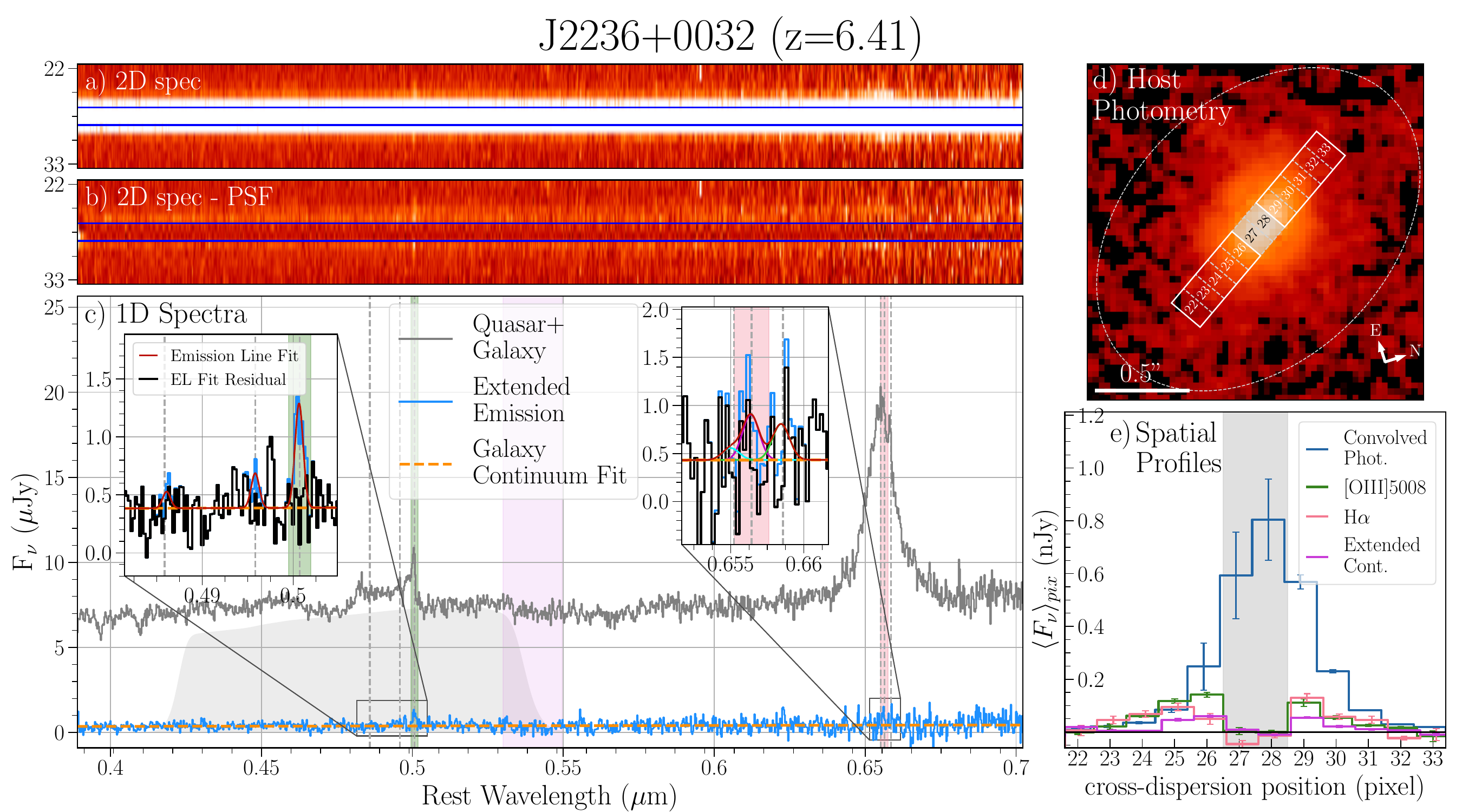}
    \caption{As in Figure \ref{fig:J0911 spectrum} for J2236+0032.}
    \label{fig:J2236 spectrum}
\end{figure*}

Figures \ref{fig:J0911 spectrum}-\ref{fig:J2236 spectrum} show, for all twelve systems in order of ascending redshift, the $(a)$ 2D spectra, $(b)$ 2D spectra with the PSF subtracted, $(c)$ the 1D spectra extracted from panels $(a)$ and $(b)$, $(d)$ the host photometry from \cite{ding2025shellqsjwstunveilshostgalaxies} with the geometry of the fixed slit aperture at the time of observation overlaid, and $(e)$ the average spatial profiles of the stellar continuum and the $\rm{[OIII]}5008$ and $\rm{H\alpha}$ emission lines. The y-limits of panels $(a)$ and $(b)$ are adjusted to eliminate the dither negatives on either side of the main trace. Obvious artifacts in the PSF-subtracted spectrum have been removed by hand, where possible. These artifacts (which contribute less than $0.01\%$ of the flux when summed over all wavelengths in any of the spectra; Figure \ref{fig:2DPSF}) were similarly masked during data analysis. 1D spectra are created from the 2D extended emission spectra with an extraction width of 12 pixels (i.e. six pixels on each side of the center of the spatial profile) and the two central pixels where the quasar dominates masked. The geometry of the extraction regions can be seen as the bold white rectangles in panel $(c)$ of Figures \ref{fig:J0911 spectrum}-\ref{fig:J2236 spectrum}. Extraction widths ranging from two to 14 pixels were tested; the width of 12 pixels provided the best balance between including extended emission and excluding the negative dithers created by the background subtraction step. Points of interest for each object are discussed in $\S$ \ref{sec:individual objects}, and a description of the components of the plots is given in the caption of Figure \ref{fig:J0911 spectrum}.

An implicit but significant assumption made in our fit is that the amplitude at the center of the spatial profile is purely the contribution of the quasar. This is not true in detail, as from photometry we see that in several cases, the brightness of the host in the central region of the slit is a significant fraction of that of the quasar \citep{Ding2023Natur.621...51D}. In an attempt to model the host galaxy light in the center, we tried to simultaneously fit the double Gaussian PSF and an exponential profile representing the galaxy. These fits failed because of the strain of adding additional parameters to an already complex fit for only ten spatial data points per wavelength. Further attempts to simultaneously fit the galaxy and quasar may require finer sub-pixel sampling to better handle the host flux at the center of the profile, a more complex handling of wavelength dependency, or the use of priors from photometric or full spectroscopic analyses \citep{Ding2023Natur.621...51D, Onoue2024arXiv240907113O}. As a result of this limitation, our measured host galaxy continuum and emission line fluxes are lower limits.

%\subsection{Model Robustness}\label{subsec:robustness}
%In order to test the robustness of our model, we compare it to the decomposed host photometry from the NIRCam F356W images analyzed in \cite{ding2025shellqsjwstunveilshostgalaxies}. Most systems with strong extended stellar continuum detected in the F356W images also show extended stellar continuum in their 2D spectra when summed over all wavelengths. Conversely, those systems with weak or no detection of stellar continuum in photometry show weak or no stellar continuum in their extracted 2D spectra. This is reassuring since it tells us that all light from the point-source quasar is accounted for by the PSF, with no non-physical residuals. An example of a system with weak detection is shown in Figure \ref{fig:J0911 spectrum}, where the residuals shown in panel (\textit{b}) are of the order of the background noise. A more thorough comparison of the photometrically and spectroscopically measured continua is given in \S \ref{magnitude corroboration}. 

\subsection{Spatial Limits of the Extended Emission}\label{subsec:spatial extent}
The three-nod dithering pattern used in data collection places hard limits on extended emission at 0.$''$9149 (9.149 pixels) on one side and 1$''$.131 (11.31 pixels) on the other side. For a redshift of $\sim6.2$, this gives an average scale of $5.86$ kpc on either side of the galaxy in which we can extract extended emission. However, we are further limited by self-subtraction during the \texttt{jwst} pipeline background subtraction stage, in which two of the dithers are averaged together and then subtracted from the third in order to remove background flux. This method places the transition from positive to negative dither at two-thirds of the distance between the closer dither and the spectrum center, or about six pixels, leaving us five pixels, or $\sim3.0$ kpc at the redshifts of our objects. %This converts to a maximum spatial extent of $2.93$ kpc for our highest redshift source (J2236+0032, z=6.407) and $3.04$ kpc for our lowest redshift source (J0911+0152, z=6.07).

Due to these limitations, we do not place limits on the sizes of these systems and simply note that, based on the apparent $\rm{[OIII]5008}$ and/or $\rm{H}\alpha$ flux near the edge of their spatial profiles, systems J0217$-$0208 and J2255+0251 may have some small amount of emission line flux beyond $0''.51$ ($\simeq3.0$ kpc) that is cut off due to the dithering pattern.  To approximate the amount of light lost due to self-subtraction, we estimate a background level from the pixels farthest from the spectral trace in the pre-background subtracted dithers, since these pixels are least likely to contain any source flux. We then subtract this estimated background from the dithers and compare the results to the background subtracted dithers from the pipeline. We find that in the case of star 1808347, the fractional light lost due to self-subtraction between dithers is $<1\%$. For our quasar+host systems, including the extended emission lines, we find a fractional light loss of $\sim3\%$ due to self-subtraction. As expected, this value is slightly higher since the extended emission is more susceptible to self-subtraction, but we conclude that a flux loss of $\sim 3\%$ is small compared to the typical noise and therefore is not significant to our analysis.

\section{Data Analysis} \label{sec:data analysis}
%\begin{equation}
%F = (0.03149 \frac{"}{pix})^{2} * pix^{2} * \sum{ap \frac{MJy}{"^{2}}}
%\end{equation}
%\begin{equation}
%m_{AB} = -2.5 * log(\frac{F_\nu}{3631 Jy})
%\label{mag}
%\end{equation}

Having produced robust 2D and 1D spectra of the host galaxies, as seen in panels (\textit{a}), (\textit{b}), and (\textit{c}) of Figures \ref{fig:J0911 spectrum}-\ref{fig:J2236 spectrum}, we now discuss the analyses performed on the spectra and specific points of interest for the individual targets.

\subsection{Stellar Continuum Fit and g-r Color}\label{subsec:continuum and color}
Before fitting for spectral lines, we fit a line ($\rm{y=a\lambda+b}$) to the host galaxy spectra in \(\rm{F}_\nu\)-wavelength space. The data quality do not allow us to do more detailed fitting, e.g., a stellar model. This is acceptable because the main purpose of this fit is to subtract the continuum prior to line fitting and to get a rough color for the galaxy. For all systems, we exclude the regions of the \(\rm{H}\beta\) and $\rm{[OIII]}$ complex \((4800 ~\mathring{\rm{A}}< \lambda_{\rm{rest}} < 5050 ~\mathring{\rm{A}})\), the $\rm{[NII]}$ and \(\rm{H}\alpha\) complex \((6450 ~\mathring{\rm{A}}< \lambda_{\rm{rest}} < 6650 ~\mathring{\rm{A}})\), and the noisy regions within $\sim 10$ pixels of the edges of the detectors. The continuum fluxes inferred from these fits at 5100$~\mathring{\rm{A}}$ are listed in Table \ref{tab:specfitvalues}. Uncertainties are calculated from the standard deviations of the fit parameters. From these values we claim robust ($>3 \sigma$) detection of extended stellar continuum in five of the twelve objects (J1512+4422, J2236+0032, J1146+0124, J1525+4303, J2255+0251), detection at the $2-3\sigma$ level in two objects (J0217$-$0208, J1425$-$0015), and non-detection ($<2\sigma$) in the remaining five objects (J0911+0152, J0918+0139, J0844$-$0052, J0844$-$0132, J1146$-$0005). All systems with robust detections of stellar continuum in the extended spectra also show robust detection (signal-to-noise ratio (SNR) $>8$) of stellar light in the quasar-subtracted F356W photometry. Conversely, all systems with weak or non-detections of the stellar continuum in the extended spectra have SNR$<8$ detections of stellar light in the photometry (see \cite{ding2025shellqsjwstunveilshostgalaxies}).  The exception is J0911+0152, which has a measured SNR of $8.4\pm2.2$ in the photometric decomposition \citep{ding2025shellqsjwstunveilshostgalaxies}. The mismatch between spectroscopic non-detection and photometric detection is likely due to the slit being misaligned with what little host galaxy light there is (see panel $(d)$ of Figure \ref{fig:J0911 spectrum}), making detection of the already faint system impossible using our current spectroscopic decomposition method. %may be because of the inferred compactness of the system, which makes the galaxy light difficult to disentangle from that of the quasar in the long-slit data.  %To verify the model, we integrate the continuum model over the F356W filter response function and convert the fluxes to modified slit magnitudes, where we define the modified slit as two $0''.2$ by approximately $0''.7$ sub-regions of the slit on either side of the quasar which are dominated by extended emission, this not including the central two pixels (see the outlined regions in Figures \ref{fig:J0918 spectrum}-\ref{fig:J0911 spectrum}). The F356W modified slit magnitudes we infer match those derived from the host galaxy photometric models in \cite{ding2025shellqsjwstunveilshostgalaxies} to $ <3 \sigma$ in all cases except one ($J2236+0032$, see \S \ref{magnitude corroboration} for further details).

In those systems showing an extended continuum, we look for stellar absorption lines, particularly in the Balmer series, i.e. $\rm{H}\gamma$ and $\rm{H}\delta$. We find no signs from visual inspection of stellar absorption features in any of the extended host galaxy spectra, even for sources J1512+4422 and J2236+0032, which show prominent stellar absorption lines in their central spectrum \citep{Onoue2024arXiv240907113O}. To test the significance of this non-detection in the quasar-subtracted extended emission, we perform a Monte Carlo analysis (N=10,000) on a noisy mock spectrum, where the noise per pixel is drawn randomly from a Gaussian distribution whose standard deviation is that measured from the continuum fit to the extended emission of each quasar in turn. The equivalent widths of the absorption lines are assumed to be the same as in the center of the galaxy, as calculated from the extracted galaxy spectra of \cite{Onoue2024arXiv240907113O}. The best-fit model robustly detected the absorption lines in $27.2 \%$ of cases in J2236+0032 and $37.8 \%$ of cases in J1512+4422, where robust detection is defined as a model within one standard deviation of the true absorption line and greater than two standard deviations away from zero. Thus, the fact that we did not detect absorption lines in the extended emission of J2236+0032 and J1512+4422 does not rule out that they have the same stellar population as in the core, despite the fact that they have the highest average extended continuum flux of all our targets (see Table \ref{tab:specfitvalues}). As a result, we find it unsurprising that we do not see absorption lines in any of our host galaxy spectra, and do not infer any physical meaning from these non-detections. 

%Of the  targets with confirmed photometric hosts, all but one show  also show evidence of galactic continuum in their spectra when summed over the wavelengths  to  between the OIII and H\(\alpha\) lines where we expect a relatively line-free spectrum. In a few cases, the continuum is strong enough to be seen without summation. An example spectrum is shown in Figure . 

We derive galaxy rest frame colors following \cite{Oke1983} for the linear continuum fits and the full 1D spectra extracted from the quasar-subtracted extended emission. We use the classical SDSS $g$ and $r$ filters and their associated response functions, centered at $4770 ~\mathring{\rm{A}}$ and $6321 ~\mathring{\rm{A}}$ rest wavelength, respectively \citep{Doi_2010}; we forego the $u$ and $i$ filters because they overlap only small portions of the G395M wavelength range. The $g-r$ colors measured from the continuum fits (Column 1) and the extracted 1D spectra (Column 2) and their associated error bars are shown in Table \ref{tab:color and dust}, below. In order to test the robustness of the continuum-derived colors, we also derived rest frame colors using a power-law fit $(y = \lambda^{-\alpha})$ to the galaxy continuum and found the derived $g-r$ colors were unchanged to within their error bars, so we do not list these colors. The colors calculated from the 1D spectra, which include the emission lines, agree with the continuum colors to within their error bars in all cases except for J0217$-$0208. This agreement implies that the emission lines in most of our systems contribute little to the total flux of the galaxy and have small equivalent widths. The exception is J0217$-$0208, which has a weak continuum and very strong emission lines, causing the galaxy to appear much bluer when the emission lines are included. This is typical of galaxies undergoing intense star formation \citep{Salzer1989ApJS...70..479S, Salzer2023AJ....166...81S}.
%All of the measured sources have $g-r$ colors similar to those of Irregular type galaxies at low redshift \citep{Einsenstein2011, Annis2014, Nair2010}. The exception is J0844$-$0132, which could not be fit for a $g-r$ color due to the apparent over-subtraction in the $r$ band resulting in an unphysical negative flux. Having fit and subtracted the continuum, we proceed to fit the emission lines present in the spectra.

\subsection{Line fits and ratios}\label{subsec:line fits and ratio}
    We fit six primary nebular lines $\rm{H}\alpha$, $\rm{H}\beta$, $\rm{[OIII]}4960$, $\rm{[OIII]}5008$, $\rm{[NII]}6551$ and \(\rm{[NII]}6585\) in the extended host galaxy spectra using the \texttt{specutils} python package \citep{specutils2023}. No other emission lines are apparent in our spectra, and any successful fit is required to include all six nebular emission lines listed above. All lines are fit using a Gaussian narrow line profile and are required to have the same redshift and velocity width. The atomic 3:1 line flux ratios were enforced for the $\rm{[OIII]}$ and $\rm{[NII]}$ doublets. The redshifts listed in Table \ref{tab:specfitvalues} are as determined for the quasar in \cite{OnoueInPrep}. The line ratios of interest, \(\rm{[OIII]}5008/\rm{H}\beta, \rm{H}\alpha/\rm{H}\beta\), and $\rm{[NII]6585}/\rm{H}\alpha$, and their associated errors can be found in Table \ref{tab:specfitvalues}. We note in particular that the $\rm{H}\beta$ line is only robustly detected in three systems (J0217$-$0208, J0844$-$0132, and J1146+0124), and the $\rm{[NII]}6551, 6585$ lines are only robustly detected in four systems (J2255+0251, J1512+4422, J0217$-$0208, and J1146+0124). In all other systems, the best fits to the $\rm{H}\beta$ and $\rm{[NII]}6551, 6585$ lines are treated as upper limits. Since the $\rm{[OIII]}5008/\rm{H}\beta$ (R3) and $\rm{[NII]6585}/\rm{H}\alpha$ (N2) line ratios are only weakly correlated with metallicity at high redshift \citep{Maiolino2024b, Harikane2023c, Kocevski2023a}, we limit ourselves to placing the line ratios on the R3N2 Baldwin-Phillips-Terlevich (BPT) diagram \citep{BPT1981PASP...93....5B} (Figure \ref{fig:BPT}), which is discussed in detail in \S \ref{subsubsec:BPT}. For a more in-depth discussion of metallicity measurements in galaxies and AGN at high redshift, please see \cite{Nakajima2023} and \cite{Mazzolari2024A&A...691A.345M}. 
\begin{table*}[h]
    \centering
    \caption{Outputs of the emission line fitting.}%Systems J0844$-$0132 and J1146$-$0005 are marked with * because they are the only two systems whose $\rm{H}\alpha$ emission is so extended that the self-subtraction in the level 2 background subtraction step may be causing errors in the creation of the 2D spectrum. We therefore urge caution when interpreting the values associated with the $\rm{H}\alpha$ line for these two systems.}
    \begin{tabular}{cccccccc}
    \hline\hline
 (1)& (2)& (3)& (4)& (5)& (6)& (7)&\\
        Name&z& $\frac{\rm{H}\alpha}{\rm{H}\beta}$& $log \frac{\rm{[OIII]}5008}{\rm{H}\beta} $& $log \frac{\rm{[NII]}6585}{\rm{H}\alpha}$&   $F_{\rm{cont}}$, 5100$\mathring A$ (nJy)&$\sigma_{\text{cont}}$ & \\
        \hline
        J0911+0152 & 6.055& $-$   & $-$   & $-0.2^{+0.2}_{-\infty}$& 20 $\pm$ 20& 1.0$\sigma$&\\
 J1425$-$0015& 6.1780& $3^{+\infty}_{-3}$& $0.8^{+\infty}_{-0.3}$& $-0.5^{+0.2}_{-0.2}$& 170 $\pm$ 50& 3.4$\sigma$&\\ 
        J0918+0139 &  6.1786& $-$   & $-$   & $-$   & 80 $\pm$ 50&1.6$\sigma$& \\  
        J1512+4422&  6.1806& $6^{+\infty}_{-5}$& $1.1^{+\infty}_{-0.3}$& $-0.4^{+0.1}_{-0.1}$& 310 $\pm$ 20&15.5$\sigma$& \\  
        J0844$-$0132&  6.18299& 3.2 $\pm$ 0.8& $0.7^{+0.2}_{-0.1}$& $-0.8^{+0.1}_{-0.1}$& 70 $\pm$ 40&1.5$\sigma$& \\ 
        J0217$-$0208 &  6.2034&  2.6 $\pm$ 0.4& $0.81^{+0.08}_{-0.07}$& $-0.70^{+0.07}_{-0.06}$& 100 $\pm$ 50&2.0$\sigma$& \\ 
        J0844$-$0052 &  6.2426& $2^{+\infty}_{-2}$& $0.6^{+\infty}_{-0.4}$& $-$   & 70 $\pm$ 70&1.0$\sigma$& \\ 
        J1146+0124&  6.2459& 2.3 $\pm$ 0.6& $0.3^{+0.2}_{-0.2}$& $-0.30^{+0.07}_{-0.07}$&   140 $\pm$ 40&3.5$\sigma$& \\ 
        J1525+4303 &  6.265& $-$   & $-$   & $-$   & 130 $\pm$ 40&3.3$\sigma$& \\ 
        J1146$-$0005&  6.30159& $3^{+\infty}_{-2}$& $-$& $-0.5^{+0.1}_{-\infty}$& 20 $\pm$ 20&1.0$\sigma$& \\ 
        J2255+0251&   6.3329&  $6^{+\infty}_{-9}$&  $1.1^{+\infty}_{-0.4}$&  $-0.2^{+0.1}_{-0.1}$&   200 $\pm$ 50&4.0$\sigma$& \\ 
        J2236+0032 &  6.4047& $2^{+\infty}_{-2}$&  $0.6^{+\infty}_{-0.3}$& $-0.5^{+0.2}_{-\infty}$&   530 $\pm$ 50&10.6$\sigma$& \\ 
        \hline
    \end{tabular}
    \tablecomments{Column 1: Object ID. Column 2: Quasar redshifts as determined from the $\rm{[OIII]}5008$ line center by \cite{OnoueInPrep}. Column 3: Balmer decrement. Columns 4 and 5: Line ratios of interest. Line ratio errors are propagated from uncertainties on the line amplitudes and widths and the noise per pixel in the 1D spectrum. Column 6: Continuum flux ($F_{\nu}$) as measured at rest frame 5100$~\mathring{\rm{A}}$, a region relatively free of emission lines. The associated errors are calculated from the uncertainties on the best fit continuum model. Column 7: The significance of the continuum detection.} \label{tab:specfitvalues}
\end{table*}

  %the $\rm{[OIII]}$ and $\rm{[NII]}$ doublets were required to be the same. The velocity widths of the Balmer series were also required to be the same within the series, but did not have to match the velocity widths of the $\rm{[OIII]}$ and $\rm{[NII]}$ doublets. This was done because the Balmer series lines were usually less well defined than the $\rm{[OIII]}$ and $\rm{[NII]}$ doublets, so their line width was allowed to vary separately to improve fitting efficiency. The best-fit line widths were consistent between the Balmer series and $\rm{[OIII]}$ and $\rm{[NII]}$ doublets to within 10 \AA\, in all cases (except J2236+0032, see Section \ref{subsubsec:J2236 reflected}). 

At low redshifts, differentiating narrow line emission powered by star formation versus AGN is usually done using line ratio diagrams such as the $\rm{log_{10}[OIII]5008/H\beta}$ vs. $\rm{log_{10}[NII]6584/H\alpha}$ (R3N2) BPT diagram \citep{BPT1981PASP...93....5B}, on which AGN and star-forming galaxies naturally separate into two distinct populations, as seen from the gray points in Figure \ref{fig:BPT} which represent the $\rm{z}\sim0$ population from the SDSS dataset \citep{Aihara2011ApJS..193...29A}. This natural divide at $\rm{z}\sim0$ is quantified by the \cite{Kauffmann2003} and \cite{Kewley2001} curves, which divide the diagram into star-forming galaxies, intermediate systems, and AGN. However, this natural divide becomes much less distinct at larger redshifts where the ionization rates are higher and the metallicities lower than in the local universe \citep{Liu2008ApJ...678..758L, Nakajima2023}. Alternatives to the R3N2 BPT diagram, such as the OHNO diagram and modified BPT diagrams using weaker emission lines such as $[\rm{SII}]6716$, $[\rm{OI}]6300$, $\rm{HeII}4686$, $[\rm{OIII}]4363$, $[\rm{NeIII}]3869$, $[\rm{OII}]3726,29$, and $\rm{CIII}]1907,09$, are more successful in distinguishing AGN from star-forming galaxies at high redshifts \citep{Mazzolari2024A&A...691A.345M, Flury2025MNRAS.tmp.1541F, Scholtz2025A&A...697A.175S}. However, our extended emission spectra show no signs of these weaker emission lines, though several do appear in the full quasar+host spectra. In addition, our quasar host galaxies are compact enough and our quasars luminous enough that, given a relatively low-density ($\rm{n_H\approx0.6}$ $\rm{cm^{-3}}$) ISM, the narrow line region of our quasars could extend throughout their host galaxies (see Appendix \ref{appendix:Stromgren} for the full calculation). As a result, we cannot definitively say whether the narrow lines measured in our extended emission spectra are powered by star formation or the AGN. In order to reflect this uncertainty, we label our objects as ``extended emission'' in the BPT diagram, and all values derived from the narrow lines, primarily star formation rates, are treated as upper limits under the assumption that all extended narrow line flux is from star formation in the host galaxy. 

\subsection{The BPT diagram}\label{subsubsec:BPT}

\begin{figure*}
    \centering
    \includegraphics[width=1\linewidth]{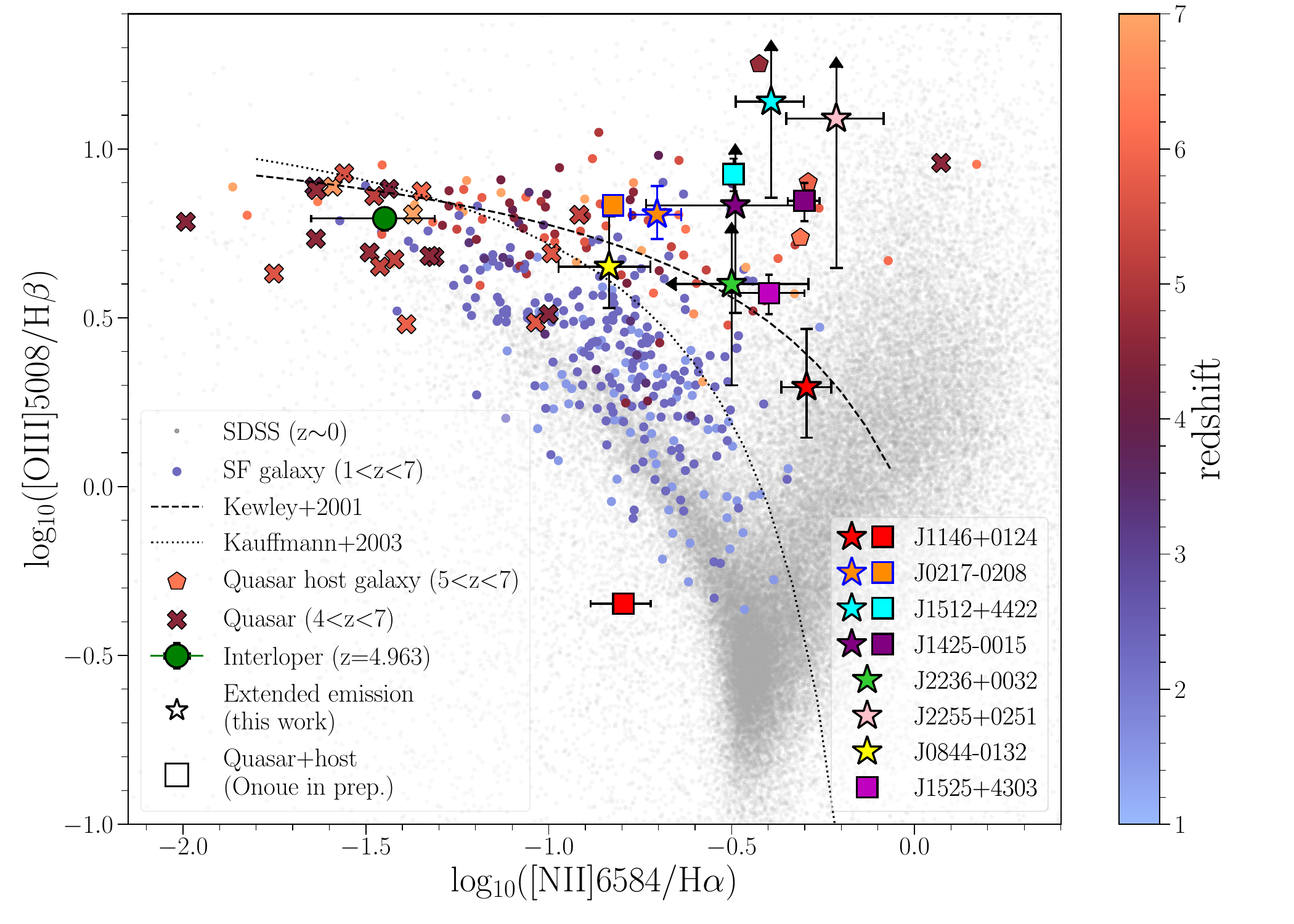}
    \caption{BPT diagram showing the line ratios of the extended emission spectra (bold stars) from this work and quasar+host spectra (bold squares) from \cite{OnoueInPrep} for which we could measure narrow components of $\rm{[OIII]}5008$, $\rm{H}\beta$, $\rm{[NII]}6585$, and $\rm{H}\alpha$. We remind the reader that the narrow lines in the extended emission may come from the narrow line region of the quasar or from star formation, depending on the spatial extent of the narrow line region (see discussion in Appendix \ref{appendix:Stromgren}). Also shown are sources from the literature, separated into star-forming galaxies (small circles) \citep{Shapley2019,Nakajima2023,Sanders2023a,Scholtz2025}, AGN (crosses) \citep{Harikane2023a,Kocevski2023b,Ubler2023,Maiolino2024a,Killi2024,Solimano2025}, and quasar host galaxies whose quasars have been removed using spectral and spatial PSF decomposition (pentagons) \citep{Decarli2024,Zamora2024,Marshall2025}. The $\rm{z}\sim0$ SDSS sample \citep{Aihara2011ApJS..193...29A} is shown in the background in light gray, representing both star-forming and AGN-dominated galaxies at low redshifts. The \cite{Kauffmann2003} and \cite{Kewley2001} curves dividing star-forming, intermediate, and AGN regions at $z \sim 0$ are plotted. The fill color of all $\rm{z}>0$ data from the literature is matched to their redshift as shown by the color bar. Error bars for the extended emission line ratios are as listed in Table \ref{tab: host properties}, and error bars for the quasar+host line ratios are as given by \cite{Onoue2024arXiv240907113O}. J0217$-$0208, which may be star-formation-dominated rather than AGN-dominated (see $\S$\ref{subsubsec:J0217AGNorGal} for full discussion), is outlined in blue. The position of the z=4.963 interloper that happened to fall along the slit of J1512+4422 (discussed in detail in Appendix \ref{appendix:J1512 companion}) is shown by a green filled circle.}
    \label{fig:BPT}
\end{figure*}
%Baldwin-Phillips-Terlevich \cite{BPT1981PASP...93....5B} plots of the ratios of various emission lines in galaxies are commonly used to distinguish between different ionizing sources. At low redshifts, galaxies naturally separate into star-forming and AGN-dominated regions on various line-ratio plots, with the $\rm{[OIII]}5008/H\beta$ and $\rm{[NII]}6585/H\alpha$ (R3N2) plot being of particular use in distinguishing obscured quasars from star-forming systems (see \cite{Veilleux1987ApJS...63..295V} and references therein for further discussion of other useful line ratios). We present in Figure \ref{fig:BPT} a version of the R3N2 BPT diagram, with the star-forming and AGN-dominated regions separated by the curves given by \cite{Kewley2001} and \cite{Kauffmann2003} and populated by the low-redshift (light gray) SDSS sample. In addition, the distinction between star-forming galaxies and AGN on the BPT diagram has long been known to break down with increasing redshift, likely because of metallicity evolution with cosmic time \citep{Liu2008ApJ...678..758L, Nakajima2023}. 
The R3N2 BPT diagram is traditionally used to understand the ionization source(s) of extragalactic emission lines. We plot our objects on the R3N2 BPT diagram in Figure \ref{fig:BPT} and compare them to data from the literature. To illustrate the evolution of the ionization line ratios of AGN and star-forming galaxies as a function of redshift, we plot star-forming galaxies (small circles) \citep{Shapley2019, Nakajima2023, Sanders2023a, Scholtz2025}, AGN (crosses) \citep{Harikane2023a, Ubler2023, Maiolino2024a, Killi2024, Solimano2025}, and quasar host galaxies observed with the NIRSpec IFU whose quasars have been removed using spectral and spatial PSF decomposition (pentagons) \citep{Decarli2024, Zamora2024, Marshall2025}. Each point is colored by its redshift, where light blue represents $z=1$ and light orange represents $z=7$. As redshift increases, star-forming galaxies move into a higher $\rm{[OIII]}5008/H\beta$ region not heavily populated at $\rm{z}\sim0$ and defined by the \cite{Kauffmann2003} curve as AGN-dominated. In contrast, AGN move into a lower $\rm{[NII]}6585/H\alpha$ region with higher redshift, meaning that by redshifts of $\sim4$ the star-forming galaxies and AGN overlap heavily on the R3N2 BPT diagram and cannot be distinguished from each other using the usual emission line ratios. 

%Other emission line ratio diagrams, such as the OHNO diagram and modified BPT diagrams using weaker emission lines such as $[\rm{SII}]6716$, $[\rm{OI}]6300$, $\rm{HeII}4686$, $[\rm{OIII}]4363$, $[\rm{NeIII}]3869$, $[\rm{OII}]3726,29$, and $\rm{CIII}]1907,09$, have been proposed as alternatives, and have proven quite successful in distinguishing AGN from star-forming galaxies at high redshifts \citep{Mazzolari2024A&A...691A.345M, Flury2024arXiv241206763F, Scholtz2025A&A...697A.175S}. However, our extended emission does not show any signs of these weaker emission lines (though several do appear in the full quasar+host spectra), leaving us to place our objects on the traditional BPT diagram with the understanding that the $z\sim0$ categorizations of AGN and star-forming galaxies likely do not apply to our systems.

Our extended emission line ratios (bold stars) tend to lie slightly above the $\rm{z}=0$ AGN regime, with $\rm{[NII]}6585/H\alpha$ and $\rm{[OIII]}5008/H\beta$ ratios higher than expected from $\rm{z}=0$ models of star-forming galaxies, though we note the large error bars on several of our targets. This co-locates our extended emission with both high redshift star-forming galaxies \citep{Shapley2019, Nakajima2023, Sanders2023a, Scholtz2025} and, notably, decomposed quasar host galaxies observed with the NIRSpec IFU \citep{Decarli2024, Zamora2024, Marshall2025}. Interestingly, the narrow line ratios derived by \cite{OnoueInPrep} from simultaneous fits to the broad and narrow components of the full quasar+host spectra also lie in this same region, rather than in the lower $\rm{[NII]}6585/H\alpha$ region populated by $\rm{z}\sim6$ AGN from the literature \citep{Harikane2023a,Kocevski2023b,Ubler2023,Maiolino2024a, Killi2024,Solimano2025}. This implies that the quasar$-$host decompositions performed in this work and by \cite{OnoueInPrep} are unable to completely separate narrow-line emission originating from the quasar and host galaxy.  

%Interestingly, our extended emission and the high-redshift quasar host galaxies from the literature \citep{Decarli2024, Zamora2024, Marshall2025} appear to have slightly higher $\rm{[NII]}6585/H\alpha$ ratios than star-forming galaxies at the same redshifts. This may indicate that the AGN hosted in these galaxies have spatially extended narrow line regions, boosting the natural $\rm{[NII]}6585/H\alpha$ ratio from star-forming regions. It could also indicate improper subtraction of the quasar PSF from the (in our case) 2D spectra or (in the literature cases) IFU cube \citep{Zamora2024,Decarli2024, Marshall2025}. We find this unlikely in our case, however, since in systems with little or no host detection in photometry (i.e. J1146$-$0005, J0911+0152, etc.), the quasar is cleanly subtracted off with no residuals, meaning our PSF model properly fits the quasar. In addition, the broad line widths measured in the inner regions of the spectrum are not seen in our extended emission spectra, indicating that all emission from the BLR has been removed by our fitting procedure. More data for such systems at high redshift are needed to distinguish between true physical differences and possible systematic errors; however, the presence of high $\rm{[NII]6585}/H\alpha$ ratios in several independent studies suggests that the difference is real.

Three of the four systems for which we measure line ratios from the PSF and extended emission (J1146+0124, J0217$-$0208, J1512+4422, and J1425$-$0015) show approximate agreement between the line ratios measured in this work and in \cite{OnoueInPrep} to within their error bars. The exception is J1146+0124, whose quasar$+$host galaxy line ratios place it in the $\rm{z}=0$ star-forming region. We note that, depending on how good the emission line fit is, narrow line ratios derived from systems with both broad and narrow lines (for example, the AGN from \cite{Harikane2023a, Ubler2023, Maiolino2024a, Killi2024, Solimano2025} and our quasar+host systems analyzed by \cite{OnoueInPrep}) may be contaminated by light from the broad-line region, causing their line ratios to move towards the $\rm{z}=0$ star-forming region of the BPT diagram. This may be the cause of the notably different placement of J1146+0124. The J0217$-$0208 system (differentiated by a blue border in Figure \ref{fig:BPT}) is possibly a star-forming system with little to no AGN contribution (see \S \ref{subsubsec:J0217AGNorGal}). This is tentatively supported by its position among the $6<\rm{z}<7$ star-forming galaxies from the literature, though, as mentioned before, this region overlaps heavily with the positions of high-redshift AGN. 

\subsection{Balmer decrement}\label{subsec:Balmer decrement}
For nebula optically thin to the Balmer lines at a temperature of \(10^{4}\) K, the ratio of the strength of H\(\alpha\) to H\(\beta\) should be 2.86 to 1; variation from this ratio is assumed to be the result of reddening due to dust and is called the Balmer decrement \citep{Osterbrock1989agna.book.....O}. The Balmer decrements measured in the quasar-subtracted 1D spectra for the nine systems with successful fits to the $\rm{H}\alpha$ and $\rm{H}\beta$ emission lines are shown in Table \ref{tab:specfitvalues} with appropriate error bars.  Using these Balmer decrements, we calculate the dust extinction as \(\rm{A_{V}} = 7.98\times log_{10}(\frac{\rm{H}\alpha}{\rm{H}\beta}*\frac{1}{2.86})\), taken from \cite{Calzetti2000}, which assumes a starburst-like reddening law. The inferred $\rm{A_{V}}$ extinction magnitudes for all systems with measurable Balmer lines are shown in Table \ref{tab:color and dust}. Since six of the $\rm{H}\beta$ measurements are upper limits due to low signal-to-noise ratios, they have limited ability to constrain the $\rm{A_V}$ magnitudes, which is reflected in their error bars. Of the three systems with robust detection of $\rm{H}\beta$, two produce nonphysical (i.e. negative) values of $\rm{A_V}$. As a result, we do not apply a reddening correction to any subsequent data. We note that \cite{Matsuoka2025ApJ...988...57M} measure $\rm{A_V}$ values of $2.56\pm0.03$ and $2.62\pm0.03$ magnitudes for systems J0844$-$0132 and J1146$-$0005, respectively. While our measured $\rm{A_V}$ for J1146$-$0005 is consistent with that found by \cite{Matsuoka2025ApJ...988...57M}, our value for J0844$-$0132 differs by 2.4$\sigma$. This may indicate that the outer regions of the galaxy are less dust-obscured than the region nearest the black hole, however, given the lower signal-to-noise ratio of our data, such a variation in dust distribution cannot be confirmed.  

\begin{table*}[h]
\centering
\caption{Optical color and inferred $A_V$ (dust extinction) for all systems.}%, J0844$-$0132 and J1146$-$0005 are marked with * to indicate their values derived from $\rm{H}\alpha$ (i.e. $A_{V}$) should be interpreted with caution.}
\label{tab:color and dust} \begin{tabular}{c ccc }
\hline\hline
 (1)& (2)& (3)&(4)\\
 Name &   Continuum fit &1D spectrum & \(A_{V}\) \\
 & $g-r$ color (mag)& $g-r$ color (mag)&(mag)\\   \hline
J0911+0152 &   0.9 $\pm$ 0.3&1.0 $\pm$ 0.3& $-$\\ 
J1425$-$0015 &   0.1 $\pm$ 0.1&$-$0.3 $\pm$ 0.1& $0^{+\infty}_{-3}$\\
 J0918+0139 & 0.9 $\pm$ 0.3& 0.8 $\pm$ 0.3&$-$   \\ 
 J1512+4422 &   $-$0.11 $\pm$ 0.03&$-$0.10 $\pm$ 0.02& $3^{+\infty}_{-3}$\\ 
 J0844$-$0132&   0.3 $\pm$ 0.2&0.5 $\pm$ 0.1& 0.4 $\pm$ 0.9\\ 
 J0217$-$0208 & $-$1.0 $\pm$ 0.2& $-$0.18 $\pm$ 0.09&unphysical \\ 
 J0844$-$0052 & 2.0 $\pm$ 0.6& 0.7 $\pm$ 0.2&unphysical\\ 
 J1146+0124 & 0.4 $\pm$ 0.1& 0.40 $\pm$ 0.08&unphysical \\ 
 J1525+4303 &   0.7 $\pm$ 0.1&0.6 $\pm$ 0.1& $-$   \\ 
 J1146$-$0005&   0.7 $\pm$ 0.4&0.8 $\pm$ 0.3& $0^{+\infty}_{-2}$\\ 
 J2255+0251 &   0.45 $\pm$ 0.08&0.47 $\pm$ 0.07& $3^{+\infty}_{-5}$\\ 
 J2236+0032 &  0.13 $\pm$ 0.03&0.13 $\pm$ 0.03&unphysical \\ 
 \hline
\end{tabular}
\tablecomments{Column 1: Object ID. Column 2: $g-r$ color determined from the linear continuum fit with error bars propagated from the fit parameter uncertainties. Column 3:  $g-r$ color determined from the 1D spectrum with emission lines included. Errors are propagated from the 1D spectrum noise. Column 4: Dust extinction in magnitudes as measured from the Balmer decrement.} 
\end{table*}

%\subsection{Metal Abundances from R3 and N2}\label{subsec:abundances}
%We reliably measure the metal lines of $\rm{[OIII]4960,5008}$ and $\rm{[NII]6551,6585}$ in most of our systems, but we see no other extended metal line emission. Therefore, we present only the R3 and N2 metrics, defined as  \((\rm{R}3 = log{\frac{L_{\rm{OIII}}}{L_{\rm{H}\beta}}})\) and  \((\rm{N}2 = log{\frac{L_{\rm{NII}}}{L_{\rm{H}\alpha}}})\), respectively. The measured values of R3 and N2 are shown in Table \ref{tab: host properties} with their associated error bars. Since R3 and N2 are only weakly correlated with metallicity at high redshift \citep{Maiolino2024b, Harikane2023c, Kocevski2023a}, we limit ourselves to placing the line ratios on the Baldwin-Phillips-Terlevich (BPT) diagram \citep{BPT1981PASP...93....5B} (Figure \ref{fig:BPT}), which is discussed in detail in \S \ref{subsubsec:BPT}. For a more in-depth discussion of metallicity measurements in galaxies and AGN at high redshift, we recommend \cite{Nakajima2023, Mazzolari2024A&A...691A.345M}. 

\subsection{Star Formation Rates}\label{subsec:star formation rates}
In order to characterize the quasar hosts, we calculate the star formation rate (SFR) from the $\rm{H}\alpha$ line luminosity within the modified slit, where the modified slit is defined as two $0''.2$ by approximately $0''.5$ sub-regions of the slit on either side of the quasar which are dominated by extended emission, thus not including the central two pixels (see the outlined regions in Figures \ref{fig:J0911 spectrum}-\ref{fig:J2236 spectrum}). Here, we assume that the quasar provides few to no ionizing photons outside the central $\sim0.5$ kpc ($\sim2 ~\text{pix}$) of the host galaxy. To test this assumption, we estimated the classical Strömgren radii for the relatively high luminosity systems J2236+0032 and J1512+4422 (see Appendix \ref{appendix:Stromgren} for the full calculation). We found that the inferred Strömgren radii range from about half an effective radius to seven times larger, depending on the assumed neutral hydrogen fraction in the host galaxies. Since we cannot constrain such a value given our dataset, we simply acknowledge that the narrow $\rm{H}\alpha$ line seen in extended emission may be excited by the quasar, star formation, or a mix of both. As a result, all SFRs calculated below should be considered upper limits in which all ionizing photons have been attributed to star formation.

\begin{table*}[h]
    \centering
    \caption{Measured host galaxy parameters.}%J0844$-$0132 and J1146$-$0005 are marked with * to indicate their values derived from $\rm{H}\alpha$ (i.e. SFR and sSFR) should be interpreted with caution due to possible interference in the level 2 self-subtraction step of the \texttt{jwst} pipeline. }
    \label{tab: host properties}
    \begin{tabular}{cccc}
    \hline\hline
 (1)& (2)& (3)&(4)\\  
        Name & $\rm{SFR_T}$ (\(\rm{M_{\odot}}\)/yr)& $\rm{log \: \rm{M_{*}}}$$\rm{/ M_{\odot}}$& sSFR ($10^{9}$/yr)\\ \hline
 J0911+0152 & $-$& $9.72^{+0.49}_{-0.38}$&$-$\\ 
 J1425$-$0015 & $37^{+5}_{-27}$& $10.54^{+0.36}_{-0.37}$ &$0.8^{+1.2}_{-0.6}$\\
 J0918+0139 & $-$ & $10.21^{+0.52}_{-0.37}$&$-$   \\ 
 J1512+4422 & $33^{+4}_{-16}$& $10.57^{+0.29}_{-0.41}$ &$0.7^{+1.2}_{-0.4}$\\ 
 J0844$-$0132& $39^{+9}_{-24}$& $10.03^{+0.53}_{-0.45}$ &$4^{+9}_{-5}$\\ 
 J0217$-$0208 & $111^{+9}_{-36}$& $10.06^{+0.28}_{-0.32}$ &$9^{+10}_{-5}$\\ 
 J0844$-$0052 & $7^{+2}_{-4}$& $9.49^{+0.52}_{-0.42}$&$3^{+7}_{-4}$\\ 
 J1146+0124 & $40^{+4}_{-12}$& $10.38^{+0.22}_{-0.30}$&$1.3^{+1.4}_{-0.7}$\\ 
 J1525+4303 & $-$ & $10.00^{+0.30}_{-0.38}$&$-$   \\ 
 J1146$-$0005& $-$ & $9.92 \uparrow$ &$-$   \\ 
 J2255+0251 & $40^{+20}_{-40}$& $10.73^{+0.47}_{-0.30}$&$0.5^{+1.0}_{-0.6}$\\ 
 J2236+0032 & $16^{+4}_{-6}$& $10.96^{+0.08}_{-0.09}$ &$0.2^{+0.1}_{-0.1}$\\ 
 \hline
    \end{tabular}
    \tablecomments{Column 1: Object ID. Column 2: Total star formation rate for the galaxy, extrapolated from the star formation rate measured in the modified slit assuming the surface mass to brightness ratio and specific star formation rate is constant across the galaxy as described in $\S$ \ref{subsec:star formation rates}. Errors are propagated from the $\rm{H}\alpha$ line flux and the photometric signal-to-noise ratio maps. Column 3: Stellar mass of the host galaxy as taken from \cite{ding2025shellqsjwstunveilshostgalaxies}. Column 4: Specific star formation rate and associated errors as calculated in $\S$ \ref{subsec:specific star formation rates}.}
\end{table*}

The exact scaling relationship between $\rm{H}\alpha$ luminosity and SFR depends on details of the interstellar medium and galactic stellar population that we cannot measure at such high redshifts, requiring us to make several further assumptions. We assume solar and Milky-Way-like metallicities for the stars and ISM of the quasar host galaxies, respectively, that the extended $\rm{H}\alpha$ emission line is produced through Case B recombination with an electron temperature of 10,000 K, that the hydrogen is being ionized by young O, B, and A type stars in strongly star-forming regions of the galaxy, and that the host stellar population follows a \cite{Salpeter1955} IMF. The applicability of these assumptions has been well established in the low-redshift universe \citep{ Kennicut1989, Brinchmann2004TheUniverse, Conroy2009}. The extrapolation of the usual Salpeter IMFs to high redshifts is common in the literature, but probably brings with it the likelihood of substantial systematic error \citep{Inayoshi2022ApJ...938L..10I, DingSilverman2022}, whose characterizations are beyond the scope of this paper. 

Given these assumptions, we calculate the average star formation rate $(\rm{SFR}_{\rm{slit}})$ within the modified slit from the luminosity of the extended $\rm{H}\alpha$ emission line as \(\frac{\rm{SFR_{slit}}}{\rm{M_{\odot}\cdot yr^{-1}} }= 7.9\times10^{-42} \frac{\rm{L(H\alpha)}}{\rm{erg \cdot s^{-1} }}\). By assuming that the surface mass-to-light ratio is uniform across the galaxy (a rough approximation, but the best we can do given the limitations of our data), we can estimate the total star formation rate in the galaxy from the proportionality $\frac{\rm{SFR_{T}}}{\rm{SFR_{slit}}}=\frac{\rm{F_{\nu,T}}}{\rm{F_{\nu, slit}}}$, where $\rm{F_{\nu,T}}$ is the total flux from the host galaxy as measured by an elliptical aperture (see panel $(d)$ of Figures \ref{fig:J0911 spectrum}-\ref{fig:J2236 spectrum} for a visualization of the geometry) overlaid on the PSF-subtracted photometry \citep{ding2025shellqsjwstunveilshostgalaxies} and $\rm{F_{\nu,slit}}$ is the product of applying the F356W throughput function to the 1D extended emission spectrum and integrating over all wavelengths. The elliptical apertures shown have axis ratios and orientations as given by the Sérsic modeling of \cite{ding2025shellqsjwstunveilshostgalaxies}, and their semimajor axes are chosen such that they include all of the galaxy flux without including too many background pixels. The inferred galactic star formation rates $(\rm{SFR_T})$ for all systems with measured H\(\alpha\) in the extended spectra are shown in Table \ref{tab: host properties} with error bars.  

\subsection{Specific star formation rates}\label{subsec:specific star formation rates}
Having measured the star formation rate within the modified NIRSpec Fixed Slit aperture from the \(\rm{H}\alpha\) emission line, we may now determine the specific star formation rate (sSFR), which is the star formation rate per unit stellar mass. As in the previous section, we assume a constant surface mass-to-light ratio across the galaxy, meaning the ratio of stellar mass in the modified slit to the stellar mass of the whole galaxy equals the ratio of flux in the modified slit $(\text{F}_{\nu,\text{slit}})$ to flux from the entire galaxy $(\rm{F_{\nu,T}})$:

\begin{equation}
\text{sSFR} = \frac{\text{SFR}_{\text{T}}}{\text{M}_{\text{T}}}.
\label{sSFR}
\end{equation}

We note again that the stellar mass values rely on an assumed IMF which is calibrated at low redshifts. Moreover, the quasar removal process is quite difficult, and the resulting signal-to-noise ratio for each photometric pixel is taken into account when calculating the stellar mass \citep{Ding2023Natur.621...51D}. In addition, our assumption of constant surface mass-to-light ratio does not account for dust obscuration or spatial variations in the stellar population. Given these caveats, we present the specific star formation rates for systems with measurable \(\rm{H}\alpha\) in Table \ref{tab: host properties}. The errors are propagated from the stellar mass, \(\rm{H}\alpha\) luminosity, and 1D spectrum noise, with the uncertainty in the stellar mass dominating the error budget.

\subsection{Asymmetry of nebular and stellar emission}\label{subsec:asymmetry}
\cite{ding2025shellqsjwstunveilshostgalaxies} measure offsets of up to $\sim 0.5$ kpc of the quasar center from the Sérsic model (i.e., host galaxy) center in several of our systems (notably J2255+0251 and J0844$-$0132), and suggest that this may be due to uneven dust attenuation or tidal interactions. Similarly, an examination of panels $(d)$ and $(e)$ of Figures \ref{fig:J0911 spectrum} - \ref{fig:J2236 spectrum} shows that several of the systems have spatially asymmetric emission lines. This is not unexpected, as galaxies in the early universe are much more irregular, on average, than those at lower redshifts \citep{Abraham1996ApJS..107....1A, Giavalisco1996ApJ...470..189G, Papovich2005ApJ...631..101P, Ravindranath2006ApJ...652..963R, Mortlock2013MNRAS.433.1185M, Huertas2016MNRAS.462.4495H}, and even in a fairly symmetric system at low redshift, a slit placed at a random orientation across it will cross a random set of $\rm{HII}$ regions, leading to apparent asymmetries in the emission lines. 

Of additional interest is the difference in spatial profiles between the nebular emission lines and the stellar continuum. We plot in panel $(e)$ of Figures \ref{fig:J0911 spectrum} - \ref{fig:J2236 spectrum} the cross-dispersion profile of the stellar light (averaged over wavelength), $\rm{[OIII]}5008$ emission line, and \(\rm{H}\alpha\) emission line for each system. The photometry is interpolated onto the NIRSpec pixels within the S200A2 slit (see panel $(d)$ of Figures  \ref{fig:J0911 spectrum} - \ref{fig:J2236 spectrum}) to create the stellar profiles. The convolved photometry is averaged over the 471 columns with filter transmission greater than half the peak transmission to create the average $F_\nu$ profile per pixel. The error bars on the convolved photometry are calculated from the SNR maps of the photometric quasar fits shown in \cite{ding2025shellqsjwstunveilshostgalaxies}. As a result, the convolved photometry error bars are largest at the centers of the images where the quasar light is strongest and the separation of quasar and host light is most uncertain. The $\rm{[OIII]}5008$ and $\rm{H}\alpha$ profiles are summations along the dispersion axis of the 5 pixels redward and blueward of the centers of the $\rm{[OIII]}5008$ and $\rm{H}\alpha$ lines, respectively. For $\rm{H}\alpha$, this includes emission from the $\rm{[NII]}$ doublet which straddles H\(\alpha\). However, we assume that the $\rm{[NII]}$ doublet arises from the same ionized gas as the Balmer series, and in any case the $\rm{[NII]}$ doublet doesn't contribute significant amounts of flux to the profiles we are calculating. The error bars on the emission lines come from the pixel-to-pixel variance of the extracted 2D spectra. %We compared this continuum profile measured from the convolved photometry to the continuum profile measured from the 2D spectrum of the extended emission in the full (i.e. not modified) slit. Their amplitudes were found to differ by a factor of $\sim1.5$, which is likely the result of calibration differences in the NIRCam and NIRSpec detectors. FOr clarity we only show the continuum profile as determined from the onvolved photoemtry

In J1425$-$0015 (Figure \ref{fig:J1425 spectrum}) and J1146+0124 (Figure \ref{fig:J1146124 spectrum}), the stellar continuum is relatively symmetrically distributed across the spatial profile, whereas the $\rm{[OIII]5008}$ and $\rm{H}\alpha$ emission lines peak sharply to the left of center, though there is still nonzero emission line flux to the right of center in both systems. This indicates a strong line-emitting source on one side of the quasar but not the other, such as an outflow or an uneven distribution of star formation. Other examples of asymmetry are J0844$-$0132, where the $\rm{[OIII]5008}$ line peaks to the right of center, the convolved photometry peaks to the left, and the $\rm{H}\alpha$ line profile is relatively symmetric, and J2255+0251, where the $\rm{H}\alpha$ and stellar continuum peak to the left of center while the $\rm{[OIII]5008}$ line peaks to the right. These differences in spatial distribution between the $\rm{[OIII]5008}$ and $\rm{H}\alpha$ emission lines may be due to actual differences in metallicity across the galaxy, different ionization sources in different regions, or uneven dust attenuation. Without higher energy line ratios with which we could distinguish ionization due to AGN versus star formation or millimeter observations showing the dust distribution, we can only speculate on the causes of these asymmetries. In any case, the asymmetry seen in multiple systems further emphasizes the broad range of morphologies present in high redshift systems.

\subsection{Points of interest for individual objects}\label{sec:individual objects}
\subsubsection{J0911+0152}\label{subsec:J0911}
The 2D residuals after PSF subtraction shown in panel $(b)$ of Figure \ref{fig:J0911 spectrum} are consistent with background noise, which means that the PSF model is a good fit to the 2D quasar$+$host spectrum in panel $(a)$. Assuming our PSF model is accurate, this implies that the system has no detectable extended emission, which is consistent with the weak detection in photometry from \cite{ding2025shellqsjwstunveilshostgalaxies}, especially considering that the slit is misaligned with what little light is detected, as seen in panel $(d)$ of the same figure. Despite the lack of residuals in panel $(b)$, the emission line fit was successful. Given the lack of extended emission in panel $(b)$, the misalignment of the slit with the starlight in panel (\textit{d}), and the poorness of the fit, especially to the $\rm{[OIII]5008, 4960}$ and $\rm{H}\beta$ lines, we conclude that our code is fitting background noise and that these line fits are not robust, except for perhaps the $\rm{H\alpha}$ line, which is marginally detected. The $\rm{H\alpha}$ line is also the only spatial profile shown in panel $(e)$ of Figure \ref{fig:J0911 spectrum} which does not agree with zero to within its error bars.

\subsubsection{J1425-0015}\label{subsec:J1425}
We note in panel $(b)$ of Figure \ref{fig:J1425 spectrum} prominent asymmetric extended emission lines along with a very weakly detected extended stellar continuum. For lines with large equivalent widths, such as the $\rm{[OIII]}$ doublet and $\rm{H}\alpha$, the emission line fits as seen in panel $(c)$ are very robust; however, for the weaker $\rm{[NII]}$ doublet and $\rm{H}\beta$ line, the emission lines have low signal-to-noise ratios and the fits are not robust. Panel $(e)$ of the same figure shows that the $\rm{[OIII]5008}$ and $\rm{H}\alpha$ spatial profiles both peak strongly to one side of the galaxy, while the continuum profile is relatively symmetric around the center, as also seen in panel $(d)$. This may imply an uneven distribution of star-forming regions along the slit or a visibly, but not physically, one-sided outflow, as discussed in $\S$ \ref{subsec:redshift corroboration}.

\subsubsection{J0918+0139}\label{subsec:J0918}
J0918+0139 has little to no extended continuum nor emission lines, consistent with the relatively weak photometric stellar continuum detected in this system by \cite{ding2025shellqsjwstunveilshostgalaxies}. This is best seen in panels $(b)$ and $(c)$ of Figure \ref{fig:J0918 spectrum}, which show that the residuals after PSF subtraction are consistent with background noise, and that no successful emission line fit was found. The spatial profiles shown in panel $(e)$ of the same figure are consistent with zero, except for the $\rm{H}\alpha$ spatial profile, which shows slight excess emission on one side of the galaxy. Although the excess is on the same side of the quasar as the detected photometry (see panel $(d)$ of the same figure), making it more plausibly real, given the consistency of the residuals in panel (\textit{b}) with background noise and the lack of a successful emission line fit, we do not claim detection of $\rm{H}\alpha$ emission from the spatial profile.

\subsubsection{J1512+4422}\label{subsec:J1512}
%strong continuum, has absorption lines in total spectrum, weird Halpha (double peaked emitter), strong extended emission lines, largest size on the sky. aligned with foreground galaxy? 
The quasar+host spectrum of J1512+4422, seen in panels $(a)$ and $(c)$ of Figure \ref{fig:J1512 spectrum}, shows a unique double-peaked $\rm{H}\alpha$ line, as well as Balmer absorption lines indicative of a post-starburst stellar population. These features are discussed in depth in \cite{Onoue2024arXiv240907113O}. The extended emission, seen in panels $(b)$ and $(c)$ of the same figure, shows no robust detection of extended $\rm{H}\beta$, and no absorption lines are detected in the extended continuum. This may indicate a difference in stellar population between the central and outer regions of the galaxy, but is just as likely an indication of the low signal-to-noise ratio of our extended emission spectra, as discussed in $\S$ \ref{subsec:continuum and color}. The host of J1512+4422 has the largest angular diameter of the targeted systems and strong stellar emission, as seen in panel $(d)$, which might have allowed us to measure rotation-driven shifts in emission line centers on opposite sides of the quasar. Unfortunately, the slit happened to be aligned close to the minor axis, and no such shifts were seen. We note that a nearby dropout galaxy happened to fall along the slit (centered at what would be pixel $\sim 35$ and subtracted from this figure), allowing us to measure its 1D spectrum as discussed in Appendix \ref{appendix:J1512 companion}. J1512+4422 shows remarkably consistent spatial profiles across the four components plotted in panel $(e)$, with the $\rm{[OIII]}5008$ profile being slightly more centrally concentrated than the $\rm{H}\alpha$ line. In addition, the average $\rm{[OIII]}5008$ and $\rm{H}\alpha$ profile fluxes are similar to those of the convolved photometry, implying that the emission lines contribute significantly to the total output flux of the system.

\subsubsection{J0844-0132}\label{subsec:J844132}
J0844$-$0132 hosts one of the most luminous quasars in our sample, and as a result the quasar+host spectrum shows very broad emission lines with many weaker ionization lines, seen in panels $(a)$ and $(c)$ of Figure \ref{fig:J844132 spectrum}. However, only the $\rm{H}\beta$, $\rm{[OIII]}$ doublet, $\rm{[NII]}$ doublet, and $\rm{H}\alpha$ lines are seen in the extended emission (panels $(b)$ and $(c)$ of the same figure). The brightness of J0844$-$0132's quasar makes PSF subtraction particularly difficult in both photometry and spectroscopy. This is likely the cause of the sharp spatial variance and negative residuals seen around the $\rm{H}\alpha$ line in panels $(b)$ and $(e)$. 

\subsubsection{J0217-0208}\label{subsec:J0217}
As seen in panels $(a)$, $(c)$, and $(e)$ of Figure \ref{fig:J0217 spectrum}, J0217$-$0208 has very strong, narrow extended emission lines but a relatively weak continuum, making proper removal of the PSF difficult. As a result, several nonphysical artifacts persisted or were introduced in the PSF removal process, including the positive residual at $\sim0.58~\rm{\mu m}$ and the negative residuals at $\sim0.64\:\rm{\mu m}$, $\sim0.66\:\rm{\mu m}$, and $\sim0.68\:\rm{\mu m}$ seen in panel $(b)$ of the same figure. Despite its relative weakness, the stellar continuum in J0217$-$0208 is detected at high significance in the photometry (panel $(d)$) because J0217$-$0208 has the highest host-to-PSF light ratio ($\sim70\%$) of the entire sample \citep{ding2025shellqsjwstunveilshostgalaxies}. The weakness of the PSF component in the photometry combined with the lack of a strong broad-line component in the PSF spectrum leads us to question the presence of an AGN in J0217$-$0208, which we discuss in detail in $\S$ \ref{subsubsec:J0217AGNorGal}. J0217$-$0208 is also the only system that shows possible extended emission around the H$\gamma$ line and whose $\rm{[OIII]5008}$ line is much stronger than the $\rm{H}\alpha$ line in the extended and PSF+host spectrum, further emphasizing its unique nature within our sample. In panel $(d)$, we note that the pixel numbers begin in the upper right and end in the lower left because the slit was oriented nearly $\sim180^\circ$ from the orientation used for most of the other systems. 

\subsubsection{J0844-0052}\label{subsec:J84452}
The spectrum of J0844$-$0052 is dominated by its quasar, with very little extended emission in the PSF-subtracted 2D spectrum, as seen in panel $(b)$ of Figure \ref{fig:J84452 spectrum}. The emission line fits, seen in panel $(c)$, are quite poor and likely do not represent actual measurable emission line properties of the host galaxy. The mismatch between the photometric detection and the spectroscopic non-detection can be explained by the slit orientation which, as can be seen in the overlay in panel $(d)$ of Figure \ref{fig:J84452 spectrum},  happens to avoid the brightest area of stellar continuum, making detection of extended emission more difficult. The lack of strong extended emission causes the emission line profiles shown in panel $(e)$ to be very inconsistent. While the convolved photometry is relatively compact, the $\rm{H}\alpha$ profile in the same panel appears to extend several kpc farther than the host starlight on one side of the galaxy. However, given we only have a marginal detection of extended $\rm{H}\alpha$ in the 1D spectrum, we do not find this compelling.

\subsubsection{J1146+0124}\label{subsec:J1146124}
J1146+0124 has strong extended emission lines but little to no extended continuum (as seen in panels $(a)$, $(b)$, and $(c)$ of Figure \ref{fig:J1146124 spectrum}), and its emission lines appear asymmetric across all wavelengths, as seen in panels $(b)$ and $(e)$ of the same figure. J1146+0124 has some of the strongest narrow lines of the twelve systems, allowing robust measurement of the $\rm{[OIII]}5008/H\beta$ and $\rm{[NII]}6585/H\alpha$ ratios used in the BPT diagram (see $\S$ \ref{subsec:line fits and ratio} and \ref{subsubsec:BPT}). While the spatial profiles, shown in panel $(e)$, of the starlight, measured by the convolved photometry (blue) and the spectroscopic continuum (magenta) are symmetric, the $\rm{H}\alpha$ (pink) and $\rm{[OIII]5008}$ profiles (green) peak strongly to one side of the quasar. In addition, the $\rm{H}\alpha$ line center measured in the extended emission is noticeably offset from that measured in the quasar+host spectrum, shown as a vertical dashed line in the inset panel of panel $(c)$, possibly indicating an outflow. This is discussed in depth in $\S$ \ref{subsec:redshift corroboration}.

\subsubsection{J1525+4303}\label{subsec:J1525}
While the quasar+host spectrum (panel $(a)$ of Figure \ref{fig:J1525 spectrum}) shows prominent broad emission lines, no emission lines are detected in the extended emission, and an extended continuum is only very weakly detected (panels $(b)$ and $(c)$ of the same figure). Similarly to system J0844$-$0052, the weakness of extended emission may be attributed to the slit orientation which, as seen in panel $(d)$ of Figure \ref{fig:J1525 spectrum}, happened to fall almost exactly along the minor axis of the galaxy, limiting the amount of extended emission that might have been measured. The spatial profiles of the $\rm{[OIII]}5008$ and $\rm{H}\alpha$ lines, shown in panel $(e)$, are slightly stronger than the extended continuum, possibly indicating the presence of some weak extended line emission. 

\subsubsection{J1146-0005}\label{subsec:J11465}
Panels $(a)$, $(b)$, and $(c)$ of Figure \ref{fig:J11465 spectrum} show that the broad and luminous lines of the quasar dominate the spectrum of J1146$-$0005, and the best-fit host galaxy continuum model is consistent with zero; \citep{ding2025shellqsjwstunveilshostgalaxies} also were unable to detect a galaxy host for J1146$-$0005, as seen in panel $(d)$ of the same figure. The possible exception is the $\rm{H}\alpha$ line, which, as seen in panel $(e)$, peaks to the left of center with significant spatial structure, unlike the starlight and $\rm{[OIII]}5008$ profiles, which are consistent with zero. This variation in the $\rm{H}\alpha$ line may be due to improper subtraction of the PSF, causing the negative values at pixels 23 and 26.

\subsubsection{J2255+0251}\label{subsec:J2255}
J2255+0251 shows extended $\rm{[OIII]4960,5008}$, $\rm{H}\alpha$, and $\rm{[NII]6551,6585}$ emission in its 2D (panel $(b)$) and 1D (panel $(c)$) spectra, shown in Figure \ref{fig:J2255 spectrum}, but does not have robust detection of extended $\rm{H}\beta$. J2255+0251 also shows asymmetric emission lines - interestingly, the $\rm{[OIII]}5008$ line peaks on the opposite side of the galaxy from the stellar continuum, which is one-sided in both the photometry and the spectral continuum, while the $\rm{H}\alpha$ line is relatively symmetric (panel $(e)$ of the same figure). This difference in spatial distribution may be due to uneven dust attenuation or different excitation mechanisms in different regions of the galaxy.

\subsubsection{J2236+0032}\label{subsec:J2236}
Along with J1512+4422, J2236+0032 has the most prominent extended stellar continuum in its 2D spectra (panels $(a)$ and $(b)$ of Figure \ref{fig:J2236 spectrum}) of all twelve sources. Although it shows some extended $\rm{[OIII]5008}$ emission, the $\rm{[NII]}$ doublet and Balmer series lines are not robustly detected (see the insets in panel $(c)$). In photometry, J2236+0032 is quite symmetric, with the slit falling almost exactly along the major axis of the galaxy, as seen in panel $(d)$. This is reflected in the spatial profile of the stellar continuum from the convolved photometry (blue) and the extracted spectral continuum (magenta) shown in panel $(e)$ of Figure \ref{fig:J2236 spectrum}. The emission line spatial profiles, while relatively symmetric, are offset from the photometric center of the galaxy. This is likely because the quasar PSF center is offset by $0.21\pm0.09$ kpc from the galaxy center, as measured by \cite{ding2025shellqsjwstunveilshostgalaxies}, causing the over-subtraction at the PSF center to be offset from the galaxy center.

\section{Results}\label{sec:results}
%Quiescence and limits on star formation history of J1512
%Possibilities and limitations of the application of this method
%rotation and dynamical mass of star forming galaxy J0217
%Collected host properties - verification of Ding2025 and Onoue 2023
%Asymmetry, morphology, and dust content of quasar hosts
%Positions on the BPT diagram - differentating quasars and star forming galaxies at low metallicity
%Comparison to the literature, and piecing together a coherent story
%\clearpage
\subsection{Placement on the Star-Forming Main Sequence}\label{subsec:SFMS}
\begin{figure*}
    \centering
    \includegraphics[width=1\linewidth]{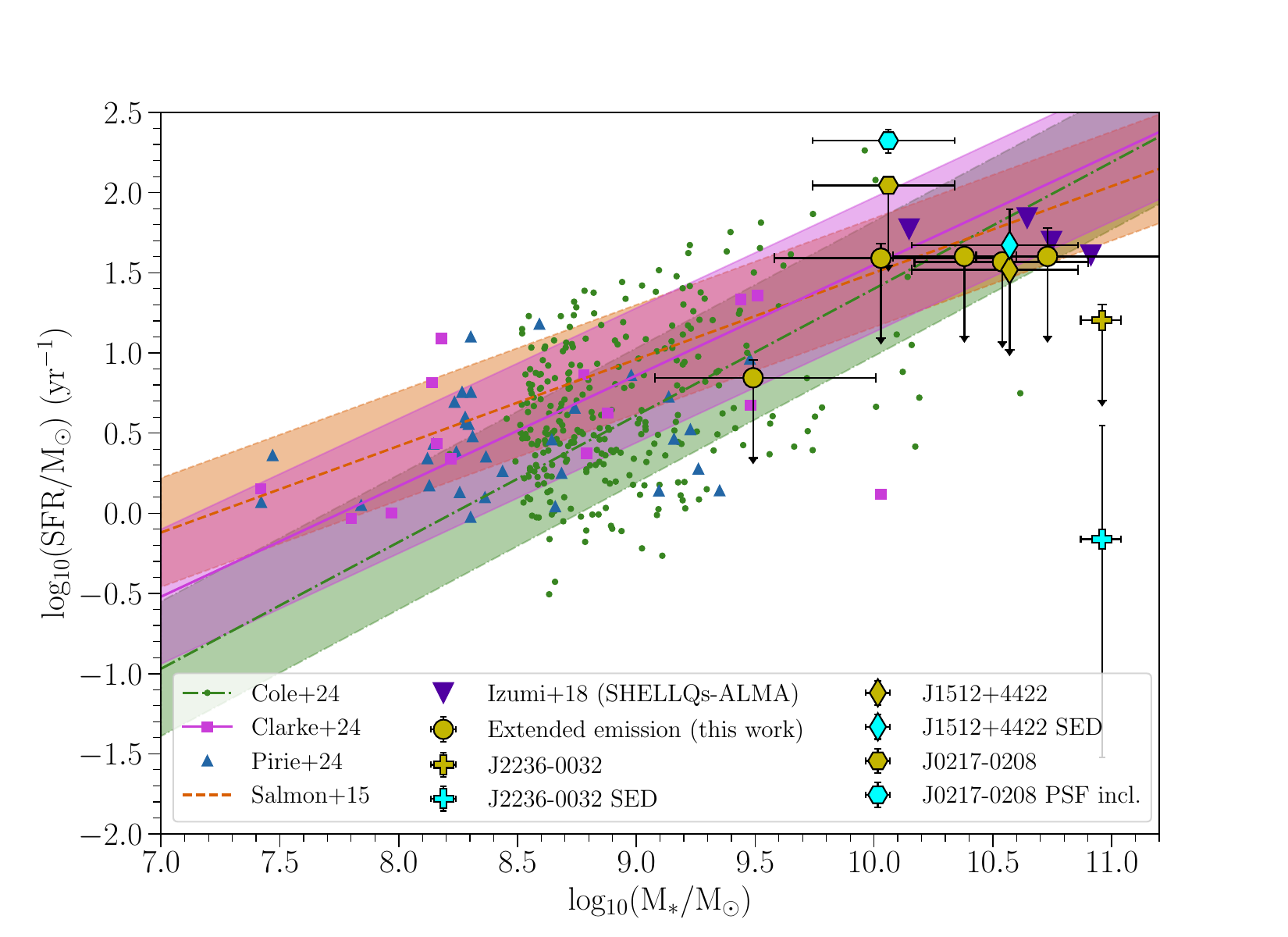}
    \caption{$\text{log}_{10}(\text{SFR})$ vs. $\text{log}_{10}(\text{M}_{*})$ for all systems with measured extended $\rm{H}\alpha$ in emission. The star-forming main sequence at $\rm{z}\approx6$ is represented by three fits and their associated scatters from the literature \citep{Cole2025, Clarke2024, Salmon2015} (green dash-dotted line, pink solid line, and orange dashed line, respectively). The individual data points for \cite{Salmon2015} (N = 266, HST-CANDELS survey) are not shown to avoid cluttering the plot. Data from \cite{Pirie2025MNRAS.541.1348P} (which was not fit by a line) is shown as blue upward-facing triangles, and data from \cite{Izumi2018}, which used ALMA to measure star formation rates in SHELLQs quasar host galaxies from far infrared dust emission, is shown as violet downward-facing triangles. Data from this work are shown in yellow, with unique shapes for three objects of note. The cyan hexagon shows J0217$-$0208's placement on the SFMS when the central PSF is included in photometric and spectral fits, assuming J0217$-$0208 is dominated by stellar emission and thus has minimal contribution from a quasar. J1512+4422 SED and J2236-0032 SED (cyan diamond and cross, respectively) are the star formation rates as measured from the SED fitting by \cite{Onoue2024arXiv240907113O}. These represent the SFR over $\sim$100 Myr timescales, while the $\rm{H}\alpha$ rates are over $\sim$10 Myr timescales. In addition, the SED fit incorporates the central region of the galaxy, which we are forced to ignore due to the way we have subtracted the PSF. All yellow data points are upper limits on star formation under the assumption that all extended narrow-line emission is produced by star formation rather than the narrow line region of the quasar. Error bars for all points are carried through from the stellar mass fits of \cite{ding2025shellqsjwstunveilshostgalaxies} and the total star formation rates of the galaxies as outlined in \S \ref{subsec:star formation rates}.}
    \label{fig:SFMS}
\end{figure*}

We place systems with measured total star formation rates on the traditional $\text{SFR-M}_{*}$ plot, alongside other $\rm{z}\approx6$ sources from the literature \citep{Salmon2015, Clarke2024,Cole2025}. \cite{Salmon2015} is a fit to SED (Spectral Energy Distribution) derived SFRs from CANDELS (Cosmic Assembly Near-IR Deep Extragalactic Legacy Survey) galaxies at $\rm{6<z<7}$, \cite{Cole2025} is a fit to CEERS (Cosmic Evolution Early Release Science Survey, \cite{Finkelstein2023}) galaxies with measured $\rm{H}\alpha$ lines from $\rm{6<z<7}$, and \cite{Clarke2024} is a fit to a combined JADES (JWST Advanced Deep Extragalactic Survey, \cite{Eisenstein2023}) and CEERS sample from $\rm{6<z<7}$. Their $1\sigma$ intrinsic scatters are shown as colored shaded regions associated with each line, and all three fits agree with each other to within their uncertainties. We also show data from \cite{Pirie2025MNRAS.541.1348P}, a narrow-band NIRCam survey of $\rm{z}\sim6.1$ $\rm{H}\alpha$ emission-line galaxies, and from \cite{Izumi2018}, an ALMA follow-up of several $\rm{z}>6$ SHELLQs quasars in which star formation rates were measured from fair infrared observations of hot dust. All of our objects, barring J0217$-$0208 and J2236+0032, fall along the star-forming main sequence to within their error bars, indicating that the quasar host galaxies are similar to the broader galaxy sample at redshift of 6, assuming the $\rm{H}\alpha$ emission we measure is excited mainly by star formation and not the AGN. 

We also plot the SFRs for post-starburst galaxies J2236+0032 and J1512+4422 as measured from their SED fitting by \cite{Onoue2024arXiv240907113O}. \cite{Onoue2024arXiv240907113O} use the \texttt{Bagpipes} code to model the star formation histories of both systems using delayed-$\tau$ and non-parametric models. In Figure \ref{fig:SFMS}, we show the star formation rates output by the delayed-$\tau$ model, as this is what is plotted in Figure 2 of \cite{Onoue2024arXiv240907113O}. Because SED fits take into account the long-term evolution of the stellar population, they are best able to measure star formation rates on $\sim$100 Myr timescales. In contrast, $\rm{H}\alpha$ emission, which mainly results from the formation of bright O and B type stars, tracks SFR on much shorter timescales of $\sim$10 Myr. By comparing the two, one can see differences in the short- and long-term stellar evolution of the system. In our systems, an additional difference is imposed because the SED fit incorporates the central region of the galaxy, which we cannot include because of the way we have subtracted the PSF. The SED-derived and $\rm{H}\alpha$-derived SFRs for J1512+4422 agree to within their error bars, with the SED-derived best fit SFR lying closer to the SFMS. This may indicate that the SFR of J1512+4422 has been relatively constant  on the scale of tens of Megayears, or that the outer regions of J1512+4422 are undergoing less star formation than the central regions. Combined with the post-starburst features in its central spectrum, this indicates a system that experienced an intense period of star formation which abruptly lessened in the last hundred Megayears and is now settled onto the star-forming main sequence. In contrast, the SED-derived SFR for J2236+0032 indicates a strongly quiescent system, well below the star-forming main sequence, while the $\rm{H}\alpha$-derived SFR lies nearly an order of magnitude above it, though still below the SFMS. This may imply that the outer regions of J2236+0032 are still star-forming while the central region is quenched, perhaps as a result of the AGN, or that J2236+0032, while quenched in the past, is not permanently so, and is moving back towards the SFMS, either as a whole galaxy, or with the outer regions of the galaxy increasing their star formation rates faster than the central regions. Either case suggests that quenching, while an important process in galaxy formation, is limited either spatially or temporally, or both, and is perhaps cyclical in at least some of these systems. 
 %The exceptions are J2236 and J0217, which each differ from Cole by  sigma. J2236, as shown in Onoue 02024, has post starburst features that indicate it was recently quenched and is no longer forming stars are as quickly as before. The system's placement on the star forming main seuqnce as measured from the SED is shpwn in a mfaded version on the plot. As can be seen, when averaged over longer time frames, the system is even more quescent thatn currently.

The other outlier is J0217$-$0208, whose upper limit lies $1\sigma$ above the main sequence, indicating that it may be a strongly star-forming system. J0217$-$0208 may actually be dominated by star formation rather than an AGN, as we discuss below in \S \ref{subsubsec:J0217AGNorGal}. Given this possibility, we show its placement on the $\text{SFR-M}_{*}$ diagram when no PSF subtraction is performed on the spectra or photometry. This causes the inferred SFR to increase by a few sigma, placing J0217$-$0208 firmly in the starburst region of the $\text{SFR-M}_{*}$ plot.  %We see that the source is even furhter aay from the star forming main seuqnce. This indicates that this soure is indeed a strongly star forming galaxy, anf , from its sSFR, can even be classified as a starburst, with SFR time/hubble time of less than 0.1. 

The upper limits of the rest of the sources with measured $\rm{H}\alpha$ lines fall along the star-forming main sequence with scatter consistent with those measured by \cite{Cole2025}, \cite{Clarke2024}, and \cite{Salmon2015}, as seen in Figure \ref{fig:SFMS}. It is interesting to note that our sample of quasar host galaxies lie on or below the SFMS, despite hosting AGN. This agrees with the trend seen in \cite{Izumi2018, Izumi2019}, where the host galaxies of low-luminosity quasars tend to lie on the SFMS while the hosts of high-luminosity quasars tend to lie well above it.  This may indicate that low-luminosity AGN are not the main source of quenching in star-forming galaxies at $\rm{z}\sim6$, since it is unlikely that we caught all of our galaxies in the transitory process of turning off of (or onto) the main sequence due to AGN feedback. \cite{OnoueInPrep} and \cite{Silverman2025arXiv250723066S} found that the black hole masses for these objects are slightly higher than expected from the $\text{M}_{\text{BH}}\text{-}\sigma_{*}$ relation, possibly implying periods of super-Eddington accretion or heavy seeding (see Figure 3 and associated discussion in \cite{Onoue2024arXiv240907113O}). As we have seen of the eight systems with measurable upper limits on star formation rates in the extended emission, only two show post-starburst features, and only one appears to lie below the star-forming main sequence. This might indicate that, while AGN feedback can quench star formation, this effect is neither immediate nor permanent and that the presence of an AGN in a galaxy does not guarantee a suppressed star formation rate at $\rm{z}\sim6$. %This implies that AGN feedback may suppress star formation on longer timescales, but on short timescales, the effects of feedback are less apparent.%We also plot the z=0 and z=2 SFMSs as well for comparison. As can be seen, our data confirm the trend of the star-forming main sequence maintaining a relatively constant slope with increasing redshift.

\subsection{Corroboration with Concurrent Works}\label{subsec:collected host properties}
Analysis of these twelve systems was also performed independently by \cite{Onoue2024arXiv240907113O}; \cite{OnoueInPrep} and \cite{Ding2023Natur.621...51D, ding2025shellqsjwstunveilshostgalaxies}, focusing on the full quasar+host spectra and the photometry, respectively. Although various products of their analyses are used in this paper, there are several properties of the target systems that were independently calculated. In this section, we compare these values to verify the robustness of our calculations and explain any differences. 
\begin{table*}
    \renewcommand{\arraystretch}{0.8}
    \centering
    \caption{Corroboration of values between this work and \cite{OnoueInPrep} (for spectral values) and \cite{ding2025shellqsjwstunveilshostgalaxies} (for photometric values).}
    \begin{tabular}{cccccccccc}
    \hline\hline
 (1)& (2)& (3)& (4)& (5)& (6)& (7)& (8)& (9)&(10)\\ 
         & && & & && $\rm{[OIII]5008}$ &  $\rm{[OIII]5008}$& \\
 Name& $\rm{m_{AB}}$&$\rm{m_{AB}}$& $\rm{\frac{\Delta m_{AB}}{\sigma}}$& $\text{z}_{\text{extended}}$& z\footnote{\cite{OnoueInPrep}}$_{\text{quasar}}$&$\frac{\Delta \text{z}}{\sigma}$& FWHM& FWHM$^{a}$ &$\rm{\frac{\Delta FWHM}{\sigma}}$\\
 & spec&phot\footnote{derived from the values given in \cite{ding2025shellqsjwstunveilshostgalaxies}}& & & $_{\text{+host}}$&& (km/s)& (km/s)&\\\hline
 J0911+0152 & 28.0 $\pm$& 28.5 $\pm$& 1.2& 6.071$\pm$& 6.055 $\pm$& 1.9& 682 $\pm$& 3531 $\pm$&4.0\\
 & 0.2& 0.4& & 0.002 & 0.008& & 162& 692&\\
 J1425$-$0015 & 25.58$\pm$& 25.85 $\pm$& 1.9& 6.1803$\pm$& 6.1780$\pm$& 2.2& 781$\pm$& 630$\pm$&1.6\\
 & 0.06& 0.12& & 0.0009& 0.0005& & 90& 20&\\
 J0918+0139 & 27.0$\pm$& 27.3$\pm$& 0.4& $-$& 6.1786$\pm$& $-$& $-$& 722$\pm$&$-$\\
 & 0.2& 0.6& & & 0.0003& & & 47&\\
 J1512+4422 & 25.03$\pm$& 25.10$\pm$& 0.9& 6.1803$\pm$& 6.1806$\pm$& 0.6& 558$\pm$& 532$\pm$&0.7\\
 & 0.02& 0.08& & 0.0003& 0.0004& & 29& 25&\\
 J0844$-$0132& 26.36$\pm$& 26.9$\pm$& 1.0& 6.1840$\pm$& 6.18299$\pm$& 5.0& 434$\pm$& 419$\pm$&0.8\\
 & 0.09& 0.5& & 0.0002& 0.00001& & 18& 2&\\
 J0217$-$0208 & 25.68$\pm$& 25.54$\pm$& 1.7& 6.2034$\pm$& 6.2034$\pm$& 0.0& 479$\pm$& 378$\pm$&8.1\\
 & 0.04& 0.07& & 0.0001 & 0.0005& & 11& 6&\\
 J0844$-$0052 & 26.52$\pm$& 26.60$\pm$& 0.2& 6.245$\pm$& 6.2426$\pm$& 2.2& 490$\pm$& 498$\pm$&0.06\\
 & 0.15& 0.33& & 0.001& 0.0004& & 121& 77&\\
 J1146+0124 & 25.95$\pm$& 25.69$\pm$& 2.2& 6.2491$\pm$& 6.2459$\pm$& 7.2& 354$\pm$& 674$\pm$&13.1\\
 & 0.06& 0.10& & 0.0002 & 0.0004& & 20& 14&\\
 J1525+4303 & 25.95$\pm$& 26.44$\pm$& 2.5& $-$& 6.2650$\pm$& $-$& $-$& 531$\pm$&$-$\\
 & 0.06& 0.19& & & 0.0005& & & 28&\\
 J1146$-$0005& 28.24 $\uparrow$& 26.38 $\uparrow$& $-$& 6.2994$\pm$& 6.30159$\pm$& 2.7& 367$\pm$& 322$\pm$&0.9\\
 & & & & 0.0008& 0.00001& & 48& 1&\\
 J2255+0251 & 25.56$\pm$&26.5$\pm$& 0.8& 6.3358$\pm$& 6.3329$\pm$&2.7& 675$\pm$& 640$\pm$&0.4\\ 
         & 0.02&1.2&&  0.0008&  0.0007&& 66&  47&\\
 J2236+0032& 24.57$\pm$&24.25$\pm$& 3.1& 6.4058$\pm$& 6.4047$\pm$&0.9& 521$\pm$& 521$\pm$&0.0\\ 
         & 0.03&0.10&&  0.0008& 0.0009&& 74& 61&\\
    \hline
    \end{tabular}
    \tablecomments{Column 1: Object ID. (2, 3) AB magnitude of the host galaxy within the S200A2 slit as measured from the 1D spectrum (Column 2) and the photometry (Column 3). Due to the difference in the magnitudes calculated by the \texttt{photutils} and \texttt{galight} codes, a $+0.16\pm0.06$ correction was applied to the photometric AB magnitudes in all systems and the uncertainty of the correction was propagated through to the resulting error bars. Column 4: Agreement of Columns 2 and 3, in standard deviations. Column 5: Redshift of the host galaxy as measured from the centroid of our best fit to the $\rm{[OIII]5008}$ emission line. Column 6: Redshift of the quasar+host galaxy as measured by \cite{OnoueInPrep}. Column 7: Agreement of Columns 5 and 6, in standard deviations. Column 8: Full Width Half Maximum (FWHM) of our best fit to the extended $\rm{[OIII]5008}$ emission line. Column 9: Full Width Half Maximum (FWHM) of the narrow component of the quasar+host $\rm{[OIII]5008}$ emission line as measured by \cite{OnoueInPrep}.  Column 10: Agreement of Columns 8 and 9, in standard deviations.}
    \label{tab:corroboration}
\end{table*}

\subsubsection{Slit AB Magnitude}\label{magnitude corroboration}
\begin{figure*}
    \centering
    \includegraphics[width=1\linewidth]{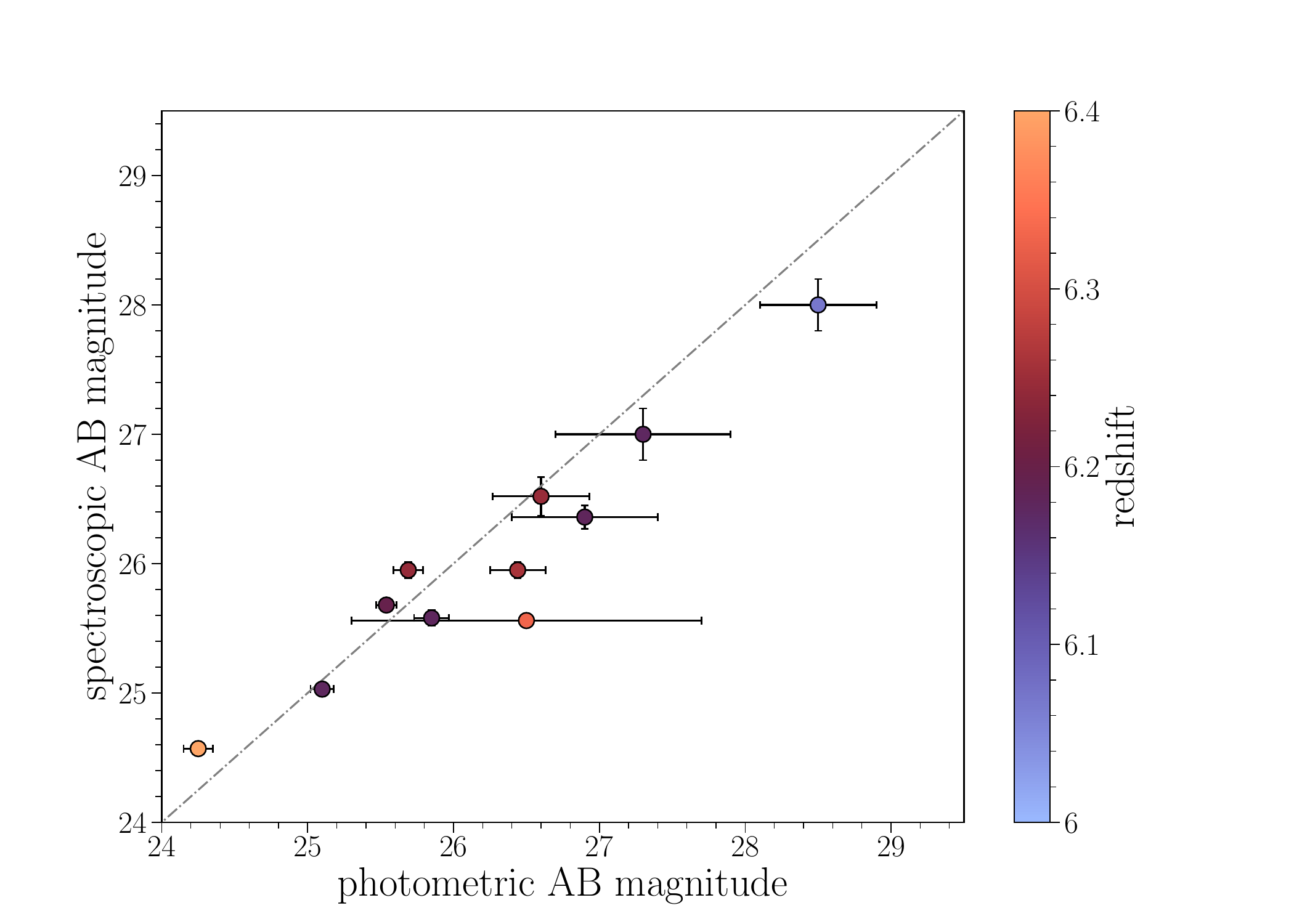}
    \caption{Plot showing the agreement of the spectroscopically and photometrically measured magnitudes of all twelve targets. Each data point is colored by its redshift according to the colorbar. A gray dash-dot line is the $x=y$ line. All systems except J2236+0032 (lower left, golden yellow) are within $3\sigma$ of the $x=y$ line, indicating good overall agreement between the PSF removal methods of this work and \cite{ding2025shellqsjwstunveilshostgalaxies}. The photometric error bars are larger than the spectroscopic ones because they include the effects of the systematic bias described in the text.}
    \label{fig:ABmags}
\end{figure*}
%Corroboration of the AB magnitudes between the spectra and photometry has been described in part in \S\S \ref{subsec:continuum and color} and \ref{subsec:specific star formation rates} above. 
In Table \ref{tab:corroboration}, we list the AB magnitude for the flux in the modified slit as measured from the photometry $(\rm{m_{AB}} \:\text{phot}, \text{Column 3})$, the AB magnitude of the spectrum convolved with the F356W filter $(\rm{m_{AB}} \:\text{spec, Column 4})$, and the difference in standard deviations between these two values $(\rm{\frac{\Delta m_{AB}}{\sigma}}\text{, Column 5})$. We use the \texttt{photutils} code library, which is a coordinated package of \texttt{astropy}, to overlay onto the PSF-subtracted photometry provided by \cite{ding2025shellqsjwstunveilshostgalaxies} two $0''.2$ by $0''.5$ rectangular apertures rotated to align with the orientation of the NIRSpec Fixed Slit at the time of observation (i.e. the modified slit, see panel $(d)$ of Figures \ref{fig:J0911 spectrum}-\ref{fig:J2236 spectrum} for a visualization of the geometry). \texttt{photutils} then calculates the magnitude of the host galaxy light within the modified slit aperture, taking into account edge effects to avoid double-counting photons. Since we use a different photometric code (\texttt{photutils}) than \cite{ding2025shellqsjwstunveilshostgalaxies} (\texttt{galight}) to produce our magnitudes, we compared the magnitudes output by \texttt{photutils} for an ellipse enclosing the entire galaxy (with the quasar PSF removed) with the host magnitudes derived using \texttt{galight} listed in Table 2 of \cite{ding2025shellqsjwstunveilshostgalaxies}. We found on average that the magnitudes output by \texttt{photutils} were systematically biased $+0.16\pm0.06$ from the magnitudes given in \cite{ding2025shellqsjwstunveilshostgalaxies}. We therefore subtract this offset from the modified slit magnitudes measured by \texttt{photutils} to correct for this systematic bias and propagate the additional systematic error through to our final photometric AB magnitudes, causing the photometric magnitudes ($\rm{m_{AB}} \:\text{phot}$) presented herein to have slightly different values and significantly larger error bars than those given in \cite{ding2025shellqsjwstunveilshostgalaxies}. 

We plot the spectroscopic and photometric AB magnitudes of each system with their error bars in Figure \ref{fig:ABmags} along with a dash-dot gray line showing the $x=y$ line. Each point is colored according to its redshift. The photometric error bars are propagated from the signal-to-noise ratio maps of the photometry and the error induced by the systematic offset, and the spectroscopic error bars come from the noise per pixel of the spectrum. The spectroscopic and photometric magnitudes agree within $3\sigma$ in all cases except J2236+0032. We note that J2236+0032 has one of the higher host-to-total light ratios ($\rm{\frac{F_{host}}{F_T}=29\%}$) as measured from photometric decomposition of \cite{ding2025shellqsjwstunveilshostgalaxies}. This indicates that there is a significant amount of stellar flux in the central region (i.e. within $\sim 0''.2$ of the quasar) of J2236+0032 which, as mentioned in \S \ref{subsec:applying PSF}, our PSF model is incorrectly attributing to the quasar. Despite this, the overall agreement in Figure \ref{fig:ABmags}, indicates that our method and the method of \cite{ding2025shellqsjwstunveilshostgalaxies} give similar results. Since the two subtraction methods were performed independently, this reinforces the accuracy and robustness of the spectral and photometric decomposition of the quasar and host in both works. 
%the the three systems with the most significant differences between their $\rm{m_{AB}}\:\text{spec}$ and $\rm{m_{AB}}\:\text{phot}$, J2236+0032, J1512+4422, and J0217$-$0208, with $\Delta m_{AB}=3.9\sigma$, $2.2\sigma$, and $1.1\sigma$, respectively, are the systems with the highest host-to-total light ratios ($\frac{F_{host}}{F_T}=29\%$, $36\%$, and $69\%$, respectively) as measured from the photometric decomposition of \cite{ding2025shellqsjwstunveilshostgalaxies}.

\subsubsection{Redshift}\label{subsec:redshift corroboration}
All spectral lines in our 1D host spectra are fit simultaneously and are required to have the same redshift as the $\rm{[OIII]}5008$ emission line (chosen since it is present in all of our spectra which have robustly detected emission lines). The best-fit redshifts are listed in Table \ref{tab:corroboration} along with the best-fit redshifts from \cite{OnoueInPrep}, which were similarly measured using simultaneous fits to all emission lines. The redshift error bars are propagated from the errors on the centroid of the $\rm{[OIII]}5008$ line in both fits. The redshifts measured from the extended emission and quasar+host spectra agree to within $3 \sigma$ in all cases (see Column 8, $\rm{\frac{\Delta z}{\sigma}}$), except two, J1146+0124 and J0844$-$0132. For the other seven systems, the agreement between the redshifts measured in the galaxy and quasar spectra indicates there are no high-velocity outflows or extremely asymmetric distributions of emission line regions which would cause a significant offset in the line centers between the quasar+host spectra and the host-only spectra. 

The $\rm{[OIII]5008}$ line center measured in the extended emission spectrum of J1146+0124 is $132\pm19\:\text{km/s}$ redward of that measured from the quasar$+$host spectrum, constituting a $7.2\sigma$ difference. This offset is large enough that it can be seen in the shift of the center of the $\rm{H}\alpha$ line between the quasar+host and host spectra in the right inset of panel $(c)$ in Figure \ref{fig:J1146124 spectrum}. J1146+0124 also has a significant difference ($13.1\sigma$) in the measured FWHM of the $\rm{[OIII]}5008$ emission line in the quasar+host and extended emission spectra. Combined with the difference in measured redshift, we suggest that J1146+0124 contains a high-velocity outflow in its central region which significantly blueshifts the $\rm{[OIII]}5008$ line measured in the full quasar+host spectrum. Given that this blueshift is undetected in the extended emission spectrum, we infer that this high velocity outflow does not extend farther than $1.14\: \text{kpc}$ (i.e. $0''.2$) from the quasar itself. %Masafusa says it has an eddingtn ratio "typical of the other cycle 1 targets" [\textit{9/19, on Toshi Kawagushi's suggestion I asked Masafusa for the Eddington ratio of this system so that we might draw a connection between high eddington ratios and OIII outflows across redshifts. Such a discussion would go here.}]%between the line centers measured from the quasar+host and pure host spectra. of $6.9\sigma$ $(930\pm130\:\text{km/s})$ and $5.1\sigma$ $(1080\pm210\:\text{km/s})$, 

For J0844$-$0132, the difference in redshifts between the $\rm{[OIII]}5008$ centroid measured from the quasar+host spectrum \citep{OnoueInPrep} and the extended emission (this work) corresponds to a redward shift of $42\pm8$ km/s ($5\sigma$). Given this is significantly smaller than the instrument resolution ($\sim170$ km/s), we do not find this as convincing an offset as that measured for J1146+0124. Similarly to J1146+0124, the central FWHM of J0844$-$0132 is measured to be $7.1\sigma$ broader than that measured in the extended emission, and, as seen in panels $(b)$ and $(e)$ of Figure \ref{fig:J844132 spectrum} and discussed in \S \ref{subsec:asymmetry}, J0844$-$0132 has somewhat spatially asymmetric emission lines that do not follow the spatial distribution of starlight inferred from the photometry. These differences in redshifts, FWHM, and emission line spatial distribution may also point to a high-velocity outflow in J0844$-$0132, though the evidence is weaker for this system than J1146+0124.   %except for J1146+0124, which has a difference of deltaz = , corresponding to a  sigma difference. A possible reason for this significant difference is an unresolved central outflow that may cause the quasar's redshift to appear offset from that of its host galaxy. In addition, error bars on these values come from the fitting functions and do not include any estimation of systematic errors. 

\subsubsection{Line width of \rm{[OIII]5008}}\label{FWHM corroboration}
We measure the $\rm{[OIII]}5008$ Full Width Half Maximum (FWHM) in km/s for all systems with extended emission lines, as shown in Column 8 of Table \ref{tab:corroboration}, and compare them to the values calculated by \cite{OnoueInPrep} for quasar+host (Column 9 of the same Table). In order to fairly compare our values with \cite{OnoueInPrep}, we do not include a correction for the instrumental resolution, which is $\sim170$ km/s at the wavelengths of the $\rm{[OIII]}5008$ line for $\rm{6<z<6.4}$. The error bars for both columns are propagated from the errors on the line widths from the best fit emission line models from this work and \cite{OnoueInPrep}. Our FWHM agree with the values calculated by \cite{OnoueInPrep} to within $3\sigma$ in all cases except for J0911+0152, J1146+0124, and J0217$-$0208 (see Column 10 of Table \ref{tab:corroboration}). For J0911+0152, given the low quality of the emission line fit to the $\rm{[OIII]5008}$ line, we attribute the difference in measured FWHM to the poor fit rather than any physical effect. For J1146+0124, the line width measured in the outer regions of the galaxy is $\Delta\text{FWHM} = -320 \pm 25$ km/s $(13.1\sigma)$ smaller than that measured in the central region, implying a spatially compact $\rm{[OIII]}5008$ outflow as discussed above in $\S$\ref{subsec:redshift corroboration}. %This implies that the central region of J1146+0124 may have an outflow that broadens the $\rm{[OIII]}5008$ line. This agrees with the offset measured in the redshift of J1146+0124 between the extended emission and quasar+host spectrum $\Delta z = +0.0031 \pm 0.0004$ $(6.9\sigma)$, since an outflow pointed towards the observer would shift the line center slightly blueward, making the outer regions of the galaxy appear slightly redshifted in relation to the center of the galaxy. Given the narrowness of the $\rm{[OIII]}5008$ line in the extended emission, we infer that this outflow does not extend into the outer regions of the galaxy. Since the full quasar+host spectra have much higher signal-to-noise ratio and smaller error bars than the extended emission, we leave any further discussion of the line widths from the center of the galaxies to \cite{OnoueInPrep}. 

In the case of J0217$-$0208, the $\rm{[OIII]}5008$ line width measured in the outer regions of the galaxy is $\Delta\text{FWHM} = +101 \pm 13$ km/s $(8.1\sigma)$ greater than that measured in the central region of the galaxy. This broadening may be caused by rotation, which would affect the lines more strongly in the outer regions of the galaxy than in the inner regions, assuming a disk-like structure and a rotation curve typical of a spiral galaxy (see, for example, \cite{Corbelli2000MNRAS.311..441C}). Rotation as the cause of the difference, as opposed to an outflow, is further supported by the spatial symmetry of the $\rm{[OIII]}5008$ line as seen in panel (\textit{e}) of Figure \ref{fig:J0217 spectrum} and the consistency of the measured line center between the PSF+host spectrum and the extended emission spectrum, for which the difference is $0\pm21$ km/s. While outflows often appear spatially one-sided (the far side is usually hidden by the bulk of the galaxy) and shift the $\rm{[OIII]}5008$ line center, a coherently rotating galactic disk with a uniform distribution of star-forming regions would appear spatially symmetric and only broaden, rather than shift, the emission lines, so long as star-forming regions both sides of the galaxy fall within the slit. In order to test this hypothesis, we spatially decompose J0217$-$0208 into the spectrum on either side of the galaxy and use the line fitting algorithm described in \S \ref{subsec:line fits and ratio} to fit the spectrum of each side separately. We find that the $\rm{[OIII]}5008$ line center is shifted by $\Delta \rm{v}= 81 \pm 9$ km/s from one side of the galaxy to the other, suggesting that J0217$-$0208 is a rotating, disk-like system with maximum speed along the line of sight of $41 \pm 6$ km/s. Approximating the system as a circular disk with an effective radius of $1.02 \pm 0.17$ kpc, we use the apparent semimajor to semiminor axis ratio to infer an inclination of $37\pm4^\circ$, where the axis ratio and effective radius come from the best fit photometric model of \cite{ding2025shellqsjwstunveilshostgalaxies}. Given the slit position is rotated from the major axis by approximately $17 \pm 5^\circ$, we infer a maximum rotation speed of $52\pm12$ km/s at the effective radius. 

For completeness, we performed an equivalent analysis on all systems with extended emission lines with a signal-to-noise ratio that was high enough on both sides of the galaxy to allow for separate line fitting. This condition was met for J1512+4422, J0844$-$0132, and J2255+0251, which had $\rm{[OIII]}5008$ line center offsets of  $\Delta\rm{v} = 16 \pm 33$ km/s, $\Delta\rm{v} = 240 \pm 80$ km/s, and $\Delta\rm{v} = 40 \pm 54$ km/s, respectively. The offsets in J2255+0251 and J1512+4422 are consistent with zero, indicating a lack of rotation \citep{Shao2022A&A...668A.121S} or fast galaxy-scale outflows, which would create similar signatures. The J0844$-$0132 line center offset is significant to $3\sigma$ and is presumably caused by the outflow discussed in $\S$ \ref{subsec:redshift corroboration}. 
%From the lack of detection of offset OIII line centers either side of the galaxy, we can assume  place weak upper limits on host rotation. Our wavelength resolution when using only half of the available data is approximatly () assum all galactic rotation is  and place much tighter constraints on the maximum outflow and/or rotation speed of our systems. 
%We see no signs of outflows or rotation from any of the emission lines, though this may be due obscuration or orientation effects. This lack of detection, assuming optimal orientation and minimal obscuration, places an upper limit on the rotation of the galaxies, since past a certain rotation speed, we would expect to see the effects of the rotation. While the low signal-to-noise makes this a weak constraint, nevertheless, we can conclude the galaxy has a maximum rotation along the line of sight of . [Describe what this means] 

\subsection{Is J0217-0208 an AGN?}\label{subsubsec:J0217AGNorGal}
J0217$-$0208 was identified, along with 11 other sources, as a quasar candidate in \cite{Matsuoka2018PASJ...70S..35M} based on the presence of a luminous and narrow Ly$\alpha$ line $\rm{(L (Ly\alpha)} \geq 10^{43} \:\rm{erg \: s^{-1}})$. \cite{Matsuoka2018PASJ...70S..35M} acknowledged that, despite the luminosity of the Ly$\alpha$ line, without a measured broad-line component the J0217$-$0208 system (as well as other narrow-line objects in this sample) would require follow-up observations to determine the presence of an AGN. These follow-up observations were published in \cite{Matsuoka2025ApJ...988...57M} and found that two of the systems were unambiguously AGN, while the rest, including J0217$-$0208, were better classified as galaxies. With these follow-up observations in hand, we now discuss the nature of the J0217$-$0208 system. 

\cite{ding2025shellqsjwstunveilshostgalaxies} found that the system shows extended emission with a possible unresolved point source at the center of the galaxy. Since the NIRCam F356W PSF at $\rm{z}=6$ corresponds to a size of $0.68$ kpc, this point source could indicate an AGN (either narrow-line or broad-line), a dense star-forming region, or simply a dense stellar core. In order to determine the presence of a broad line region, the 1D spectrum of the central region (i.e. not including the extended emission) of J0217$-$0208 was fit by \cite{OnoueInPrep} and \cite{Matsuoka2025ApJ...988...57M} using independent line fitting codes. \cite{OnoueInPrep} use the \texttt{PyQSOFit} code and finds that the best fit model includes a faint broad $\rm{H}\alpha$ component, whereas \cite{Matsuoka2025ApJ...988...57M} use a custom fitting code. The combined fits for both models are very similar with nearly equivalent residuals, making the presence of a broad line region in J0217$-$0208 ambiguous. We also note that the spectrum does not show detectable higher energy ionization lines associated with AGN, such as $\rm{HeII}$, nor are stellar absorption lines seen in the full spectrum. 

Without conclusive detection of the broad-line region, we must attempt to differentiate a compact star-forming region from an AGN. At low redshifts, this would be the purview of the BPT diagram; however, as discussed in \S \ref{subsubsec:BPT}, the R3N2 BPT diagram is unable to distinguish star formation and AGN activity at high redshifts. As an alternative, we perform the spectral analysis with and without the PSF component removed from the 2D spectrum and compare their positions on the BPT diagram. If the central PSF was powered by an AGN, we might expect the line ratios to change significantly when the PSF is removed. Conversely, if the central PSF is powered by star formation with the same stellar and ISM metallicity as the outer regions of the galaxy, we would expect the line ratios to be unchanged when the PSF is removed. As seen in Figure \ref{fig:BPT} (where the points representing J0217$-$0208 have blue borders), removing the PSF component does not significantly change the ionization line ratios. However, this stability in ionization line ratios is also seen in two broad line systems, J1512+4422 and J1425$-$0015, for which we measure emission line ratios in the central and extended spectra (as seen in the upper right of Figure \ref{fig:BPT}), implying that line ratio stability does not necessarily indicate a lack of a central AGN. 

Given these measurements, we conclude that the J0217$-$0208 is likely a system dominated by star formation with a possible subdominant AGN contribution. The presence of an unobscured broad line region is observationally ambiguous, but we do not rule it out. Reflecting this uncertainty, we calculate the star formation rate of J0217$-$0208 with and without the central PSF removed and report both values in the relation between stellar mass and SFR (Figure \ref{fig:SFMS}). In Figure \ref{fig:SFMS}, the position of J0217$-$0208 moves further from the SFMS and towards the starburst regime when the central PSF is included (represented by the cyan hexagon). This implies that, if the central PSF is not a quasar, it is likely a strongly star-forming core of a young galaxy.  

\section{Conclusions} \label{sec:conclusion}

We analyze the extended emission of twelve low-luminosity $(\rm{M_{1450}}>-24)$ quasars at $\rm{z}=6$ using the NIRSpec Fixed Slit observing mode with the G395M/F290LP grism/filter configuration. In order to separate quasar and host emission, we use a calibration star (ID 1808347) to characterize the Fixed Slit spatial PSF, which varies significantly with wavelength. We model the PSF as a sum of two linked Gaussians plus smoothed residuals, and find that this model leaves residuals $<0.05\%$ when applied to the calibration star. Quasar systems with marginal or no host emission detected in photometry, such as J0911+0152 and J1146$-$0005, also have no extended emission in the spectrum, confirming the robustness of the technique. We fit the PSFs using this two-Gaussian model and subtract them from the 2D spectra, revealing extended continuum emission from the host galaxy and extended narrow emission lines, which may be from the galaxy or the narrow-line region of the quasar. Using this method, we see extended continuum and/or line emission in ten of the twelve systems and, assuming that the extended emission is not heavily contaminated by the NLR of the quasar, are able to measure several spectral properties of the host galaxies, including ionization line ratios, star formation rates, and $A_V$ magnitudes. We compare these values with systems from the literature to track the evolution of quasar host galaxies with redshift.

The main findings of this work are as follows.
\begin{itemize}
\item We successfully model and then remove quasar emission from the 2D spectra of our targets (see Figures \ref{fig:J0911 spectrum}-\ref{fig:J2236 spectrum}), revealing extended emission lines in eight of twelve cases, proving the effectiveness of the NIRSpec Fixed Slit in separating the PSF from extended emission at $z \sim 6$.

\item We measure robust $(>3\sigma)$ stellar continuum in five of twelve systems, weak $(2-3\sigma)$ stellar continuum in two of twelve systems, and non-detections $(<2\sigma)$ in the remaining five systems. These detection strengths match those found by \cite{ding2025shellqsjwstunveilshostgalaxies} (i.e. our strongest detections match their strongest detections, and vice versa), verifying both works. 

\item We extract 1D spectra and measure ionization line ratios for the extended emission of eight of the twelve systems, placing those with robust line measurements on the traditional BPT diagram alongside various sources from the literature (Figure \ref{fig:BPT}). We find that the ionization line ratios of the extended emission reside at slightly higher $\text{log}_{10}(\rm{[OIII]}5008/H\beta)$ values than low-redshift AGN, in agreement with other high-redshift quasar host galaxies and star-forming galaxies \citep{Harikane2023a, Marshall2025}. 

\item Using the extended $\rm{H}\alpha$ line flux and total mass as measured by \cite{ding2025shellqsjwstunveilshostgalaxies}, we measure upper limits on the total star formation rates for eight of the twelve systems and place them on the star-forming main sequence plot (Figure \ref{fig:SFMS}). Assuming no AGN contamination of the extended narrow lines, all of our AGN host galaxies lie along the star-forming main sequence except for J2236+0032, which shows post-starburst features and lies below the star-forming main sequence, and J0217$-$0208, whose ionization lines may be powered by intense star formation rather than a central AGN, causing it to lie above the SFMS. This is in agreement with \cite{Izumi2018}, who measured SFRs for SHELLQs quasars with ALMA and found that they tend to lie along the SFMS, unlike the host galaxies of high luminosity AGN. %Since $\rm{H}\alpha$ is only produced by O, B, and A type stars, which have typical lifetimes of 10 Myr, and the majority of our host galaxies lie on or below the SFMS, we infer that feedback mechanisms from our low-luminosity AGN do not act to increase star formation rates on shorter.

\item  Spatial asymmetry of emission lines is common but not extreme in our systems, with several showing one-sided emission lines. 

\item Two systems with asymmetric emission lines, J1146+0124 and J0844$-$0132, show velocity shifts of $132\pm19\:\text{km/s}$ $(7.2\sigma)$ and $42\pm8\:\text{km/s}$ $(5.0\sigma)$, respectively, in their $\rm{[OIII]}5008$ line centers between their quasar+host and extended emission spectra. In addition, the $\rm{[OIII]}5008$ line profiles of the extended emission component of J1146+0124 are narrower by $-320 \pm 25$ km/s $(13.1\sigma)$ than the narrow $\rm{[OIII]}5008$ profiles of the quasar+host component. From these measurements we infer an $\rm{[OIII]}5008$ outflow in J1146+0124, while we find the evidence too weak to make such a claim in J0844$-$0132.

\item We measure an $\rm{[OIII]}5008$ line offset of $\Delta \rm{v} = 81 \pm 9$ km/s on opposite sides of J0217$-$0208, giving a rotation speed of $52\pm12$ km/s relative to the center assuming a circular disk inclined by $37\pm4^\circ$. We also find that the line ratios and star formation rate of J0217$-$0208 imply a system dominated by star formation rather than an AGN. 

\item Due to chance alignment, we are able to measure the spectrum of a $z=4.96$ foreground galaxy in the field of J1512+4422, and find it has line ratios typical of a high redshift star-forming galaxy, and a star formation rate in the slit of $19.4\pm0.3 \: \rm{M_{\odot}/yr}$.

\item Independently from \cite{ding2025shellqsjwstunveilshostgalaxies} and \cite{OnoueInPrep}, we measure the AB magnitudes, redshift, and emission line FWHM of the extended emission, which we attribute to the host galaxy with the caveats given in the text. When compared to these works, our values agree within $3\sigma$ in seven of 12 cases, with the five exceptions (J2236+0032, J1146+0124, J0844$-$0132, J0911+0152, and J0217$-$0208) accounted for by low signal-to-noise ratios, outflows, or rotation, as explained above. This agreement between independent analyses validates the robustness of the quasar-host decomposition in all three papers. \end{itemize}

Our results indicate that quasar host galaxies have properties similar to those of star-forming galaxies at $\rm{z}\sim6$. Given the prevalence of SMBH in the centers of galaxies at low redshift and the increasingly apparent need for periodic and rapid SMBH growth in the early universe \citep{Onoue2024arXiv240907113O}, this consistency is perhaps a natural consequence of the majority of galaxies in the early universe hosting an active galactic nucleus approximately once every 100 $\rm{Myr}$. In this picture, most galaxies host an AGN at least once, if not several times, during their evolution, meaning those we see with AGN are a random sample of the galaxy population, rather than a separate subpopulation. However, the environments of quasars are known to differ from those of regular star-forming galaxies (for example, they tend to reside in high-mass dark matter halos \citep{Shen2007AJ....133.2222S, timlin2018ApJ...859...20T, arita2025MNRAS.536.3677A}), implying that some inherent differences between star-forming and AGN-hosting galaxies may be present. Larger samples at high redshifts would need to be studied in order to see population-wide trends in AGN hosts compared to regular star-forming galaxies. 

Such surveys are ongoing in the community (for example, JADES \citep{Isobe2025MNRAS.541L..71I}, CEERS \citep{Cole2025}, and others), and many include some level of spectral follow-up, either with the JWST NIRSpec multi-shutter array (MSA) or the integral field unit (IFU). The methods described herein for modeling and subtracting the spatial PSF in the Fixed Slit mode are quite powerful for extracting quasar host galaxy information without the use of the more time-intensive IFU, and such methods may be usable in concordance with creative methods such as MSA slit-stepping \citep{Ju2025ApJ...978L..39J}. In addition, the major weakness of our current method is the inability to model the quasar and galaxy simultaneously, causing the galaxy to be oversubtracted. We have begun developing updated code which resolves this issue, and we hope to apply our improved decomposition method to these larger samples of galaxies to better probe the AGN host population at high redshifts. It is also possible that the proper subtraction of the quasar using the spatial information of the long slit spectrum provides an avenue for separating narrow-line quasars into their AGN and stellar components at high redshifts where the BPT R3N2 diagnostic diagram does not work. This may be particularly useful for the sample of quasar candidates with narrow $\rm{Ly}\alpha$ explored by \cite{Matsuoka2025ApJ...988...57M}, whose work is already bridging the gap between high- and low-luminosity AGN systems at high redshift. We emphasize the need for medium- and high-resolution spectroscopy at high redshifts in order to probe the connection of quasars and galaxies, and to determine the timescales on which AGN feedback acts in the early universe.

\section*{Acknowledgements}
We would like to thank our peers at Princeton University and Kavli IPMU for enlightening conversations which helped greatly in the development of this work, especially David Setton, Shaunak Modak, Phillipe Yao, Yilun Ma, Helena Treiber, Sarah Thiele, Rodrigo Cordova-Rosario, and Nora Linzer. 
This work is based (in part) on observations made with the NASA/ESA/CSA James Webb Space Telescope. The data were obtained from the Mikulski Archive for Space Telescopes at the Space Telescope Science Institute, which is operated by the Association of Universities for Research in Astronomy, Inc., under NASA contract NAS 5-03127 for JWST. These observations are associated with program $\#1967$.
Support for program $\#1967$ was provided by NASA through a grant from the Space Telescope Science Institute, which is operated by the Association of Universities for Research in Astronomy, Inc., under NASA contract NAS 5-03127. M.O. is supported by the Japan Society for the Promotion of Science (JSPS) KAKENHI Grant Number 24K22894. M.S. was supported by JST SPRING, Grant Number JPMJSP2108. CLP and MAS were supported by the following NASA grants through STScI: JWST-GO-01967.002-A, JWST-GO-03859.002-A, and JWST-GO-03417.002-A. 

\clearpage
\appendix

\section{Classical Strömgren sphere estimation for J2236+0032 and J1512+4422}\label{appendix:Stromgren}
In order to determine whether the narrow lines measured in the extended emission of our systems come from the quasars or the host galaxies, we roughly calculate the maximum ionizing radius of the two systems, J1512+4422 and J2236+0032, for which we have the quasar luminosity \textit{without} host contribution, as given in \cite{Onoue2024arXiv240907113O}. We assume a uniform ISM of neutral hydrogen number density $\rm{n_H}$ and temperature $10^4 ~\rm{K}$, quasar lifetimes longer than the reionization time $\rm{(t_Q>\frac{1}{\alpha_Bn_H}})$, where $\rm{\alpha_{B}}$ is the classical Case B recombination coefficient, and quasar UV spectra consistent with the fiducial composite UV spectra given in \cite{Cai2023NatAs...7.1506C}.  Taking into account the decrease in the hydrogen ionization cross-section with increasing energy above $13.6~\rm{eV}$, we find that the classical Strömgren radius for J1512+4422 is $6.9$ kpc $(7\times \rm{R_{eff}})$ for $\rm{n_H}=0.6 ~\text{cm}^{-3}$ (warm neutral medium, \citeauthor{Draine2011piim.book} \citeyear{Draine2011piim.book}) and $0.51$ kpc $(0.5\times \rm{R_{eff}})$ for $\rm{n_H}=30 ~\text{cm}^{-3}$ (cold neutral medium, \citeauthor{Draine2011piim.book} \citeyear{Draine2011piim.book}). We repeat the calculation for J2236+0032 and find Strömgren radii of $4.4$ kpc $(7\times \rm{R_{eff}})$ and $0.33$ kpc $(0.5\times \rm{R_{eff}})$ for $\rm{n_H}$ of $0.6 ~\text{cm}^{-3}$ and $30 ~\text{cm}^{-3}$, respectively. The effective radii are measured by \cite{ding2025shellqsjwstunveilshostgalaxies}, and the inferred Strömgren radii range from about half an effective radius to seven times larger, depending on the neutral hydrogen fraction in the host galaxies. Since we cannot constrain such a value given our dataset, we simply acknowledge that the narrow $\rm{H}\alpha$ line seen in extended emission may be excited by the quasar, star formation, or a mix of both. As a result, all SFRs calculated in the text should be considered upper limits in which all ionizing photons have been attributed to star formation.

\section{J1512+4422 interloper}\label{appendix:J1512 companion}
\begin{figure}[hb]
    \centering
    \includegraphics[width=0.6\linewidth]{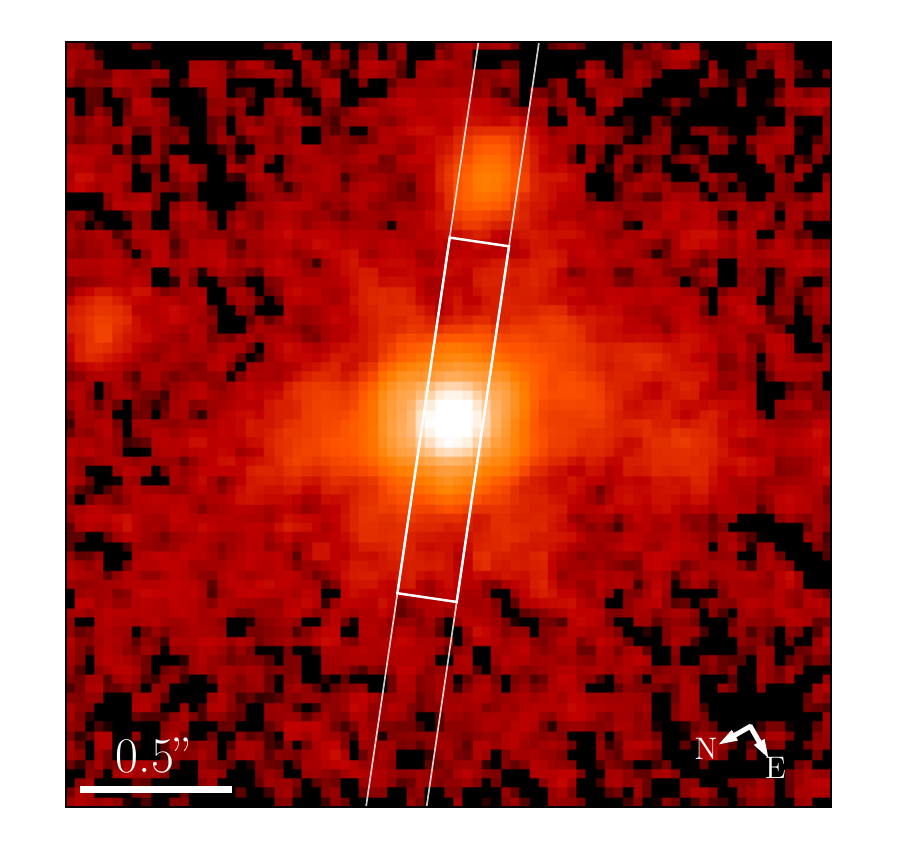}
    \caption{Photometry of J1512+4422 with the quasar and foreground galaxy included. The spatial positioning of the full NIRSpec Fixed Slit is shown, with the region used for the spectral analysis of J1512+4422 outlined by a shorter rectangle within the full slit length.}
    \label{fig:interloper_phot}
\end{figure}

\begin{figure*}[b]
    \centering
    \includegraphics[width=1\linewidth]{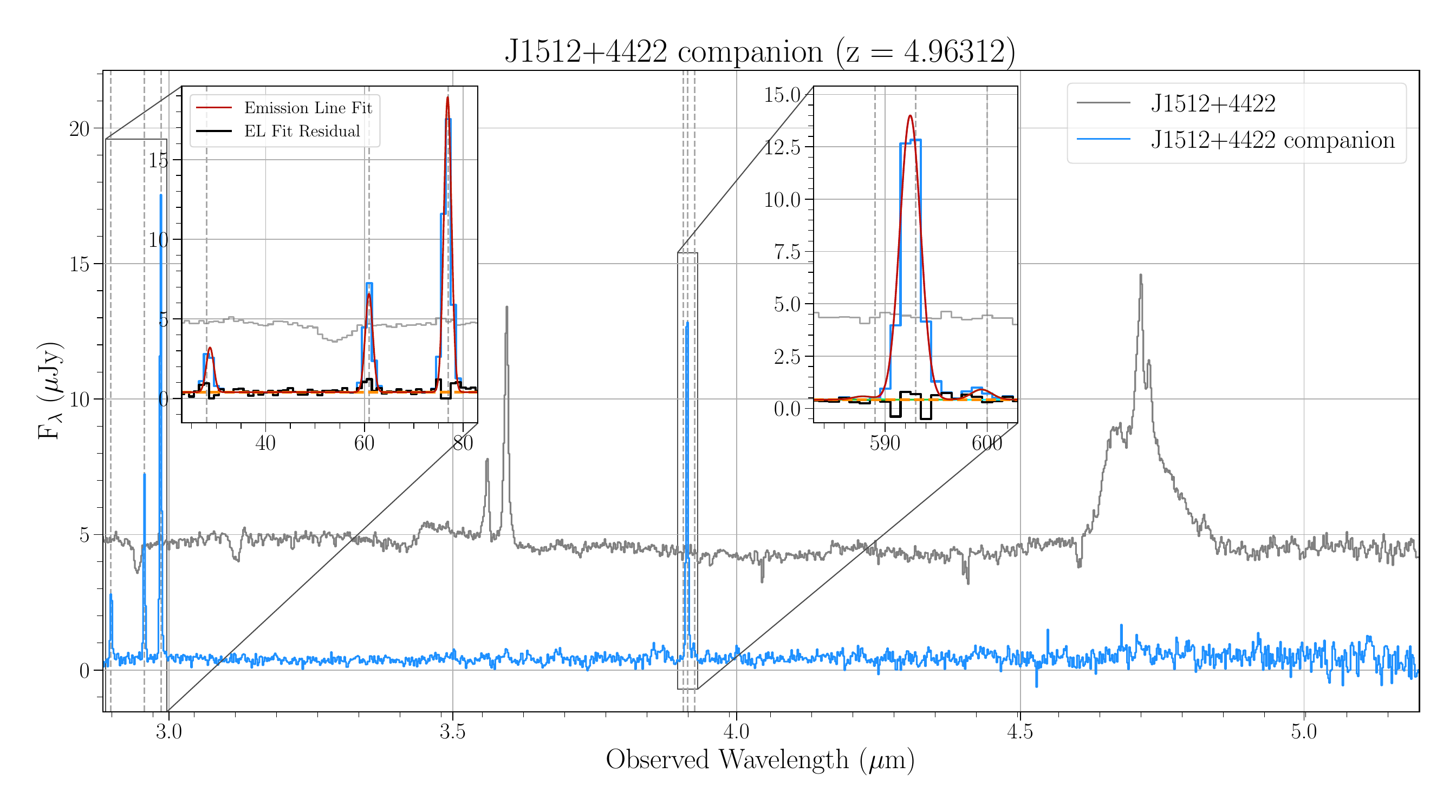}
    \caption{1D extracted spectrum of the interloper (light blue) near J1512+4422, along with the full quasar+host spectrum of J1512+4422 itself (gray). Best fit emission line profiles are overlaid in the inset plots and wavelengths of prominent emision lines at $z=4.96312$ are marked by vertical dashed lines. }
    \label{fig:interloper_spec}
\end{figure*}

Figure 1 of \cite{ding2025shellqsjwstunveilshostgalaxies} shows that the J1512+4422 system appears to have a companion of magnitude $\rm{m_{AB}}=25.3$, effective radius $\rm{R_{\text{eff}}}=0.52$ kpc, and Sérsic index $\text{n}=1$, separated by approximately $0''.75$ from the position of the PSF center. The companion galaxy is seen in the F356W filter but not the F150W filter and was not seen in HSC ground-based observations. By chance and good fortune, the NIRSpec slit was oriented so that it captured this companion galaxy, allowing us to extract the spectrum shown in Figure \ref{fig:interloper_spec}. We use the same line-fitting code as described in \S \ref{subsec:line fits and ratio} to fit the emission lines of this spectrum. 

We find that the galaxy is not a true companion, but a foreground object at $z=4.96312\pm0.00003$. From the best-fit emission line model, we measure line ratios of $\rm{[OIII]}5008/H\beta$ = $6.22\pm0.50$ and$\rm{[NII]}6585/H\alpha$ = $0.036\pm0.013$ and line widths of $139\pm2$ km/s. We measure a Balmer decrement of $\rm{H}\alpha/\rm{H}\beta$=$2.66\pm0.21$, corresponding to an extinction of $-0.25\pm0.28$ (i.e. consistent with zero). The continuum at $5100~\mathring{\rm{A}}$ is $3.98\pm 0.38\times10^{-7}$ Jy, which corresponds to an apparent magnitude of m$_{\rm{AB}}=24.9 \pm 0.1$ and constitutes a $10.6\sigma$ detection. The galaxy colors measured from the 1D spectrum including emission lines are $g-r=-0.44\pm0.02$ and $r-i= -0.06\pm0.02$. When the emission lines are excluded, the colors are $g-r=0.03\pm0.03$ and $r-i= 0.04\pm0.02$, indicating that the emission lines are strongly boosting the flux in the blueward filters. The star formation rate within the slit as estimated from the $\rm{H}\alpha$ luminosity is $19.4\pm0.3$ $\rm{M_{\odot}/yr}$. Since approximately $\sim80\%$ of the flux from the companion galaxy falls within the slit, this corresponds to a total SFR of $24.3\pm0.4$ $\rm{M}_{\odot}/\rm{yr}$ for the whole galaxy, assuming a constant stellar mass-to-surface brightness ratio. The approximate size of the semimajor axis of the galaxy on the detector, measured by eye, is $\sim0''.25$, which corresponds to $\sim1.5$ kpc at $\rm{z}=4.96$. Comparing to galaxies at a similar redshift from the COSMOS2020 survey \citep{Weaver2022}, this companion is somewhat bluer than the average $\rm{z}\approx5$ galaxy, but is otherwise completely average in its approximate size, morphology, and star formation rate. The strong blue $g-r$  color is likely due to the substantial equivalent width of the $\rm{[OIII]}$ doublet $(\sim350~\mathring{\rm{A}}, \lambda_{\text{rest}})$ when compared to the portion of the spectrum covered by the $g$ band $(\sim94~\mathring{\rm{A}}, \lambda_{\text{rest}})$. We conclude that this object is a moderately sized, blue-ish, star-forming galaxy at $\rm{z}\sim5$ with minimal dust and no visible structure or deformation. It is not particularly unusual in any way that we can measure, and we leave further investigation of this object to future works.

\clearpage
\bibliography{Exem2025}{}
\bibliographystyle{aasjournal}

\end{document}